\documentclass{article}
\usepackage{amsfonts}
\usepackage{amsmath}
\usepackage{amssymb}
\usepackage{bm}
\usepackage[compress]{cite}
\usepackage{dcolumn}
\usepackage{graphicx}
\usepackage{graphics}
\usepackage{latexsym}
\usepackage{rotating}
\usepackage{varioref}
\usepackage{color}
\usepackage{textcomp}
\usepackage{subfig}
\usepackage{float}
\usepackage{booktabs,multirow,array}
\usepackage{float}
\usepackage{xspace} 
\RequirePackage[colorlinks,citecolor=blue,urlcolor=magenta,linkcolor=blue]{hyperref}
\allowdisplaybreaks
\def\scc{strong cosmic censorship conjecture}

\def\gr{general relativity}
\def\RN{Reissner-Nordstr\"{o}m}

\def\KN{Kerr-Newmann }

\def\qnm{quasi-normal }

\def\NE{near extremal }
\def\dS{de Sitter }

\labelformat{section}{Section #1} 
\labelformat{subsection}{Section #1} 
\labelformat{subsubsection}{Section #1}
\labelformat{subsubsubsection}{Section #1}
\labelformat{equation}{Eq.~(#1)} 
\labelformat{figure}{Fig.~#1} 
\labelformat{subfigure}{Fig.~\thefigure#1} 
\labelformat{table}{Table.~#1} 
\labelformat{appendix}{Appendix #1}
\title{Strong cosmic censorship conjecture with NUT charge and conformal coupling}  
\author{Mostafizur Rahman\footnote{mostafizur@ctp-jamia.res.in} $^{1}$, Soumodeep Mitra\footnote{soumodeepphysics@gmail.com} $^{2}$ 
and Sumanta Chakraborty\footnote{sumantac.physics@gmail.com} $^{2,3}$\\
{$~^{1}$\small{Center for Theoretical Physics, Jamia Millia Islamia, New Delhi-110025, India}}\\
{$~^{2}$\small{School of Physical Sciences, Indian Association for the Cultivation of Science, Kolkata 700032, India}}\\
{$~^{3}$\small{School of Mathematical and Computational Sciences}}\\
{\small{Indian Association for the Cultivation of Science, Kolkata 700032, India}}}
\begin{document}

\maketitle

\begin{abstract}

Strong cosmic censorship conjecture is central to the deterministic nature of general relativity, since it asserts that given any generic initial data on a spacelike hypersurface, the future can be uniquely predicted. However, recently it has been found that for charged black holes in asymptotically de Sitter spacetimes, the metric and massless scalar fields can be extended beyond the Cauchy horizon. This spells doom on the strong cosmic censorship conjecture, which prohibits precisely this scenario. In this work we try to understand the genericness of the above situation by studying the effect of NUT charge and conformally coupled scalar field on the violation of strong cosmic censorship conjecture for charged asymptotically de Sitter black holes. We have shown that even in the presence of the NUT charge and a conformally coupled scalar field strong cosmic censorship conjecture in indeed violated for such black holes with Cauchy horizon. Moreover, the presence of conformal coupling makes the situation even worse, in the sense that the scalar field is extendible across the Cauchy horizon as a $C^{1}$ function. On the other hand, the strong cosmic censorship conjecture is respected for conformally coupled scalar field in rotating black hole spacetimes with NUT charge. This reinforces the belief that possibly for astrophysical black holes, strong cosmic censorship conjecture is respected, irrespective of the nature of the scalar field. 

\end{abstract}
\section{Introduction}

Existence of a well posed initial value problem is central to the success story of general relativity. This means, by specifying physically reasonable initial data on a spacelike hypersurface $S$, also known as the Cauchy hypersurface, the dynamics of the same can be uniquely determined from Einstein's field equations within its domain of dependence $D(S)$ \cite{Wald:1984rg,Hawking:1973uf}. It is generally expected that in all physically reasonable spacetimes, $D(S)$ covers the entire spacetime region. In other words, by specifying the initial data on $S$, we should able to predict the entire evolution of the universe. However, for some black hole spacetimes e.g., \RN\ or Kerr black holes, $D(S)$ fails to cover the entire spacetime region. This a priori leads to pathological situations, since even with complete knowledge on the initial conditions at a given Cauchy hypersurface $S$, we are unable predict the future evolution of a gravitating object \cite{Hawking:1973uf}. The boundary of $D(S)$, beyond which the spacetime can be extended, resulting into loss of predictability of the gravitational theory, is called the Cauchy horizon. In this regard, the strong cosmic censorship conjecture can be regarded as a central pillar, asserting the deterministic nature of any gravitational theory, which states that the evolution of the spacetime geometry and matter field beyond the Cauchy horizon is impossible \cite{Wald:1984rg,PhysRevLett.14.57,nla.cat-vn3002454}. The formulation of this conjecture is motivated from the observation that the Cauchy horizons suffer from blue shift instability, i.e., the stress energy tensor for various in-falling matter fields diverge at the Cauchy horizon \cite{1973IJTP....7..183S, PhysRevD.41.1796, Dafermos:2003wr, Dafermos:2012np}. Poisson and Israel computed back-reaction of these diverging stress energy tensor on the internal geometry of a \RN\ black hole in sufficiently late times \cite{PhysRevD.41.1796} and have shown that Kretschmann scalar diverges at the Cauchy horizon, leading to a singular behaviour at the Cauchy horizon. The above exercise illustrates that it is not possible to extend the spacetime geometry across the Cauchy horizon with a $C^{2}$ metric \cite{Costa:2014yha,Costa:2017tjc}. However, this does not rule out weaker extension of spacetime geometry across the Cauchy horizon. In particular, following the argument of Ori it follows that the singularity at the Cauchy horizon is rather weak since tidal forces there may not be strong enough to tear an extended object apart and thus possibly allowing a safe passage through it \cite{PhysRevLett.67.789}. As a result a stronger condition on the regularity of the metric is necessary to prevent such passage of an extended body across the Cauchy horizon. Christodoulou's version of \scc\ provides such a stronger criterion, which states that no weak extension of the metric is possible beyond the Cauchy horizon with locally square integrable Christoffel symbols \cite{Christodoulou:2008nj,Dafermos2014}. This discards any extension of the solutions of Einstein's field equations beyond Cauchy horizon, even in the weaker sense \cite{Costa:2014yha,Costa:2017tjc,Dafermos:2012np}. A related aspect is the inextendibility of a massless scalar field $\Phi\in H^{1}_{\rm loc}$, where $H^{1}_{\rm loc}$ corresponds to Sobolev space, acting as a toy model to the extendibility of spacetime metric beyond the Cauchy horizon \cite{Franzen:2014sqa,Franzen:2019fex}. This version of \scc\ is proven to be true for various asymptotically flat spacetimes including both \RN\ \cite{Luk:2015qja} and Kerr black holes \cite{Dafermos:2015bzz}. 

However, the situation changes drastically when we consider the effect a positive cosmological constant has on the violation of the \scc \cite{Chambers:1997ef}. Most importantly, residing within the framework of the scalar field model, it has been shown that near the event horizon, scalar field perturbations decay exponentially (the associated decay time scale $\alpha$ corresponds to the longest lived quasi-normal modes \cite{Brady:1996za,Dyatlov:2013hba,Bony2008,Dyatlov2012,Cardoso:2017soq}) in comparison to power law decay in the asymptotically flat case \cite{PhysRevD.5.2419, Dafermos:2014cua, Angelopoulos:2016wcv}. So, there is a possibility that this exponential decay at the event horizon can dominate the effect of the amplification at the Cauchy horizon by an exponential blue shift (the amplification time scale is governed by the surface gravity $\kappa_{-}$ at the Cauchy horizon) \cite{PhysRevD.19.2821, PhysRevLett.67.789, HISCOCK1981110, PhysRevLett.80.3432}. In fact, it has been recently shown that a weak extension of a scalar field beyond the Cauchy horizon of a Reissner-Nordst\"{o}m-\dS black hole is indeed possible if the following condition is satisfied \cite{Dyatlov:2013hba,Bony2008,Dyatlov2012,Cardoso:2017soq}
\begin{equation}\label{SCC_violation}
\beta\equiv\frac{\alpha}{\kappa_{-}}=\frac{\rm{Inf}_{\ell,n}\left(-\textrm{Im} \omega_{\ell,n}\right)}{\kappa_{-}}>\frac{1}{2}~,
\end{equation}
where $\alpha$ is the minimum value for $-\textrm{Im}\omega$ among all possible quasi-normal modes. Note that $\beta>(1/2)$ ensures, the scalar field has a finite energy at the Cauchy horizon, while $\beta>1$ says that the scalar field is extendable across the Cauchy horizon as a $C^{1}$ function, even though the curvature still diverges. Thus $\beta>(1/2)$ leads to a possible violation of the Christodoulou's version of \scc. This has opened up a plethora of possibilities and a quantitative way to see whether \scc\ is violated in a certain spacetime or not. For example, the above assertion was for a scalar field living in the \RN-\dS spacetime, which naturally brings up the question, what happens when other fundamental fields are taken into account, or, when some different spacetime geometry is being considered. Note that the condition presented above is generic and is independent of the geometry or the nature of the field, i.e., the same criteria holds for different black hole spacetimes as well as for different fundamental fields.  

Following the scenario for \RN-\dS spacetime, subsequent computation for gravitational as well as scalar perturbations on Kerr-\dS spacetime demonstrates that the \scc\ is indeed respected \cite{Dias:2018ynt}. This gives hope that possibly the \scc\ is respected for astrophysical black holes. Subsequently, there have been several attempts to discuss the violation of \scc\ in various other contexts, these include: (i) for charged fields in \RN-\dS spacetime \cite{Dias:2018ufh,Cardoso:2018nvb}, (ii) implications for not-so-smooth initial data \cite{Dias:2018etb}, (iii) influence of higher spacetime dimensions \cite{Rahman:2018oso,Liu:2019lon} (iv) effect of both charged and uncharged Fermionic fields \cite{Ge:2018vjq,Destounis:2018qnb,Rahman:2019uwf,Liu:2019rbq} as well as the effect of higher curvature gravity \cite{Destounis:2019omd,Gan:2019jac,Mishra:2019ged,Gan:2019ibg}. We would like to emphasize that most of these results assume that the perturbations at late times decay exponentially, which has been rigorously proven only for Reissner-Nordstr\"{o}m-de Sitter and Kerr-de Sitter spacetime in \cite{Hintz:2016gwb,Hintz:2016jak}. However, the works presented above \emph{assumes} that such an assertion is true even for other black holes in modified gravity theories, which are asymptotically de Sitter. In this work also, in absence of any rigorous mathematical result, we follow such a heuristic argument that any asymptotically de Sitter black hole must have exponential decay of the late time tail.

In the midst of such an extensive study elaborated above, regarding the violation of \scc, the effect of conformal coupling has not been discussed so far, except for the study of certain specific \qnm modes in \cite{Gwak:2018rba, Guo:2019tjy}. The necessity to study the effect of conformal coupling is further motivated by the fact that for massive fields one arrives at a stronger violation of \scc, i.e., $C^{1}$ extendibility of the massive field across the Cauchy horizon \cite{Cardoso:2018nvb,Destounis:2018qnb} and in the presence of conformal coupling, the field equation for the scalar field mimics that of a massive field. Thus it will be of significant interest if the presence of conformal coupling can also generate $C^{1}$ extendibility of the metric across the Cauchy horizon, since this will constitute even a larger violation of \scc. Whether this violation of \scc\ continues in rotating spacetimes as well, must be explored as well. We would also like to emphasize that in the context of \scc, the effect of NUT charge has not been considered so far. The presence of a NUT charge in a spacetime is interesting in its own right and it appears naturally when asymptotically non-flat solutions are considered. Besides having several interesting properties, including separability of Hamilton-Jacobi and Klein-Gordon equation, the solutions with NUT charge has a duality property, namely, the spacetime structure is invariant under the transformation $\textrm{mass}\leftrightharpoons\textrm{NUT~charge}$ and $\textrm{radius}\leftrightharpoons\textrm{angular~coordinate}$. This suggests that NUT charge may be interpreted as a measure of gravitational magnetic charge, for more discussion, see \cite{Chen:2006xh,Dadhich:2001sz,Argurio:2009xr,LyndenBell:1996xj,Turakulov:2001jc,Mukherjee:2018dmm}. There are several investigations to search for observational evidence of the gravitomagnetic mass, these include investigation of the geodesics in Kerr-NUT spacetime \cite{Jefremov:2016dpi,Mukherjee:2018dmm}, effect on thin accretion disk \cite{GarciaReyes:2004qn,Chakraborty:2017nfu,vanderKlis:2000ca,Motta:2013wga}, weak field tests, e.g., perihelion precession, Lense-Thirring effect etc. \cite{Chakraborty:2013naa} (for a taste of these weak field tests in theories beyond general relativity, see \cite{Mukherjee:2017fqz,Bhattacharya:2016naa,Chakraborty:2013ywa,Chakraborty:2012sd}). As we will see, the presence of NUT charge requires the field to be conformally coupled with the curvature, in order for the angular perturbation equation in a rotating spacetime to be solvable by standard means. 

Following this motivation, in this paper we have studied the response of external scalar perturbations with conformal coupling in Reissner-Nordstr\"{o}m-NUT-de Sitter as well as Kerr-NUT-\dS black hole spacetimes and its effect on strong cosmic censorship conjecture. Both of these black hole geometries involve additional hair, namely the NUT charge. Our motivation is to check whether the presence of NUT charge as well as conformal coupling of the perturbing scalar field has anything to say about the deterministic nature of general relativity, in particular, whether the presence of the such a conformal coupling term can make the parameter $\beta$ not only larger than $(1/2)$, but also greater than unity, leading to severe violation of the \scc.

The paper is organized as follows: In \ref{lyapunov} we have discussed the Lyapunov exponent and \qnm modes associated with Reissner-Nordstr\"{o}m-de Sitter-NUT black hole. Then in \ref{kerr_modes} we discuss the geometry of Kerr-de Sitter-NUT black hole spacetime, which has been used in \ref{SecScalar} to determine the scalar perturbation equation in this black hole background. We have discussed both the angular and radial part of the perturbation equation analytically and have used numerical techniques to solve these equations and obtain the associated \qnm modes. We finish with a brief discussion on the results obtained. 

\emph{Notations and Conventions:} We will work with mostly positive signature convention. The roman letters $a,b,c,\cdots$ denote spacetime indices. We will further assume the fundamental constants $G=1=c$. 
\section{Lyapunov exponent, photon sphere and quasi-normal modes in Reissner-Nordstr\"{o}m-de Sitter-NUT spacetime}\label{lyapunov}

In this section, we present the case of a spherically symmetric black hole spacetime with a NUT charge. However, in order to ensure existence of a Cauchy horizon and asymptotic de Sitter nature we took the black hole to be Reissner-Nordstr\"{o}m-de Sitter-NUT \cite{grenzebach2014photon}. We start by writing down the line element associated with this static and spherically symmetric black hole, which takes the following form,
\begin{equation}
ds^{2}=-\frac{\Delta}{\Sigma}dt^{2}-4N\cos \theta \left(\frac{\Delta}{\Sigma}\right)dtd\phi
+\frac{\Sigma}{\Delta}dr^{2}+\Sigma d\theta^{2}
+\frac{1}{\Sigma}\left\{\Sigma^2\sin^2\theta-4N^{2}\cos^{2}\theta\Delta\right\} d\phi^2~,
\end{equation} 
where the quantities $\Delta$ and $\Sigma$ introduced above are functions of the radial coordinates alone and have the following expressions, 
\begin{align}
\Sigma=r^{2}+N^{2}~;
\qquad
\Delta= r^{2}-2Mr-N^{2}+Q^{2}- \Lambda \left(-N^{4}+2N^{2}r^{2}+\frac{1}{3}r^{4}\right)~.
\end{align}
Here, $M$ stands for the mass of the black hole, $Q$ is the electric charge associated with the electromagnetic field, $\Lambda$ is the cosmological constant and $N$ is the NUT charge, whose effect on the \scc\ is one of our prime aim of study in this section. 

\begin{figure}
	\centering
	\minipage{0.32\textwidth}
	\includegraphics[width=\linewidth]{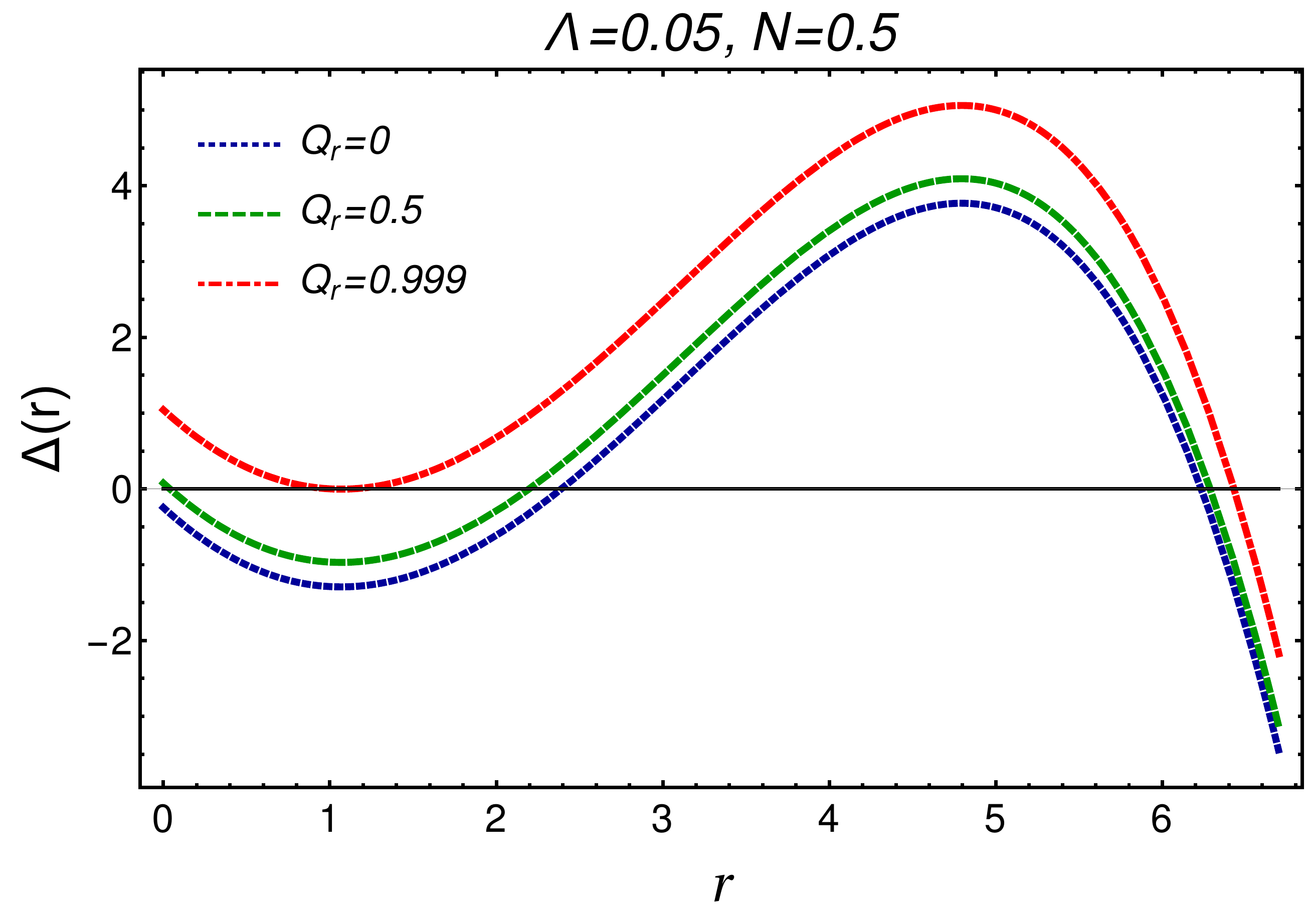}
	\endminipage\hfill
	\minipage{0.32\textwidth}
	\includegraphics[width=\linewidth]{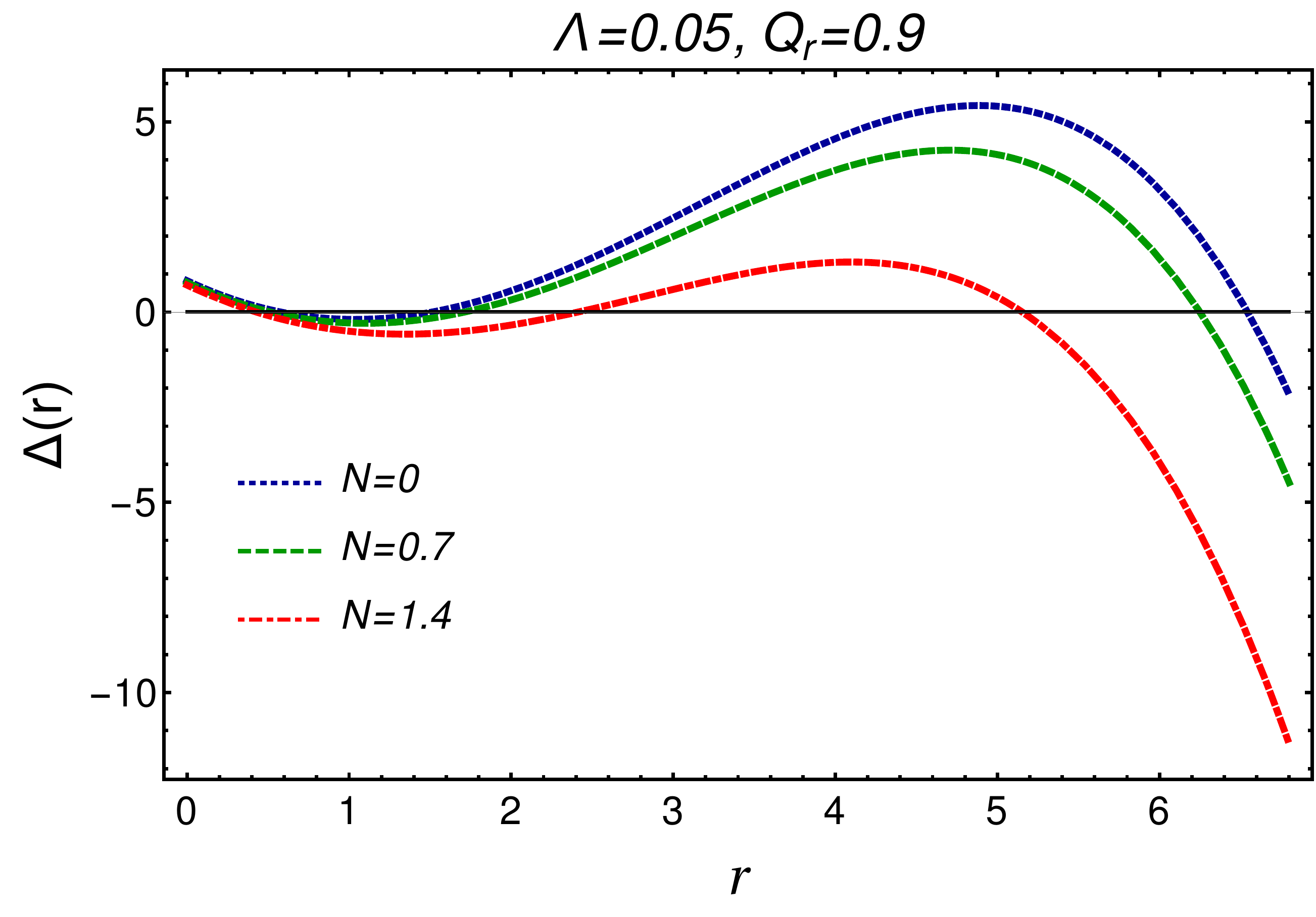}
	\endminipage\hfill
	\minipage{0.32\textwidth}%
	\includegraphics[width=\linewidth]{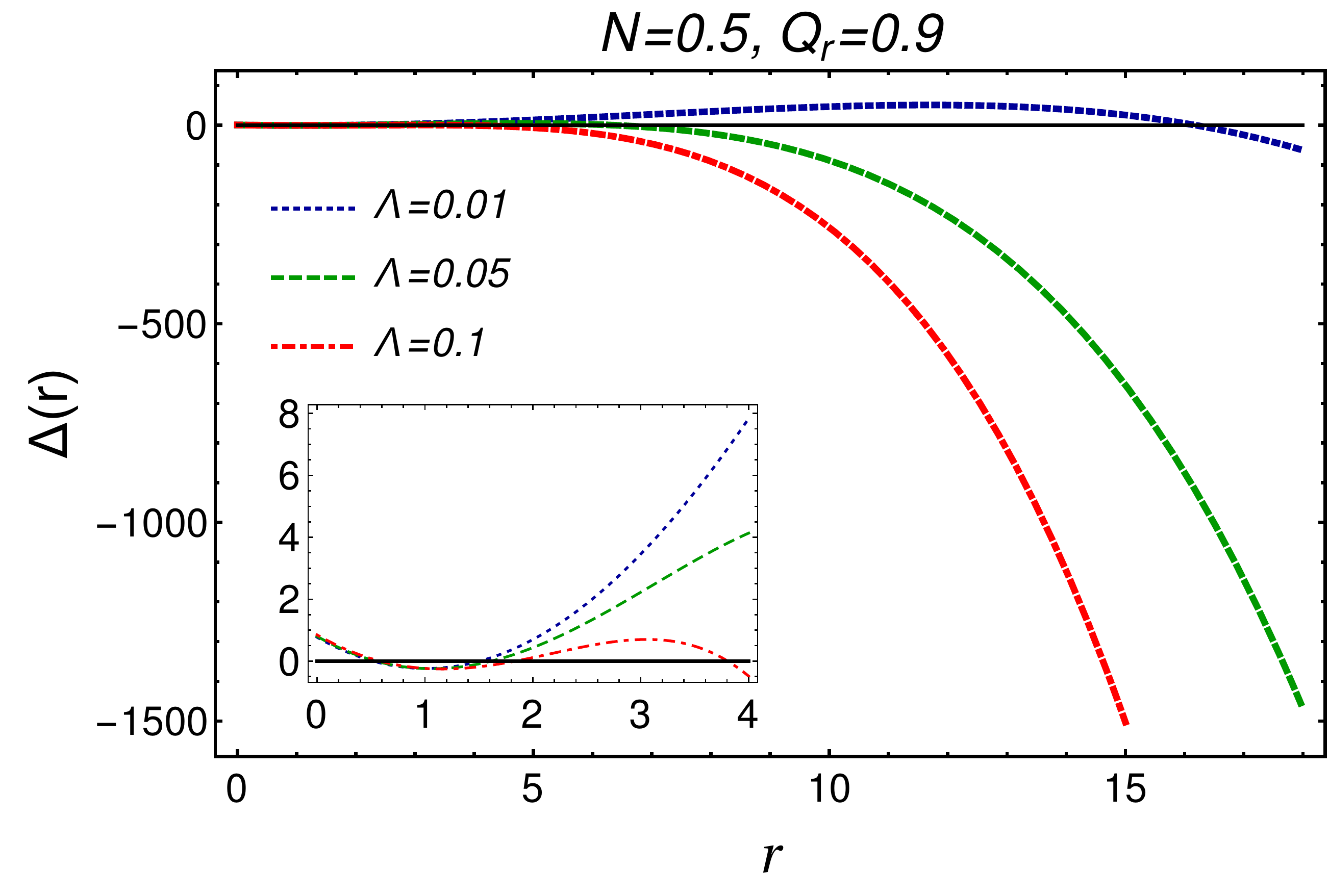}
	\endminipage
	\caption{The variation of the function $\Delta(r)$ with $r$ is presented for different values of \RN-NUT-\dS\ black hole parameters $N$, $Q$ and $\Lambda$ with black hole mass set to unity. In the above plots $Q_{r}$ denotes the charge ratio $(Q/Q_{\rm max})$, where $Q_{\rm max}$ denotes the extremal value of the black hole charge.}\label{horizons_q}
\end{figure}

Horizons associated with the above black hole solution corresponds to the real and positive roots of the equation $\Delta(r)=0$. This being a quartic equation has four real roots for certain choices of the black hole parameters, among which one root is negative and the other three denotes Cauchy, event and cosmological horizon respectively. Since analytical expressions for these horizons in terms of the black hole parameters are difficult to write down, we have presented the plot of $\Delta(r)$ as function of $r$ for different values of black hole parameters $\Lambda$, $Q$ and $N$ in \ref{horizons_q} with the black hole mass $M$ set to unity. Note that our primary interest is to study black holes which are asymptotically de Sitter with Cauchy horizon and hence we will consider the parameter space within which all the three horizons exist. The position of the cosmological, the event and the Cauchy horizon is denoted by $r_{c}$, $r_{+}$ and $r_{-}$ respectively, such that $r_{c}\geq r_{+}\geq r_{-}$. By taking derivative of the metric elements, the surface gravity associated with these horizons can be determined explicitly and are given by the following expression,
\begin{align}
\kappa_{X} =\bigg|\frac{\Delta_{r}'(r)}{2\Sigma}\bigg|_{r=r_X}= \bigg|\frac{3(r_{X}-M)-\Lambda(2r^{3}_{X}+6N^{2}r_X)}{3 \left(r^2_X+N^2\right)}\bigg|~,\qquad X\in \{c,+,-\}~.
\nonumber
\end{align}
Thus given any horizon and its location as a function of the black hole parameters, the surface gravity of that horizon as a function of black hole parameters can also be determined from the above expression.

Let us now discuss the other central ingredient to our work, namely the \qnm modes associated with a conformally coupled scalar field living in this spacetime and thereby perturbing the same. The evolution of such a conformally coupled scalar field of mass $m$ in curved spacetime is governed by the following Klein-Gordon equation, 
\begin{equation}\label{kg_eqn}
\left(g^{ab}\nabla_{a}\nabla_{b}-\xi_C \mathfrak{R}\right)\Phi(x)=0,
\end{equation}
where, $\xi_C$ denotes the coupling strength of the scalar field with the Ricci scalar $\mathfrak{R}$ of the background spacetime. Such non-minimal couplings are expected to arise in various settings, including renormalization of scalar fields in classical curved background  \cite{FREEDMAN1974354,Birrell:1982ix, PhysRevD.22.322,Faraoni:1996rf}. Note that the minimally coupled case corresponds to vanishing coupling constant, i.e., requires $\xi_C=0$. As emphasized earlier, the study of possible violation of \scc\ has mostly been performed in the context of minimally coupled scalar field, i.e., with vanishing choice for $\xi_C$. However, as \cite{Cardoso:2018nvb} hints, existence of a mass term or equivalently the conformal coupling can lead to an even stronger violation of \scc. This is what we want to explore in the present context. It follows from \cite{Sonego_1993} that the value of the coupling constant $\xi_C$ has to be fixed to $(1/6)$ in order to satisfy the equivalence principle. This particular choice for the conformal coupling $\xi_C$ also originates from the invariance of the scalar field Lagrangian, from which \ref{kg_eqn} is derived, under conformal transformation: $\Phi\to\Omega^{-1}\Phi$ and $g_{ab}\to\Omega^{2}g_{ab}$, where $\Omega$ is an arbitrary function of the spacetime coordinate \cite{Wald:1984rg}.

The natural question to ask in this context is, whether the analytical estimation for the \qnm mode frequencies can still be provided through the Lyapunov exponent calculation, even in the presence of conformal coupling. We have elaborated on this issue in  \ref{App_QNM}, where we have explicitly demonstrated that even in the presence of conformal coupling the scalar field experiences the potential identical to that of a radial null geodesic in the eikonal limit. Thus the photon sphere modes are still determined by the Lyapunov exponent associated with instability of photon circular orbits.

Even though we will present all the results using numerical techniques, it is instructive to briefly discuss about the theoretical avenue towards the same, namely using the Lyapunov exponent and instability of photon circular orbit. For this purpose, we concentrate on the equatorial plane, given by $\theta=\pi/2$ and hence the $dtd\phi$ cross-term drops out from the line element. Thus the reduced metric on the equatorial plane becomes,
\begin{equation}
\label{2}
ds_{\rm eq}^2=-\frac{\Delta}{r^{2}+N^{2}}dt^{2}+\frac{r^{2}+N^{2}}{\Delta}dr^{2}+\left(r^{2}+N^{2}\right)d\phi^{2}
\end{equation}
Even though the structure of this metric is overwhelmingly similar to a static and spherically symmetric metric, due to the extra $N^2$ term in the expression for $g_{\phi\phi}$, the algebraic equation governing the photon sphere and the Lyapunov exponent will change accordingly in this system of coordinates. Note that one can always redefine the radial coordinate, such that $R^{2}=r^{2}+N^{2}$ and the $g_{\phi \phi}$ will have the same structure as that of a static and spherically symmetric metric. However, we will work with the above choice of the coordinate system.  The Lagrangian associated with the particle motion in the above metric within the equatorial plane is given by,
\begin{equation}
L=\frac{1}{2}\left\{-\frac{\Delta}{r^{2}+N^{2}}{\dot{t}}^{2}+\frac{r^{2}+N^{2}}{\Delta}{\dot{r}}^{2}+\left(r^2+N^2\right){\dot{\phi}}^2\right\}
\end{equation}
As the Lagrangian is independent of t and $\phi$ coordinates, it immediately follows that the associated conjugate momenta are the constants of motion, namely the energy $E$ and the angular momentum $L$. In particular, $-p_{t}=E$ is the conserved energy and $p_{\phi}=L$ is the conserved angular momentum. Given these conserved quantities and the fact that for null geodesics, $p_{a}p^{a}=0$, one immediately arrives at the following equation,
\begin{equation}\label{eq_photon}
{\dot{r}}^{2}=E^{2}-f(r)\left(\frac{L^2}{r^2 +N^2}\right);\qquad f(r)\equiv \frac{\Delta}{r^{2}+N^{2}}~,
\end{equation}
where, the right hand side of the above equation defines the effective potential a radial null geodesic experiences, namely $V_{\rm eff}(r)\equiv E^{2}-f(r)\{L^{2}/(r^{2}+N^{2})\}$. The photon circular orbit(s) can be determined by solving the following equations,
\begin{equation}
V_{\rm eff}(r=r_{\rm ph})=0=V'_{\rm eff}(r =r_{\rm ph})
\end{equation}
These two conditions, when expressed explicitly in terms of the function $f(r)$ and its derivative yields, (i) $(E^{2}/L^{2})=\{f(r_{\rm ph})/(r_{\rm ph}^{2}+N^{2})\}$ as well as (ii) $2r_{\rm ph}f(r_{\rm ph})=(r_{\rm ph}^{2}+N^{2})f'(r_{\rm ph})$. The first equation provides an expression for the specific angular momentum, while the second equation provides an algebraic equation for $r_{\rm ph}$, which must be solved in order get the expression for the photon circular orbit in terms of the black hole hairs $(M,Q,N)$. 

The next job is to compute the Lyapunov exponent associated with the instability of the photon sphere. This exponent measures the rate of departure of a nearby null geodesic from the photon circular orbit. This is obtained by considering a small perturbation over the photon circular orbit, i.e., one considers $r\rightarrow r_{\rm ph}+\delta r$. Then from \ref{eq_photon}, using the small perturbation $\delta r$ about the photon circular orbit, it immediately follows that the perturbation scales as, $\delta r\sim \exp(\pm \lambda_{\rm ph}t)$. The quantity $\lambda_{\rm ph}$ depends explicitly on $V_{\rm eff}''(r_{\rm ph})$ and is defined as the Lyapunov exponent associated with photon circular orbit. It is possible to express the Lyapunov exponent explicitly in terms of the metric components, as,
\begin{equation}
\lambda_{\rm ph}=\sqrt{\frac{f(r_{\rm ph})}{2}\left[\frac{2 f(r_{\rm ph})}{r_{\rm ph}^{2}+N^{2}}-f''(r_{\rm ph})\right]}
\end{equation}
This gives an estimation of the minimally damped \qnm mode frequency, since under WKB approximation, one can explicitly demonstrate that imaginary part of the \qnm mode frequency is proportional to the Lyapunov exponent associated with photon circular orbits. Since the spacetime also admits a Cauchy horizon, it follows that one can compute the surface gravity $\kappa_{-}$ associated with the same and hence obtain the parameter $\beta$, which is simply $\{\lambda_{\rm ph}/2\kappa_{-}\}$. This will certainly depend on the NUT charge $N$, electromagnetic charge $Q$, cosmological constant $\Lambda$ and mass of the black hole $M$. Thus one can easily verify whether the parameter $\beta$ can become larger than $(1/2)$, leading to violation of the \scc. 

We should also point out that, the above illustrates the procedure to assess the violation of the \scc\ for the photon sphere modes. There are also two other \qnm modes living in this spacetime, which are of importance. The first one corresponds to near extremal mode, which originates, as the Cauchy horizon and the event horizon of a black hole come close to one another. In this context, as \ref{AppN} explicitly demonstrates, the near-extremal modes have the following form, 
\begin{equation}
\omega_{\rm NE}=-i\left(n+\ell+1\right)\kappa_{+}=-i\left(n+\ell+1\right)\kappa_{-}~.
\end{equation}
The last equality follows from the fact that in the near extremal regime, the Cauchy and the event horizons coincide and hence one can consider them to have identical surface gravity. As the \qnm frequency for the \NE modes suggests the dominant contribution will come from $n=0=\ell$ mode, for which $\{-\textrm{Im}(\omega_{\rm NE})/\kappa_{-}\}=1$. It may appear that due to the above result, $\beta_{\rm NE}$ can never cross unity. But we would like to emphasize that such may not be the case, since the numerical analysis can modify the analytical result, as various approximations are involved in the analytical result, which may not hold for all the parameter space.  

The other mode, namely the \dS mode, arises when we are interested in the asymptotic structure of the spacetime, such that only the cosmological constant term contributes. In this case, for conformally coupled massless scalar field, the \qnm frequency associated with the \dS mode has the following structure in four spacetime dimensions \cite{LopezOrtega:2006my},
\begin{equation}
\omega_{\rm dS}=-i\left(\ell+2n+1\right)\kappa_{\rm c}
\end{equation}
As evident the dominant contribution again comes from $\ell=0=n$ mode and hence is coincident with the \NE mode described earlier. However, we have $-\textrm{Im}(\omega_{\rm NE})=\kappa_{-}$ and $-\textrm{Im}(\omega_{\rm dS})=\kappa_{c}$ and in the near extremal region, $\kappa_{-}\rightarrow 0$. Thus the imaginary part of the \qnm mode frequency is smallest for the \NE mode and hence will provide the dominant contribution. Therefore, unlike the case of a massless, minimally coupled scalar field, for massless and conformally coupled scalar field, the \dS and \NE modes coincide. However, as the \NE mode has the dominant contribution, for massless and conformally coupled scalar field, there are no \dS modes present. This is further corroborated from the numerical analysis presented below. 

\begin{figure}
\minipage{0.32\textwidth}
  \includegraphics[width=\linewidth]{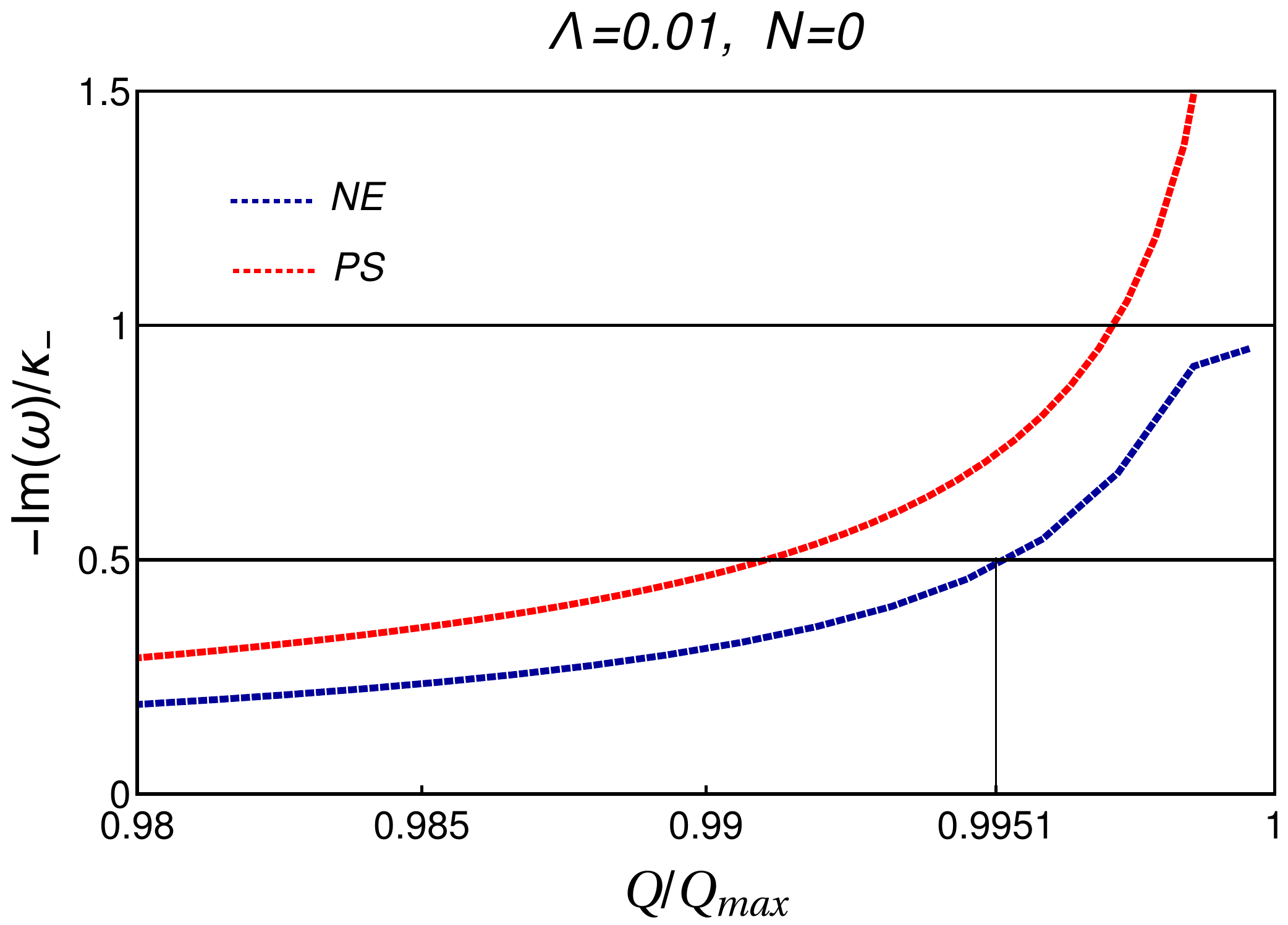}
\endminipage\hfill
\minipage{0.32\textwidth}
  \includegraphics[width=\linewidth]{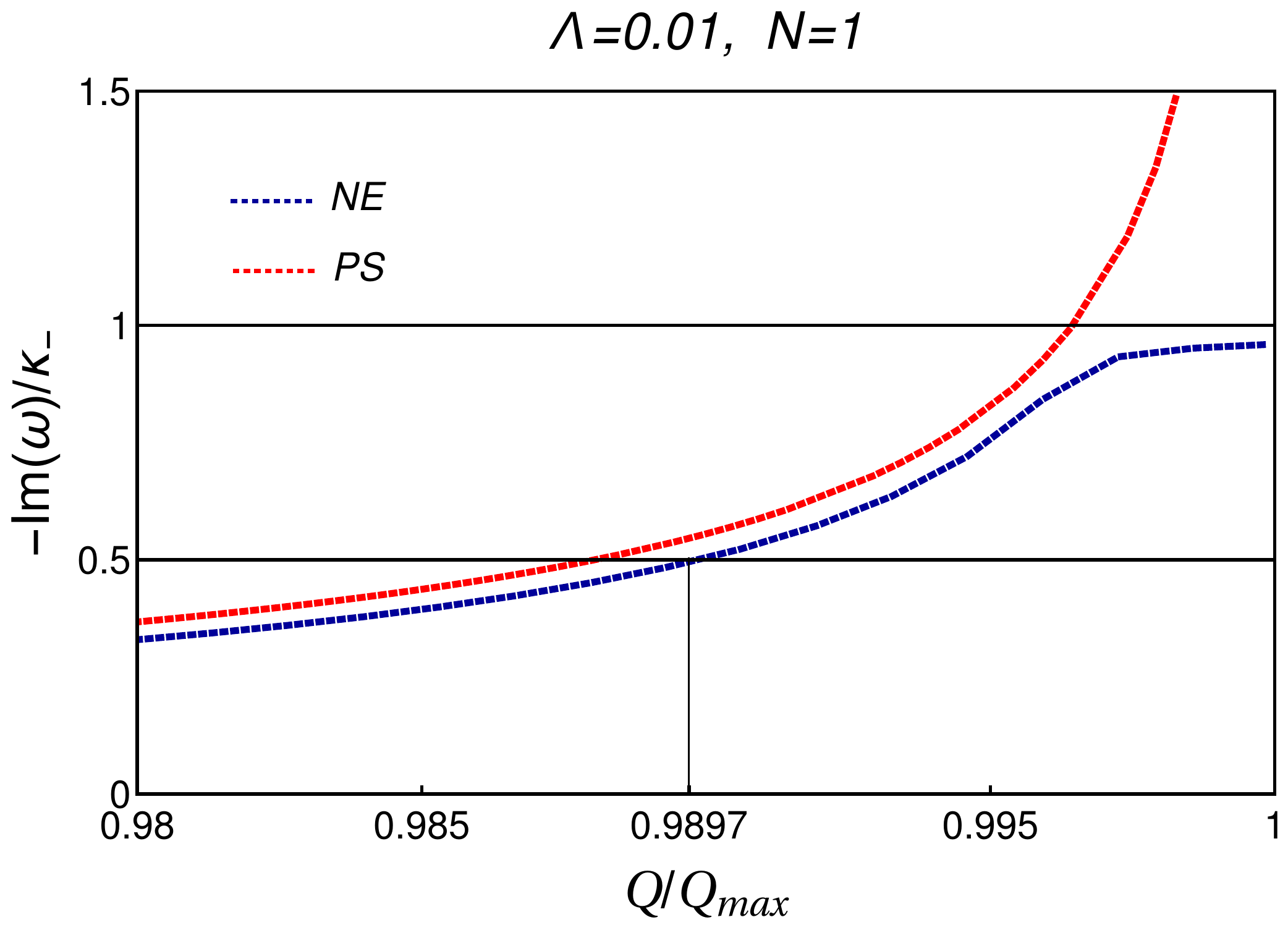}
\endminipage\hfill
\minipage{0.32\textwidth}%
  \includegraphics[width=\linewidth]{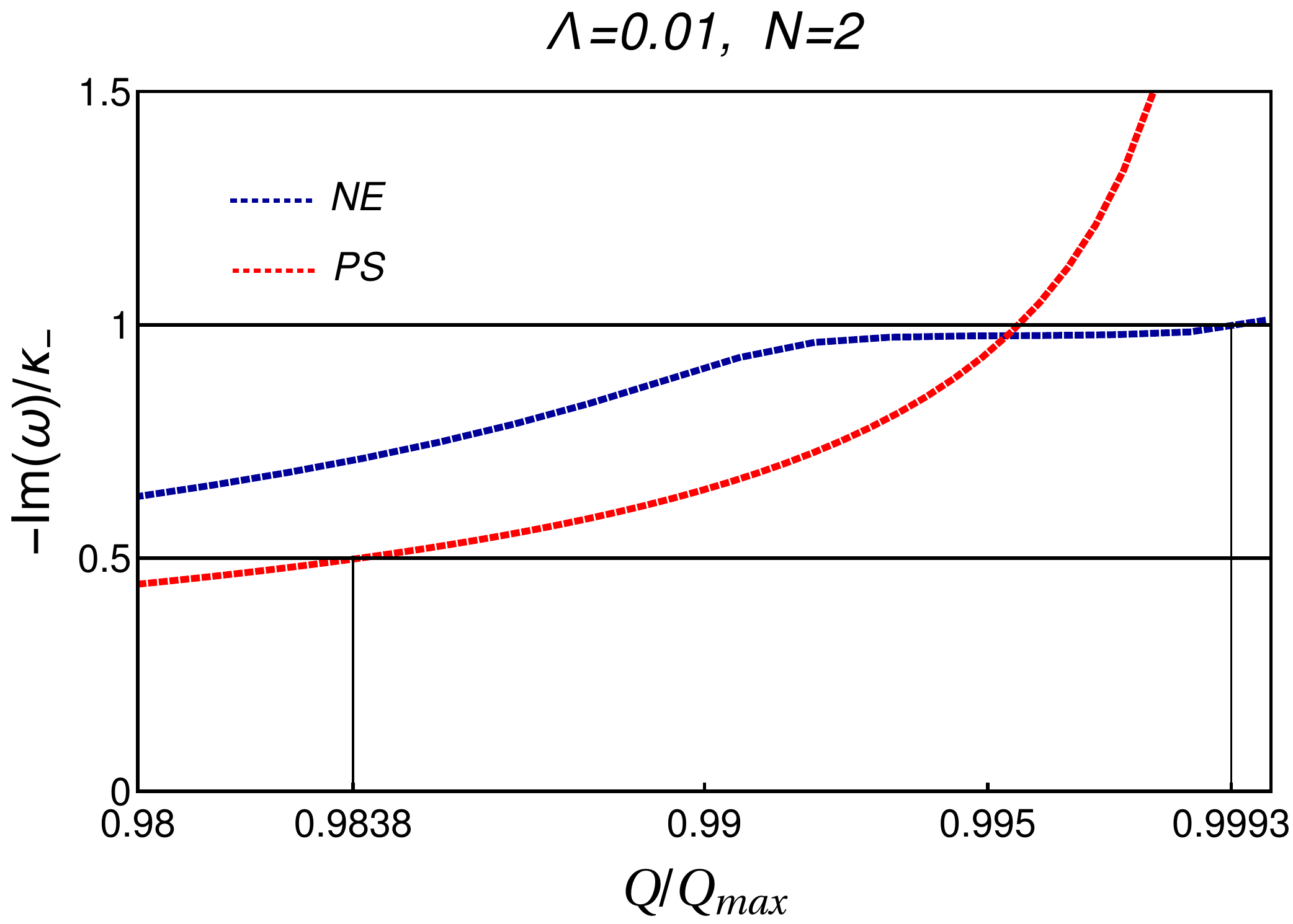}
\endminipage\hfill
\minipage{0.32\textwidth}
  \includegraphics[width=\linewidth]{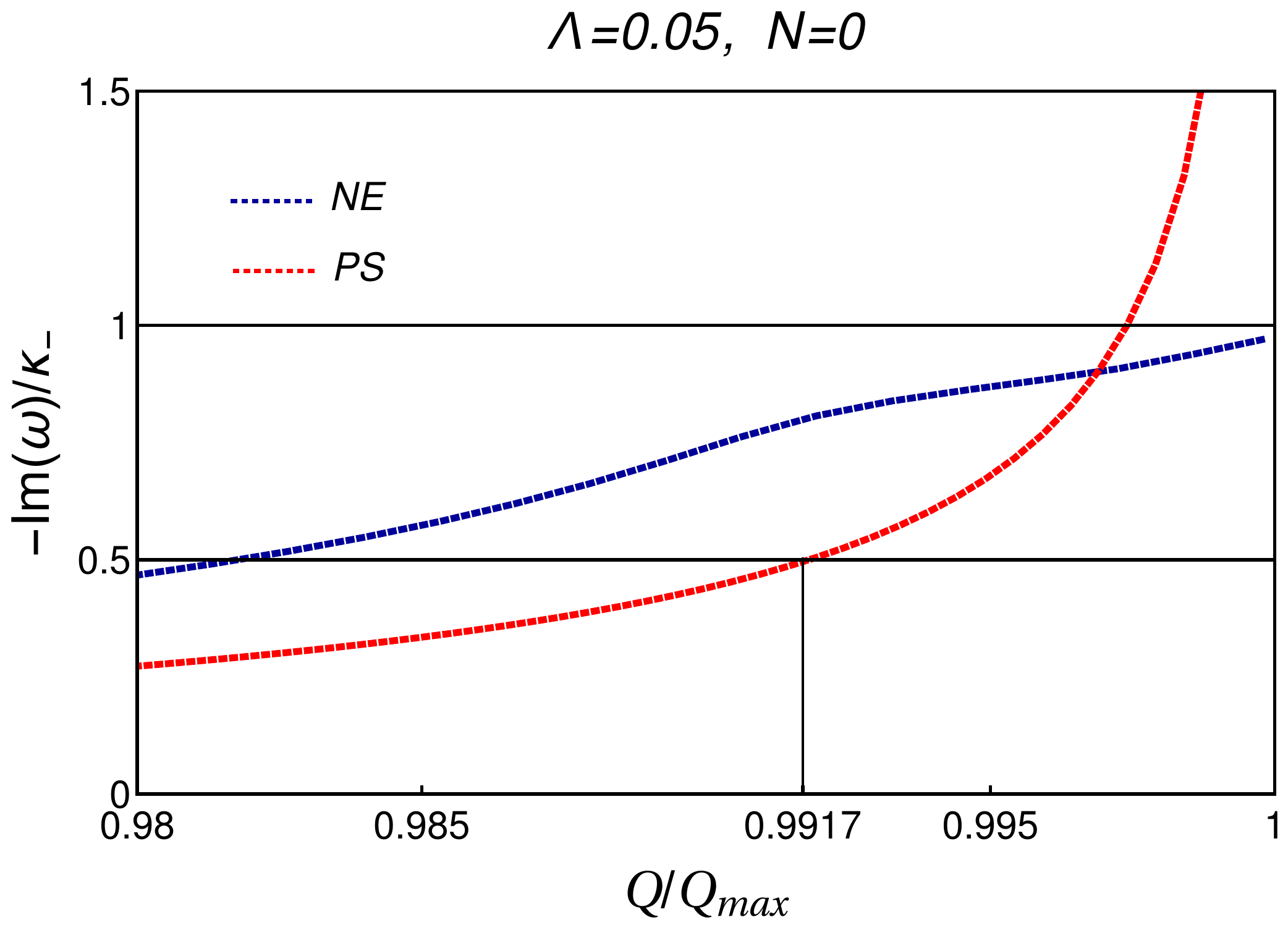}
\endminipage\hfill
\minipage{0.32\textwidth}
  \includegraphics[width=\linewidth]{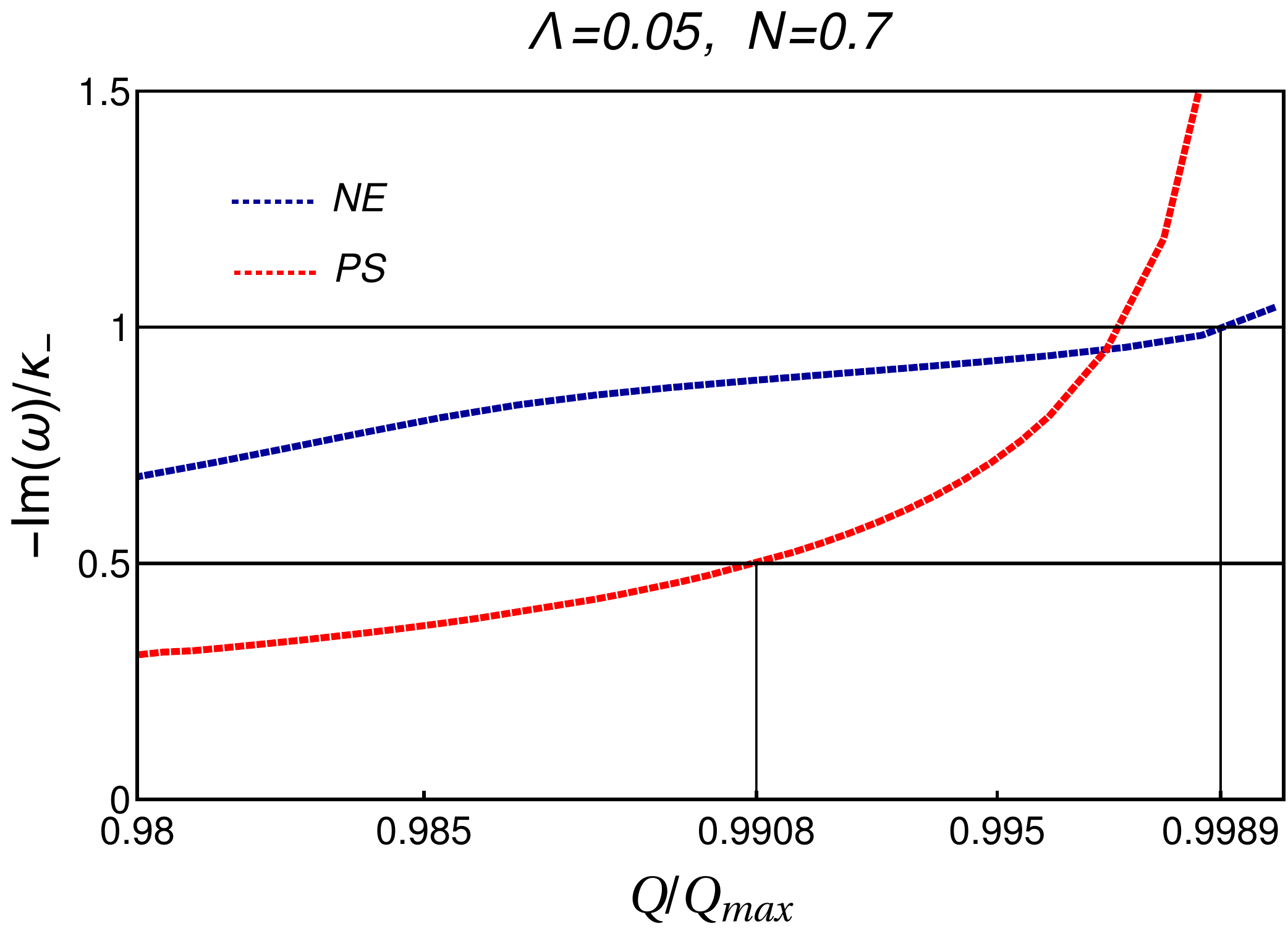}
\endminipage\hfill
\minipage{0.32\textwidth}%
  \includegraphics[width=\linewidth]{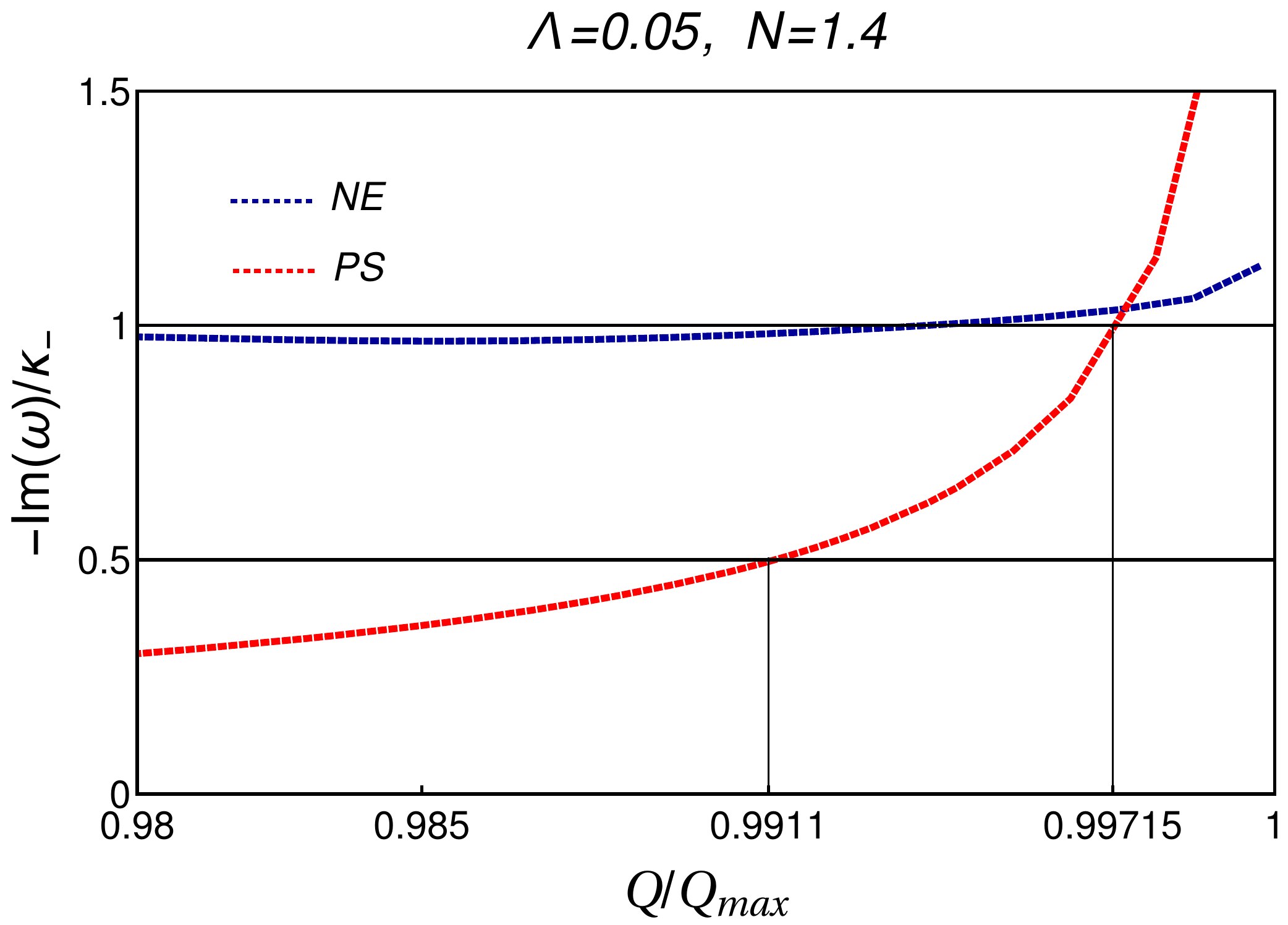}
\endminipage\hfill
\minipage{0.32\textwidth}
  \includegraphics[width=\linewidth]{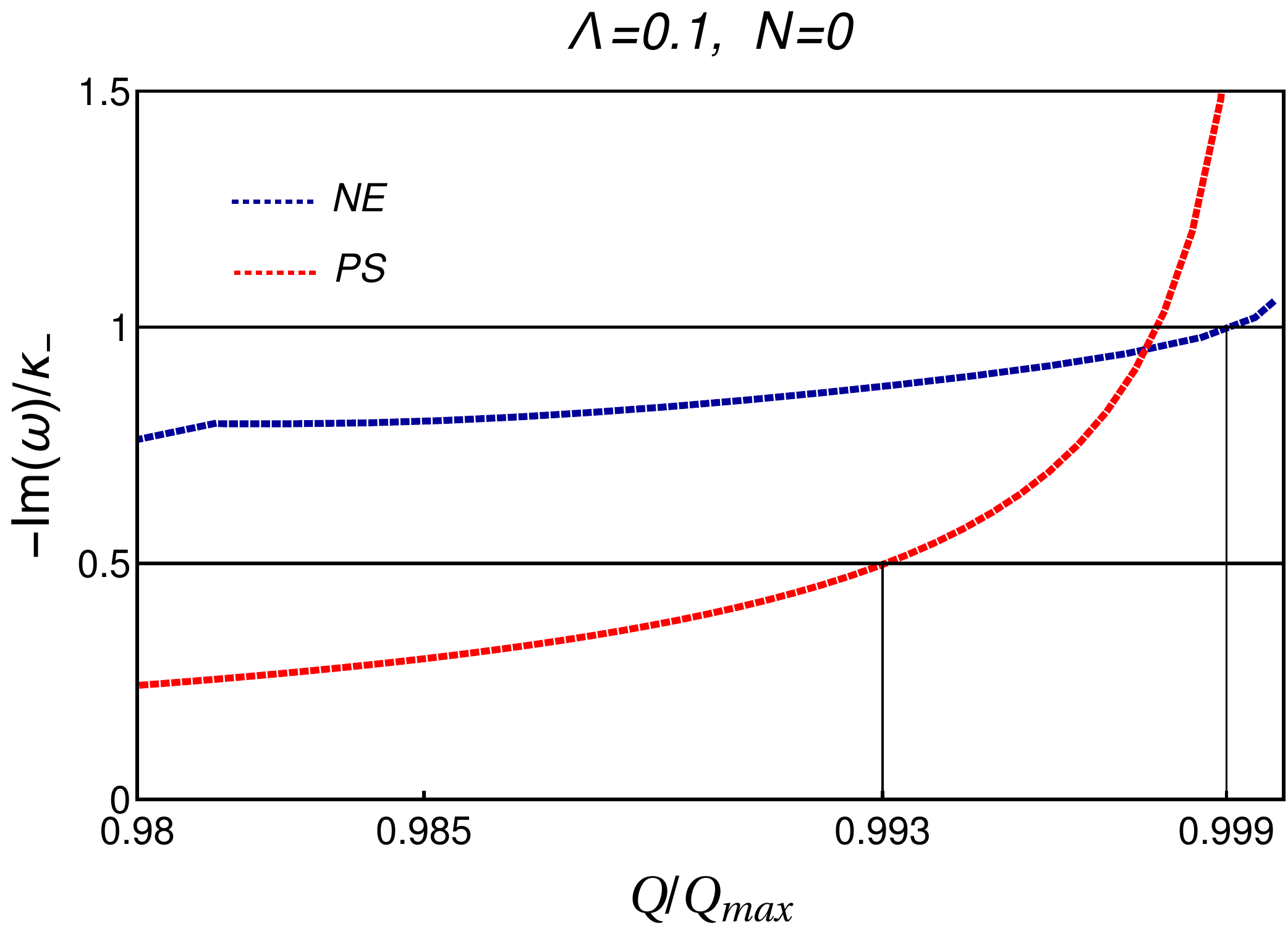}
\endminipage\hfill
\minipage{0.32\textwidth}
  \includegraphics[width=\linewidth]{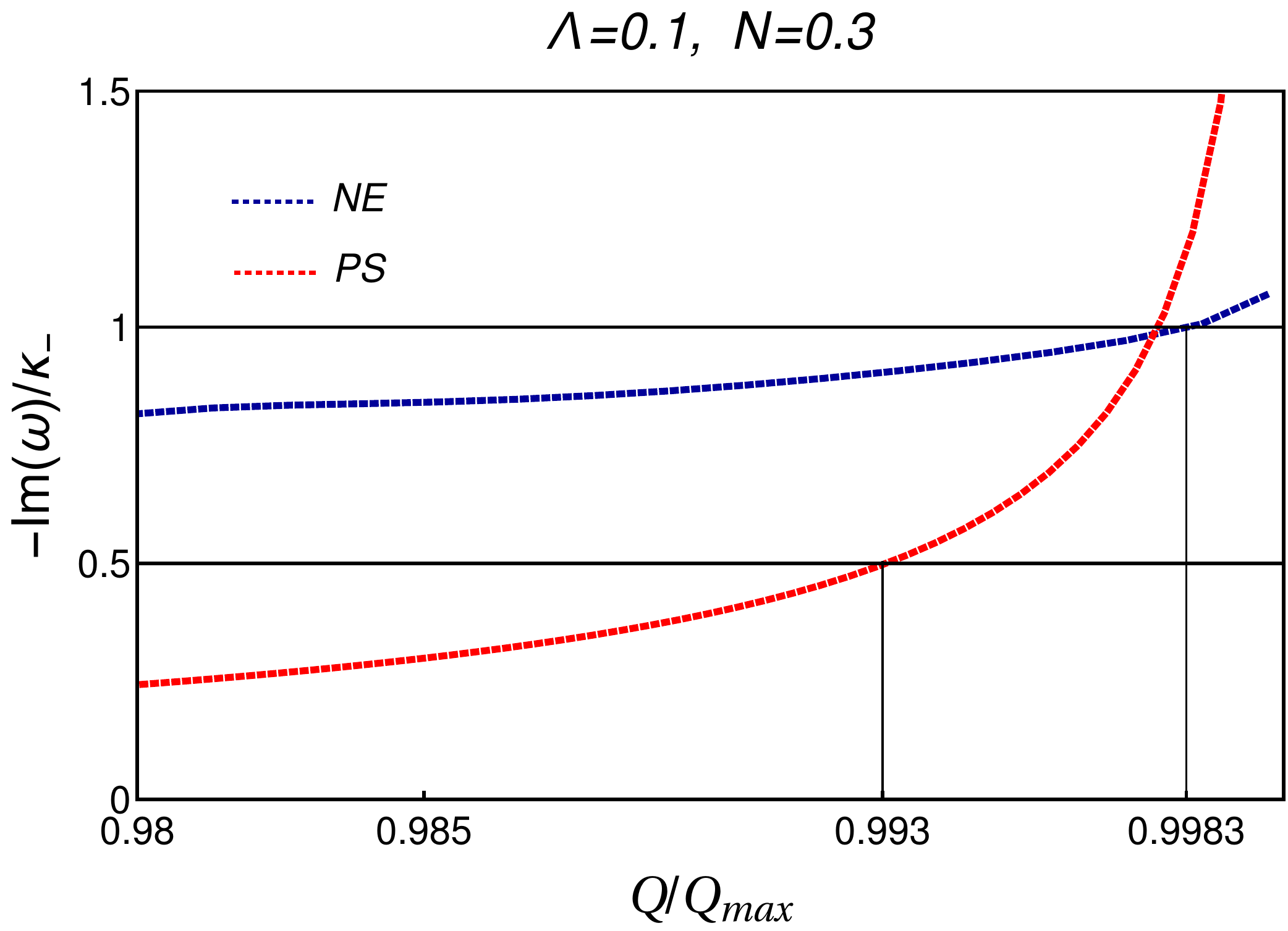}
\endminipage\hfill
\minipage{0.32\textwidth}%
 \includegraphics[width=\linewidth]{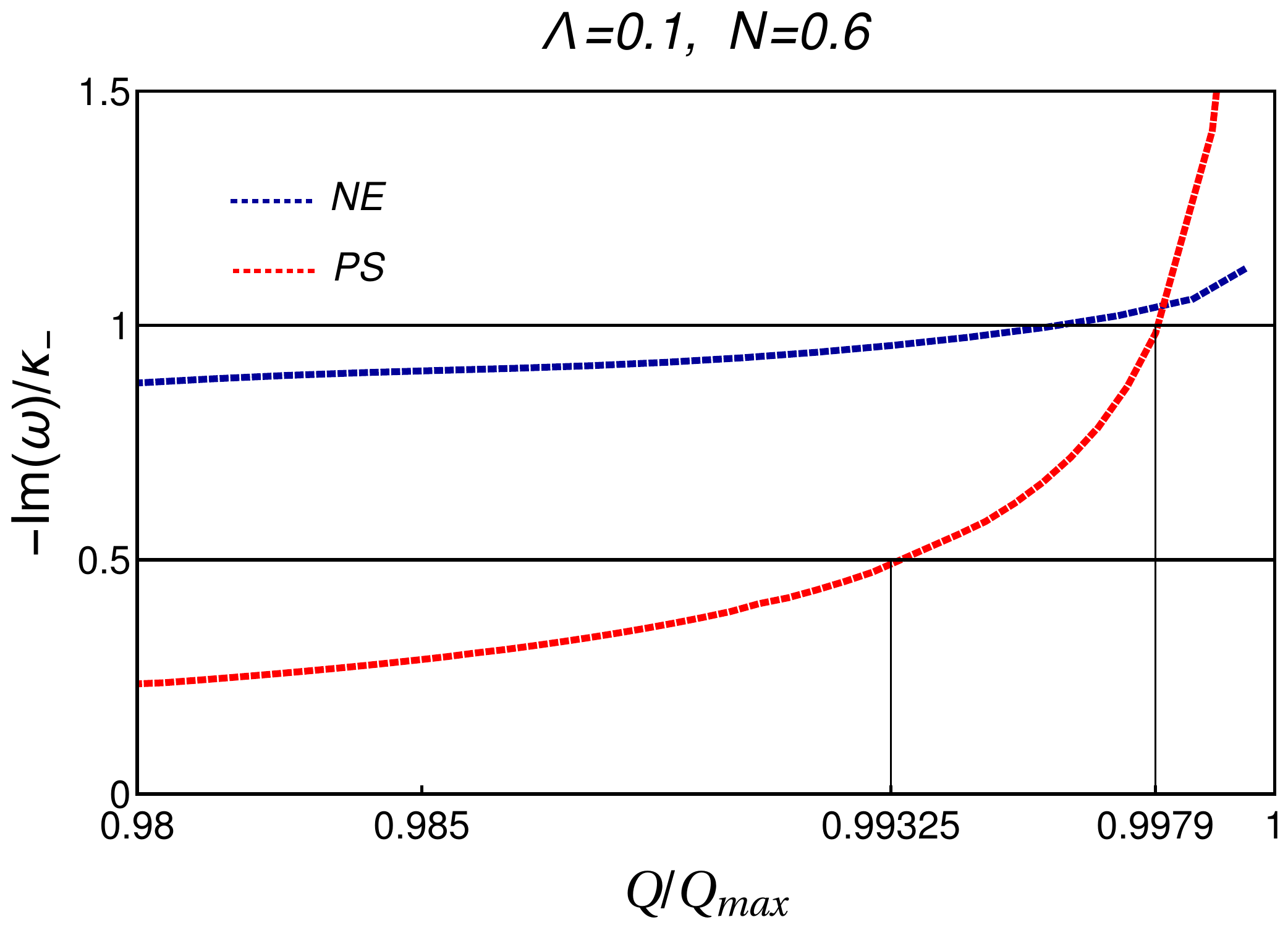}
\endminipage%
\caption{The plot of $-\{\textrm{Im}(\omega)/\kappa_{-}\}$ as a function of $(Q/Q_{\rm max})$ for different values $N$ and $\Lambda$ in a \RN-NUT-\dS black hole is presented. Here, $Q_{\rm max}$ corresponds to the maximum allowed value of the charge, associated with the \RN-NUT-\dS extremal black hole. The value of $\beta$ is determined by the lowest lying quasi-normal mode, i.e., the mode for which the value of $-\textrm{Im}(\omega)$ is the smallest. In the plot, the blue lines denote the near extremal modes (corresponds to $\ell=0$) whereas the red lines signifies the photon sphere modes (corresponds to large $\ell$ value). The plots suggest that with the \scc is indeed violated for \RN-NUT-\dS black holes and the violation is stronger with increase of NUT charge. Most importantly, for larger $\Lambda$ and $N$, the value of $\beta$ can even be larger than unity, suggesting breakdown of the $C^{2}$ version of \scc. The first vertical line in the plots suggests the value $(Q/Q_{\rm max})$ for which the Christodoulou's version of the strong cosmic censorship gets violated, while the second vertical line depicts the value of $(Q/Q_{\rm max})$ for which even the $C^{2}$ version of \scc\ is violated. See text for more discussion.}\label{fig_rn_numric}
\end{figure}
Having discussed the analytical results pertaining to the \qnm modes, we now elaborate on the numerical scheme employed in this work to determine the \qnm modes. Since the \RN-\dS-NUT spacetime admits two Killing vectors as well as a second rank Killing-Yano tensor, the scalar field equation given by \ref{kg_eqn} permits a separation of variables of the form $\Phi = e^{i\omega t} e^{-im\phi} \Theta(\theta) R(r)$ \cite{Frolov:2006pe, Kubiznak:2006kt}. Inserting the above form for the scalar field $\Phi$ in \ref{kg_eqn}, we find the angular part $\Theta(\theta) $ satisfies the following equation,
\begin{equation} \label{ang_rnds}
\frac{d}{dx}\left[\left(1-x^2\right)\frac{d\Theta}{dx}\right]+k\Theta - \frac{(m+2N\omega x)^2}{(1-x^2)} \Theta - \frac{2\Lambda N}{3}\Theta =0~,
\end{equation}
whereas, the radial part of the scalar field equation can be expressed as follows
\begin{equation}\label{radial_rnds}
\Delta(r)\frac{d^{2}R(r)}{dr^{2}}+\Delta'(r)\frac{dR(r)}{dr}+\left(\frac{\omega ^2 \left(r^2+N^2\right)^2}{\Delta(r)}-\frac{2}{3} \Lambda  r^2\right)R(r) =k R(r)~, 
\end{equation}
where we have introduced a new variable $x=\cos\theta$ and set the value of the coupling constant as $\xi_{C}=(1/6)$. The separation constant $k$ can be obtained analytically by solving \ref{ang_rnds}. Defining a new constant, $\lambda=k-(2\Lambda N/3)-2N\omega$, the angular equation can be rewritten in the form of a generalized spheroidal wave equation \cite{Berti:2005gp}, such that
\begin{equation}
\label{k_calc_1}
\frac{d}{dx}\left[\left(1-x^2\right) \frac{d\Theta}{dx}\right]+\left[\lambda + 2N \omega - \frac{(m+ 2 N \omega x)^2}{1-x^2}\right]\Theta=0~.
\end{equation}
The parameter $\lambda$ can further be related to the angular momentum $\ell$ associated with the spheroidal harmonic, such that $\lambda=\ell(\ell+1)-2N \omega(2N \omega+1)$. Using the expression for $\lambda$ in terms of the separation constant $k$, we obtain,
\begin{equation} \label{k_calc_2}
k=\ell(\ell+1)+\frac{2N \lambda}{3}-4N^2 \omega^2~.
\end{equation}
Note that for vanishing NUT charge, i.e., with $N=0$, the spheroidal wave equation turns to spherical wave equation and the separation constant becomes $\ell(\ell+1)$. Formally, the quasi-normal modes are defined as the solution of the above perturbation equation subject to the boundary conditions that at the cosmological horizon $r_{\rm c}$, the modes are purely outgoing, while at the event horizon $r_{+}$, we have only ingoing modes. Taking this into account, we redefine the radial function $R(r)$ in terms of a new radial function $y(r)$ as follows
\begin{equation}\label{Frobenious_series}
R(r)=\left(\frac{1}{r}\right)e^{iB_{1}(r)}\bigg(\frac{r-r_{-}}{r-r_{+}}\bigg)^{2iB_{2}(r_{+})}y(r)~,
\end{equation}
where, $B_{1}(r)$ is a function of the radial co-ordinate alone and is given by the expression $(dB_{1}/dr)=[V_{r}(r)/\Delta(r)]$ and $B_{2}(r_{+})=[V_{r}(r_{+})/\Delta'(r_{+})]$, with the following expression for the potential: $V_{r}(r)=(r^{2}+N^{2}) \omega$. In the Frobenius method, one expands this new radial function $y(r)$ as a power series expansion around $r=r_{+}$. For this purpose, one introduces a new variable $\varrho$, which is related to $r$ as follows, $ \varrho=[(r-r_{+})(r_{c}-r_{-})/(r-r_{-})(r_{c}-r_{+})]$. Using this new variable in \ref{radial_rnds} one can write down the second order differential equation for the unknown function $y(\varrho)$. Subseqently one uses the expansion $y(\varrho)=\sum_{k}b_{k}\varrho^{k}$ in the above differential equation and obtain the recursion relation for the coefficients $b_{k}$, which can be used to determine the \qnm modes. To summarize, in this work we have first solved the angular equation, leading to spheroidal harmonics. This in turn determines the separation constant $k$ in terms of the angular momentum $\ell$, NUT charge $N$ and \qnm mode frequency $\omega$. This separation constant has been used in the radial equation, which has been solved numerically with purely ingoing boundary condition at the event horizon and outgoing boundary condition at cosmological horizon using the package \textit{QNMspectral} \cite{Jansen:2017oag}. We have further checked the numerical estimation of the \qnm modes obtained by the above procedure, using the semi-analytical WKB method as well, following \cite{Konoplya:2003ii}. Interestingly the results matches with each other to an excellent accuracy.

The result of such a numerical analysis is presented in \ref{fig_rn_numric}, where the variation of the parameter $\beta$ against $(Q/Q_{\rm max})$ has been plotted for different choices of $\Lambda$ and $N$. Here $Q=Q_{\rm max}$ corresponds to the Maxwell charge associated with the extremal configuration of the black hole. As emphasized earlier, there are only two non-trivial \qnm modes in this spacetime, one corresponds to the near-extremal mode ($\ell=0$), while the other one refers to the photon sphere mode and the \dS modes are absent for a massless and conformally coupled scalar field. In \ref{fig_rn_numric} the photon sphere modes are denoted by red curves, and it is evident that these modes dominate the \qnm mode spectrum for smaller values of $(Q/Q_{\rm max})$. On the other hand, for larger values of $(Q/Q_{\rm max})$, the near extremal mode starts to dominate the spectrum of \qnm modes and is denoted by blue curves in \ref{fig_rn_numric}. It is clear that the value of $\beta$ goes beyond $(1/2)$ for all the modes in this spacetime. In particular, the value of $(Q/Q_{\rm max})$, for which the \qnm mode crosses the value $(1/2)$ decreases as the NUT charge increases, which suggests that the violation of \scc becomes stronger in the presence of NUT charge. This is because, the presence of the NUT charge makes the spacetime asymptotically non flat and hence the decay of the \qnm mode to the Cauchy horizon is further suppressed. More interestingly, the presence of a conformal coupling leads to $\beta>1$ for larger values of $\Lambda$ and/or $N$, which has an even severe consequence, namely it violates the $C^{1}$ version of the \scc. This suggests that for a conformally coupled scalar field with large $\Lambda$ and/or large $N$, even the scalar field can be continued across the Cauchy horizon, leading to a serious breakdown of \scc\ and deterministic nature of general relativity is at stake. However, the recent result of \cite{Hollands:2019whz} suggests that possibly quantum corrections will help to cure the scenario. To summarize, as \ref{fig_rn_numric} explicitly demonstrates, the Christodoulou's version of the \scc\ is indeed violated in this spacetime, since both the near-extremal as well as photon sphere modes cross the $\beta=(1/2)$ line in the near extremal region. Even the $C^{1}$ version of the \scc\ is being violated here, as the parameter $\beta$ crosses unity. In this context NUT charge plays a very important role. Hence there is indeed a finite parameter space in the \RN-NUT-\dS spacetime for which the violation of both Christodoulou's version of \scc\  as well as more sacred $C^{1}$ version of \scc\ occurs in this spacetime. 
\section{Kerr-de Sitter-NUT spacetime and strong cosmic censorship conjecture}\label{kerr_modes}

\begin{figure}
	\centering
	\minipage{0.32\textwidth}
	\includegraphics[width=\linewidth]{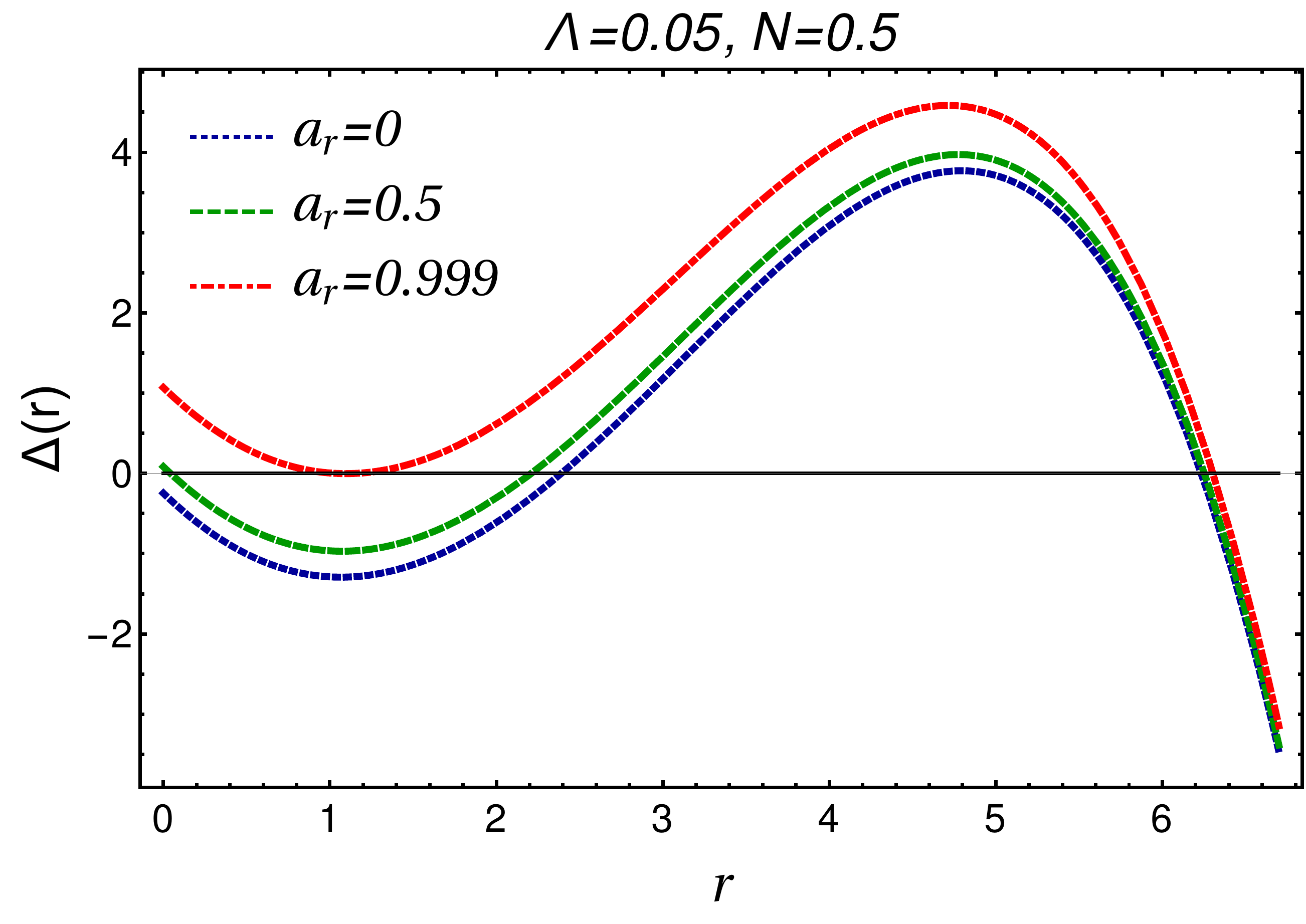}
	\endminipage\hfill
	\minipage{0.32\textwidth}
	\includegraphics[width=\linewidth]{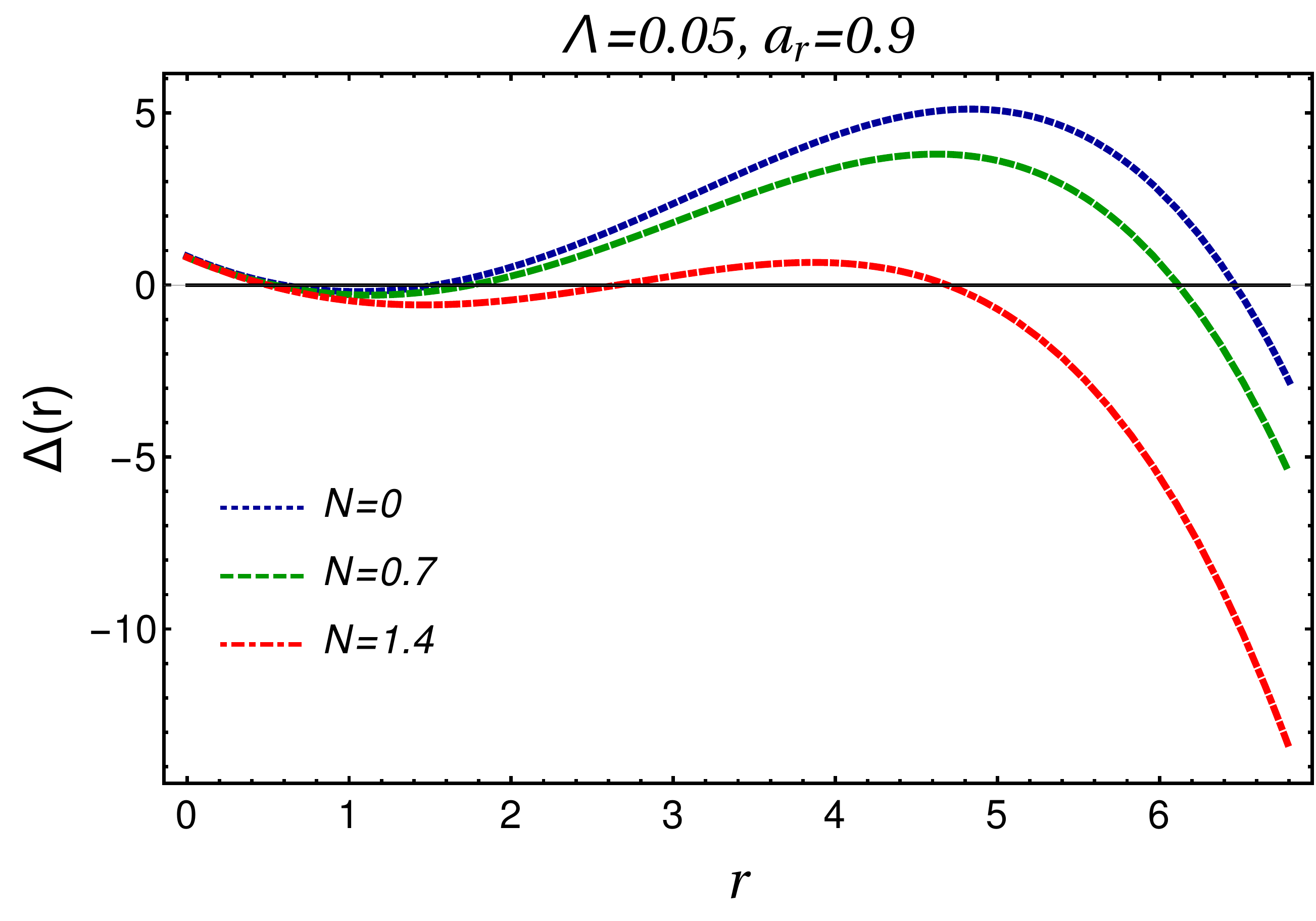}
	\endminipage\hfill
	\minipage{0.32\textwidth}%
	\includegraphics[width=\linewidth]{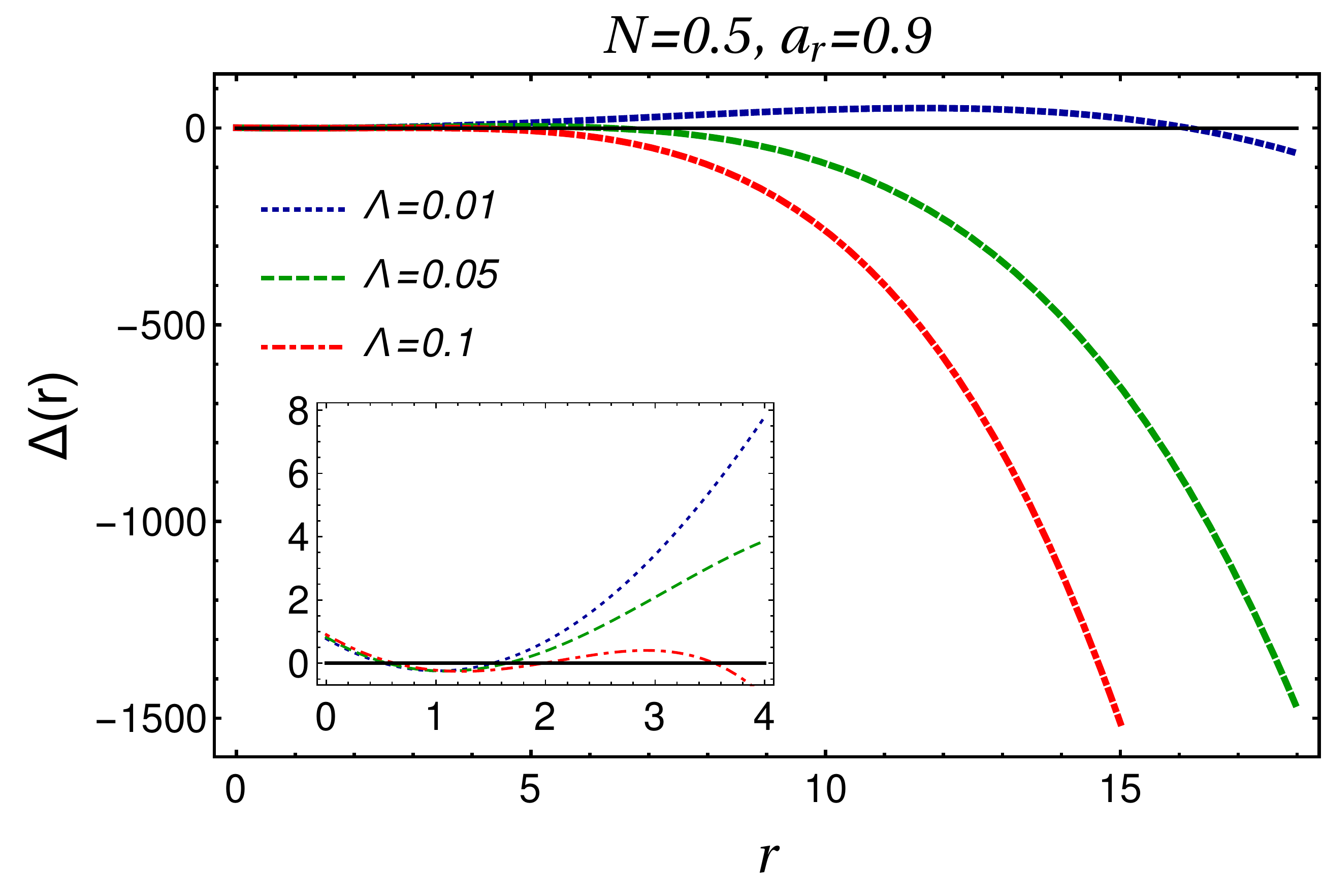}
	\endminipage
	\caption{The variation of the function $\Delta(r)$ with $r$ has been presented for different values of Kerr-NUT-\dS black hole parameters $N$, $a$ and $\Lambda$, with black hole mass $M$ set to unity. Here, $a_{r}$ is the ratio $(a/a_{\rm max})$, where $a_{\rm max}$ denotes the extremal value of black hole rotation parameter for a given value of $N$ and $\Lambda$.}\label{horizons_a}
\end{figure}

In this section we will see what effect NUT charge has on the violation/restoration of \scc\ for rotating black hole spacetimes. We start by writing down the metric for the Kerr-NUT-\dS spacetime, which can be expressed in the following form \cite{grenzebach2014photon}, 
\begin{align}\label{29}
ds^{2}&=-\frac{1}{\Sigma}\left(\Delta_r - a^2 \Delta_{\theta} \sin^2\theta \right) dt^2  
+\Sigma\left(\frac{1}{\Delta_r} dr^2 + \frac{1}{\Delta_{\theta}} d\theta^2\right) 
+ \frac{2}{\Sigma}\left\{\Delta_r \chi - a\left(\Sigma + a\chi \right)\Delta_{\theta} \sin^2\theta \right\} dt d\phi
\nonumber
\\
&+\frac{1}{\Sigma}\left\{\left(\Sigma+ a\chi\right)^2 \Delta_{\theta} \sin^2\theta - \Delta_r\chi^2\right\} d\phi^2~,
\end{align}
where various quantities introduced above have the following expression in terms of the black hole hairs, i.e., mass $M$, rotation parameter $a$ and nut charge $N$, such that,
\begin{align}
\Sigma &\equiv r^2 + \left(N+a \cos\theta\right)^2~;\qquad
\chi \equiv a \sin^2\theta -2N \left(\cos\theta + C\right)~,
\nonumber
\\
\Delta &\equiv r^2 -2mr + a^2 - N^2~;\qquad
\Delta_r \equiv \Delta - \Lambda \left\{\left(a^2-N^2\right) N^2 + \left(\frac{1}{3}a^2 +2 N^2\right)r^2 + \dfrac{1}{3} r^4\right\}~,
\nonumber
\\
\Delta_{\theta} &\equiv 1 + \Lambda \left(\frac{4}{3} aN \cos\theta + \frac{1}{3} a^2 \cos^2\theta\right)~.
\end{align}
As in the case of \RN-NUT-\dS black hole, the horizons of the Kerr-NUT-\dS black hole also correspond to real and positive roots of the equation $\Delta(r)=0$. Since analytical expressions for the horizons in terms of black hole parameters are involved, we have plotted $\Delta(r)$ as a function of $r$ for different values of black hole parameters $\Lambda$, $a$ and $N$ in \ref{horizons_a}. Here we have set the value of the black hole mass $M$ to be unity. Similar to the spherically symmetric case, we focus only on the black hole configuration which possesses three horizons, namely the cosmological, the event and the Cauchy horizon, denoted by $r_{c}$, $r_{+}$ and $r_{-}$ respectively, such that $r_{c}\geq r_{+}\geq r_{-}$. The surface gravity at the respective horizons are given by the following expression
\begin{align}
\kappa_{X} =\bigg|\frac{\Delta_{r}'(r)}{2\left(\Sigma+a\chi\right)}\bigg|_{r=r_X}=\bigg|\frac{3(r_{X}-M)-2\Lambda r^3_{X}-\Lambda r_{X}\left(a^2+6N^2\right)}{3 \left(r^2_X+a^2+N^2\right)}\bigg|~,\qquad X\in \{c,+,-\}~.
\nonumber
\end{align}
Thus given location of any horizon as a function of black hole parameters, one can determine the corresponding surface gravity as a function of black hole parameters as well. Also note that for $\Lambda=0$, the above expression for surface gravity coincides with the one derived in \cite{Mukherjee:2018dmm}.

In the Kerr spacetime one normally considers photon orbits on the equatorial plane and hence compute the parameter $\beta$ associated with perturbation around photon circular orbits. However, in presence of NUT charge one can explicitly demonstrate that even though $\theta=(\pi/2)$ is a solution of the equation $\dot{\theta}=0$, it will not satisfy the equation $\ddot{\theta}=0$. Thus even if a photon starts initially on the equatorial plane, eventually they will end up moving away from the equatorial plane. Thus except for very specific choices of the impact parameter $(L/E)$, in general there are no circular photon orbit on the equatorial plane. Still for small values of the NUT parameter it follows that the photon will hover around the equatorial plane for a longer time. Following which we will discuss the orbital dynamics of a photon on the equatorial plane for completeness, but one should keep in mind that strictly speaking such orbits will eventually move away from the equatorial plane.     

The orbital dynamics on the equatorial plane can be determined by substituting $\theta=(\pi/2)$ in the expression for the Kerr-NUT-\dS metric presented in \ref{29}. Using the ADM decomposition performed in \ref{AppA}, we can now relate the metric coefficients in \ref{29} to the general formalism developed in \cite{Rahman:2018oso}, and hence determine the parameter $\beta$. These yield,
\begin{align}\label{30}
g_{tt}&\equiv \mathbb{A}=-\frac{\Delta_r - a^2 \Delta_{\theta}}{\Sigma}~;\qquad
g_{t\phi}\equiv \mathbb{B}=\frac{\Delta_r \chi - a\left(\Sigma + a\chi\right)\Delta_{\theta} \sin^2\theta}{\Sigma}
\nonumber
\\
g_{\phi\phi}&\equiv \mathbb{C}=\frac{\left(\Sigma+ a\chi\right)^2 \Delta_{\theta}  - \Delta_r\chi^2}{\Sigma}~;\qquad
g^{rr}\equiv \mathbb{D}=\frac{\Delta_{r}}{\Sigma}~.
\end{align}
Given this one can explicitly write down the geodesic equation for null trajectories in terms of these four quantities defined above. In addition, it will also depend on the energy $\mathcal{E}$ and angular momentum $\mathcal{L}$ associated with the trajectory of the null geodesic, such that we obtain,
\begin{align}\label{25}
\dot{r}^2 =\mathcal{E}^{2} \mathbb{D}\left[\frac{\mathbb{C}-2\mathbb{B}\tilde{\ell}+\mathbb{A}\tilde{\ell}^2}{\mathbb{B}^2 - \mathbb{A}\mathbb{C}}\right]~,
\end{align}
where we have defined the specific angular momentum to be $\tilde{\ell}=(\mathcal{L}/\mathcal{E}$). The quantity on the right hand side corresponds to the effective potential, the null geodesic experiences. Setting this effective potential to be vanishing we will obtain an expression for the specific angular momentum $\tilde{\ell}$ in terms of the metric parameters. Further, setting the derivative of the potential to be vanishing, we obtain the algebraic equation, the solution of which correspond to the radius of the photon circular orbit, in the following form,
\begin{equation}
\left[\mathbb{A} \frac{d\mathbb{C}}{dr} - \frac{d\mathbb{A}}{dr} \mathbb{C}\right]^2 = 4 \left[\mathbb{A}\frac{d\mathbb{B}}{dr}- \frac{d\mathbb{A}}{dr}\mathbb{B}\right]\left[\mathbb{B}\frac{d\mathbb{C}}{dr} - \frac{d\mathbb{C}}{dr}\mathbb{B}\right]~.
\end{equation}
Subsequently, one can relate the double derivative of the effective potential to the instability timescale associated with the motion of the photon on this circular orbit. That can also be presented in terms of the metric coefficients $\mathbb{A}$, $\mathbb{B}$, $\mathbb{C}$ and $\mathbb{D}$ along with their derivatives and the specific angular momentum $\tilde{\ell}$ following \cite{Rahman:2018oso}. Similarly, the surface gravity at the Cauchy horizon can also be determined in terms of the metric elements presented above. This analysis explicitly shows how the parameter $\beta$ can be determined for photon orbits and hence one can comment on the validity of \scc. This again brings us back to the question, whether the \qnm modes indeed experience the potential of a radial null geodesic. Following which, we have explicitly demonstrated in \ref{App_QNM} that even for Kerr-NUT-\dS spacetime the \qnm modes in the eikonal approximation experience the potential of a radial null geodesic and hence the instability of the photo sphere gets related to the associated \qnm modes. This will provide us a theoretical backdrop to compute the \qnm modes, which can be compared with the numerical analysis. 

In addition to the photon sphere modes there are two additional modes which may contribute to the \qnm mode spectrum. The first one corresponds to the near extremal mode, which appears to dominate the \qnm mode spectrum in the extremal limit and the other one is the de Sitter mode which contributes as the event and cosmological horizons coincides. In the case of \RN-NUT-\dS black hole we could provide an analytical expression for the near extremal modes, however, in the present context, due to complicated nature of the metric elements it is not possible to give an analytic expression, rather we have presented the near extremal modes through numerical analysis which we will present below. On the other hand, for the de Sitter modes the same consideration as in the previous section will apply, since asymptotic structure of the metric will be identical and one may argue that  that it may still coincide with the near extremal modes. As we will see such is really the scenario here. Thus we will not discuss the \dS modes further in our analysis. Below we present the simplification of the Teukolsky equation for Kerr-\dS-NUT spacetime, which can be used in the numerical analysis in order to determine the associated \qnm modes.
\section{Teukolsky equation for a scalar field in Kerr-de Sitter-NUT spacetime}\label{SecScalar}

In order to proceed with the numerical computation of the \qnm modes associated with the scalar perturbation of the Kerr-dS-NUT spacetime, we need the scalar perturbation equation to decouple into angular and radial part. Subsequently, the angular part of the perturbation equation must be expressed in a tractable form if one hopes to solve for the \qnm mode frequencies. To start with one expresses the scalar field equation $\Box \Phi = \xi_C \mathfrak{R} \Phi$ in the Kerr-dS-NUT background, which can be written as (see \ref{scalar_eq_a1} in \ref{AppB}),
\begin{align}\label{scalar_eq_01}
&\left\{\frac{\Delta_r \chi^2 - (\Sigma + a\chi)^2 \Delta_\theta \sin^2 \theta}{\Delta_r \Delta_{\theta} \sin\theta} \right\}\partial_t^{2}\Phi 
+\sin\theta\partial_r(\Delta_r\partial_r \Phi)
+\partial_{\theta} (\sin\theta~\Delta_\theta\partial_\theta \Phi)
\nonumber
\\
&\qquad+\left\{\frac{\Delta_{r}-a^2\Delta_\theta \sin^2\theta}{\Delta_r \Delta_{\theta} \sin\theta}\right\}\partial_{\phi}^{2}\Phi
+ 2\left\{\frac{\Delta_r \chi - a(\Sigma + a\chi)\Delta_{\theta} \sin^2\theta}{\Delta_r\Delta_{\theta} \sin\theta}\right\}\partial_t\partial_\phi \Phi
-\frac{2\Lambda}{3}\Sigma \sin \theta \Phi=0~.
\end{align}
Owing to the symmetries of the problem, the metric components of the Kerr-dS-NUT spacetime neither depend on $t$ nor on $\phi$. This suggests the following anasatz for the scalar field $\Phi$, namely $\Phi = e^{i\omega t} e^{-im\phi} \Theta(\theta) R(r)$. Substituting the above ansatz in the above equation, we obtain two separate equations for $R(r)$ and $\Theta(\theta)$, yielding (for a derivation see \ref{AppB}),
\begin{align}
\frac{d}{dr}\left(\Delta_r \frac{dR}{dr}\right)&+\frac{m^2 a^2}{\Delta_r} R - \frac{2 m \omega a (\Sigma +a \chi)}{\Delta_r }R 
+\frac{\omega^2 (\Sigma + a\chi)^2}{\Delta_r}R-\frac{2\Lambda}{3}r^{2}R=kR~,
\label{scalar_radial_01}
\\
\dfrac{1}{\sin\theta}\frac{d}{d\theta}\left(\Delta_\theta \sin\theta \frac{d\Theta}{d\theta}\right)&-\frac{m^2}{\Delta_\theta \sin^2 \theta} \Theta 
+ \frac{2 m \omega \chi}{\Delta_\theta \sin^2 \theta} \Theta - \frac{\omega^2 \chi^2}{\Delta_\theta \sin^2 \theta} \Theta 
-\frac{2\Lambda}{3}\left(N+a\cos\theta\right)^{2}\Theta=-k\Theta~.
\label{scalar_angular_01}
\end{align}
One may wonder why \ref{scalar_radial_01} is being referred to as the radial part, even though both $\Sigma$ and $\chi$ depends on the angular coordinate $\theta$ explicitly. This is because, one has the following identity obtained from $\Sigma$ and $\chi$, as,
\begin{equation}\label{42}
\left(\Sigma + a\chi \right)=r^{2}+\left(N+ a \cos\theta\right)^{2}+a^{2}\sin^{2}\theta-2aN\left(\cos\theta+C\right)=r^{2}+N^{2}+a^{2}-2aNC~,
\end{equation}
where $N$ stands for the NUT charge. Thus as evident $(\Sigma + a\chi)$ is a function of the radial coordinate alone and hence \ref{scalar_radial_01} depends solely on the radial coordinate. In the above set of equations $k$ serves as the separation constant between the radial and angular equation, which is often referred to as the Carter constant. In the next section we will rewrite the angular equation in a suitable form, which can be solved using the series solution method without any further complication. After the angular equation has been studied we will discuss the radial equation and numerical techniques to find out the \qnm modes.
\subsection{Angular equation in the Kerr-de Sitter-NUT spacetime}\label{SecAng}

In this section we will exclusively consider the angular equation for the scalar perturbation presented above. To start with we re-express the angular equation by introducing a new co-ordinate $x$, in place of the angular coordinate $\theta$, such that $ x=\cos\theta$. Using the fact that $\sin\theta d\theta=-dx$, \ref{scalar_angular_01} can be written as,
\begin{align}\label{angular_01}
\frac{d}{dx}\left[\left(1-x^{2}\right)\Delta_\theta(x)\frac{d\Theta}{dx}\right] + k\Theta 
-\frac{\left[\omega \chi(x) - m\right]^{2}}{\Delta_\theta(x)\left(1-x^{2}\right)}\Theta
-\frac{2\Lambda}{3}\left(N+ax\right)^{2}\Theta=0~.
\end{align}
The quantities appearing in the above equation, e.g., $\Delta_{\theta}$, $\chi$ has the following expressions in terms of the variable $x$,
\begin{align}
\chi(x)&=a\left(1-x^{2}\right)-2N\left(x+C\right)= a-2NC-ax^{2}-2Nx~,
\nonumber
\\
\Delta_{\theta}(x)&=1+\left(\dfrac{4}{3}aN\Lambda\right)x+\dfrac{\Lambda a^2 x^2}{3}\equiv 1+\gamma x+\delta x^{2}~,  
\end{align}
where, we have introduced the following two quantities $\gamma$ and $\delta$, such that $\gamma=(4/3)aN\Lambda$ and $\delta=(1/3)\Lambda a^2$. In terms of these definitions and expansions of various quantities appearing in \ref{angular_01}, the same can be expressed as,
\begin{align}\label{angular_02}
&\frac{d}{dx}\left[\left(1 - x^2\right)\left(1+ \gamma x + \delta x^2\right)\frac{d\Theta}{dx}\right] 
+k\Theta 
-\frac{\left[\omega \left(a - 2NC\right)-m-a\omega x^{2}-2\omega Nx\right]^2}{\left(1 - x^{2}\right)\left(1+ \gamma x + \delta x^2\right)}\Theta
-2\delta\left(x+\frac{N}{a}\right)^{2}\Theta=0
\end{align}
The above equation can be expressed in a more suggestive form by defining two new constants, $\xi\equiv \omega a$ and $\eta\equiv \omega N$, which transforms \ref{angular_02} to the following form,
\begin{align}\label{angular_03}
\frac{d}{dx}\left[\left(1-x^2\right)\left(1+ \gamma x + \delta x^2\right)\frac{d\Theta}{dx}\right] 
+k\Theta  
-\frac{\left[\xi\left(1-x^2\right)-\left(m+2NC\omega\right)-2N\omega x\right]^2}{\left(1-x^2\right)\left(1+ \gamma x + \delta x^2\right)}\Theta
-2\delta\left(x+\frac{N}{a}\right)^{2}\Theta=0
\end{align}
Note that, in the limit of vanishing NUT charge, the parameter $\gamma$ and $\eta$ defined above identically vanishes and the angular equation simplifies to that of a scalar field in Kerr-dS spacetime. The potential appearing in the above angular equation, i.e., \ref{angular_03} can be simplified and hence the angular equation reads (see \ref{AppB}),
\begin{align}\label{angular_04}
\frac{d}{dx}\left[\left(1 - x^2\right)\left(1+ \gamma x + \delta x^2\right)\frac{d\Theta}{dx}\right]&
+\Bigg[k+ \frac{\xi^2}{\delta}+\frac{1}{(1+ \gamma x + \delta x^2)}\left\{\frac{-(\delta + 1) \xi^2}{\delta} + 2\xi(m + 2C\eta) + 4 \eta^2\right\}
\nonumber
\\
-\frac{x}{(1+ \gamma x + \delta x^2)}\left\{\frac{\gamma \xi^2}{\delta}-4\xi \eta\right\}&
-\frac{(m + 2C\eta)^2 + 4 \eta^2}{(1-x^2) (1+ \gamma x + \delta x^2)} - \frac{ 4\eta x (m + 2C\eta)}{(1-x^2) (1+ \gamma x + \delta x^2)}
-2\delta\left(x+\frac{N}{a}\right)^{2}\Bigg] \Theta = 0 
\end{align}
As evident, the above ordinary differential equation has 4 regular singular points at the solutions of the following algebraic equation: $\left(1-x^2\right)\left(1 + \gamma x + \delta x^2\right)=0$, which are located at $x=\pm 1$ and $x=x_\pm$, where the quantities $x_{\pm}$ are given by,
\begin{align}\label{angular_roots}
x_\pm=\frac{-\gamma \pm \sqrt{\gamma^2 - 4\delta}}{2 \delta}
= -\frac{\gamma}{2 \delta} \pm \frac{i}{\sqrt{\delta}} \sqrt{1-\dfrac{4\gamma^2}{\delta}}
\end{align} 
Note that for $N=0$, i.e., for Kerr-dS spacetime, we have $\gamma=0=\eta$ and hence except for the regular singularities at $x=\pm 1$, the other singularities are present at $x_\pm^{\rm KdS}=\pm(i/\sqrt{\delta})$.

At this stage it is customary to introduce another new co-ordinate $z$, which is related to the old co-ordinate $x$ in the following manner,
\begin{equation}\label{ang_coord}
z\equiv \frac{ x+1}{2} \frac{1- x_+}{x - x_+}
\end{equation}
Besides, it is useful to introduce the following definitions, 
\begin{align}\label{definition_new}
z_\infty&=-\frac{x_+ -1}{2}=\frac{1}{2}\left(1+\frac{\gamma}{2\delta}-\frac{i}{\sqrt{\delta}} \sqrt{1-\frac{\gamma^2}{4\delta}}\right)~,
\nonumber
\\
z_s &=- \frac{(1+ x_-)(1 - x_+)}{2(x_+ - x_-)}= \frac{i\sqrt{\delta}}{4 \sqrt{1-\frac{\gamma^2}{4\delta}}} \left[\left(1 - \frac{i}{\sqrt{\delta}} \sqrt{1-\dfrac{\gamma^2}{4\delta}}\right)^2 - \frac{\gamma^2}{4\delta^2}\right]~.
\end{align}
Thus the angular equation can be rewritten in terms of this new variable $z$ along with the quantities $z_{s}$ and $z_{\infty}$ defines above as (see \ref{AppB} for a derivation),
\begin{align}
\label{Ang_Eq_02}
\frac{d^2 \Theta}{dz^2}+\left(\frac{1}{z}+\frac{1}{z - 1}+\frac{1}{z - z_s}-\frac{2}{z - z_\infty}\right)\frac{d\Theta}{dz}
+\frac{4 (x - x_+)^4}{\delta (1 - x^2) (x - x_+)(x - x _-)(x_{+}^2 - 1)^2}V(x)\Theta=0~,
\end{align}
where the potential $V(x)$ introduced above has the following form,
\begin{align}\label{Ang_Eq_03}
V(x)&=k+ \frac{\xi^2}{\delta}+\frac{1}{(1+ \gamma x + \delta x^2)}\left\{\frac{-(\delta + 1) \xi^2}{\delta} + 2\xi(m + 2C\eta) + 4 \eta^2\right\}
\nonumber
\\
&-\frac{x}{(1+ \gamma x + \delta x^2)}\left\{\frac{\gamma \xi^2}{\delta}-4\xi \eta\right\}
-\frac{(m + 2C\eta)^2 + 4 \eta^2}{(1-x^2) (1+ \gamma x + \delta x^2)} - \frac{ 4\eta x (m + 2C\eta)}{(1-x^2) (1+ \gamma x + \delta x^2)}
-2\delta\left(x+\frac{N}{a}\right)^{2}
\end{align}
Having derived a more tractable form of the angular equation, depicting all the regular singular points explicitly, let us briefly digress to consider how the above scenario works out in the case of vanishing NUT charge, since then we can relate it to the existing literature on Kerr-dS spacetime. As already mentioned, since $\gamma=0$, for Kerr-dS spacetime the singularities are at $x_{\pm}=(\pm i/\sqrt{\delta})$ and hence from \ref{definition_new} it immediately follows that, 
\begin{align}
z^{\rm KdS}_{s}=-\frac{i}{4\sqrt{\delta}}\left(1+i\sqrt{\delta}\right)^{2}~;\qquad
z_{\infty}^{\rm KdS}=-\frac{i}{2\sqrt{\delta}}\left(1+i\sqrt{\delta}\right)~.
\end{align}
The potential also gets simplified, since $\eta=0$ when the NUT charge vanishes. Thus we obtain,
\begin{align}
V(x)=k+ \frac{\xi^2}{\delta}+\frac{1}{(1+\delta x^2)}\left\{\frac{-(\delta + 1) \xi^2}{\delta}+2m\xi\right\}
-\frac{m}{(1-x^2)(1+\delta x^2)}-2\delta\left(x+\frac{N}{a}\right)^{2}
\end{align}
One can explicitly verify that the results presented above matches exactly with the literature of Kerr-dS spacetime \cite{Suzuki:1998vy}. This is important to ensure correctness of the analysis presented here. 

After this slight detour, let us concentrate on the decomposition of the potential $V(x)$ appearing in the above angular equation as well as the associated pre-factor. The strategy is as follows, one first starts by replacing the variable from $x$ to $z$, using \ref{ang_coord}. It will turn out that all the regular singular points can be transformed to $z=0,1,z_{s}$ and $z_{\infty}$ as the respective points in the $z$ coordinate system. The next job is to consider combinations of these terms and separate them appropriately, such that each term is singular at a specific $z$ coordinate. Adopting this strategy we can rewrite the potential term appearing in \ref{Ang_Eq_03} in terms of various singular points in the $z$ coordinate, such that the angular equation can be expressed as (see \ref{AppB} for a derivation),
\begin{equation}\label{Ang_Eq_05}
\begin{split}
\frac{d^2\Theta}{dz^2}&+\left[\frac{1}{z} + \frac{1}{z - 1} + \frac{1}{z - z_s} - \frac{2}{z - z_\infty}\right]\frac{d\Theta}{dz} 
+\Bigg[\frac{2}{(z- z_\infty)^2}+\left\{\frac{2(3x_+ -x _-)}{{x_+}^2 - 1}\right\}\frac{1}{z-z_\infty}
\nonumber
\\
&+ \frac{\textrm{C1}}{z^2}
+\left\{\textrm{C4}+ \frac{2\{1-(N/a)\}^2}{(1-x_+)(1+x_-)}\right\} \frac{1}{z} 
+\frac{\textrm{C2}}{(z-1)^2} 
+\left\{\textrm{C5}+ \frac{2\{1+(N/a)\}^2}{(x_+ +1)(x_- -1)}\right\}\frac{1}{z-1}
\nonumber
\\
&+\frac{\textrm{C3}}{(z-z_s)^2}
+\left\{\textrm{C6}-\frac{4(x_+ -x_-)}{x_{+}^2 - 1}\left(1+\frac{\{1+(N/a)\}^2}{x_{-}^2 - 1} + \frac{2(N/a)}{x_-+1}\right)\right\} \frac{1}{z-z_s}\Bigg] \Theta = 0
\end{split} 
\end{equation}
where the constant coefficients $C1$ to $C6$ introduced above have very extended analytical expressions, which for brevity has been deferred to the appendix. The interested reader may take a look at \ref{Ang_Coeff_Def} in \ref{AppB}. To re-express this equation in a tractable form, we make the following redefinition of $\Theta(z)$, which yields, 
\begin{equation}
\begin{split}
\Theta(z) = z^{A_1} (z-1)^{A_2} (z-z_s)^{A_3} (z-z_\infty) f(z)
\end{split} 
\end{equation}
where $A_{1}$, $A_{2}$ and $A_{3}$, as of now are arbitrary constants, which will be determined later. Substitution of $\Theta(z)$ in \ref{Ang_Eq_05}, yields the following differential equation for $f(z)$, which reads,
\begin{align}\label{50}
\frac{d^2 f}{dz^2}&+\left[\frac{2A_1 +1}{z} +\frac{2A_2+1}{z-1}+\frac{2A_3 +1}{z-z_s}\right]\frac{df}{dz}
+\Bigg[\frac{{A_1}^2 +\textrm{C1}}{z^2}+\frac{{A_2}^2 +\textrm{C2}}{(z-1)^2}+\frac{{A_3}^2 +\textrm{C3}}{(z-z_s)^2}
\nonumber
\\
&+ \frac{1}{z}\left\{-\left(2A_1A_2 +A_1+A_2\right) -\frac{2A_1A_3 +A_1+A_3}{z_s} -\frac{1}{z_\infty} 
+\textrm{C4}+\frac{2\{1-(N/a)\}^2}{(1-x_+)(1 + x_-)}\right\}  
\nonumber
\\
&+\frac{1}{z-1}\left\{\left(2A_1A_2 +A_1+A_2\right) -\frac{2A_2A_3 +A_3+A_2}{z_s -1}
-\frac{1}{z_\infty -1}+\textrm{C5} +\frac{2\{1+(N/a)\}^2}{(x_+ +1)(x_- -1)}\right\}
\nonumber
 \\
&+\frac{1}{z-z_s}\Bigg\{\frac{2A_1A_3 +A_1+A_3}{z_s} +  \frac{2A_2A_3 +A_3+A_2}{z_s -1}- \frac{1}{z_\infty -z_s} 
+\textrm{C6}
\nonumber
\\
&-\frac{4 (x_+ - x_-)}{x_{+}^2 - 1}\left(1+\frac{\{1+(N/a)\}^2}{x_{-}^2 - 1} + \frac{2(N/a)}{x_-+1}\right)\Bigg\}
+\frac{1}{z-z_\infty}\left\{\frac{1}{z_\infty} + \frac{1}{z_\infty -1} + \frac{1}{z_\infty -z_s} +\frac{2(3x_+ -x_-)}{x_{+}^2 - 1}\right\}\Bigg]f = 0  
\end{align}
From the definition of $x_{+}$, we obtain, $(x_{+}-1)=-2z_{\infty}$, such that, $x_{+}^2-1=4z_\infty(z_\infty -1)$. Further, we have the following relation connecting $x_{+}$ with $x_{-}$ as, $x_{-}=x_{+}-(1/2)\{(x_{+}^2 - 1)/(z_s - z_\infty)\}$. Thus using the above identities involving $z_{s}$, $z_{\infty}$ as well as $x_{\pm}$, we obtain,
\begin{equation}
\begin{split}
\frac{2 (3x_+ -x_-)}{x_{+}^2 - 1}&=\frac{2}{4z_\infty (z_\infty -1)} 
\left[3 (1 -2 z_\infty) -1 - \frac{2z_\infty(1-z_s)}{z_s - z_\infty}\right] 
\nonumber
\\
&= \frac {z_s -2 z_\infty +3 {z_\infty}^2 -2 z_s z_\infty}{z_\infty (z_\infty -1)(z_s - z_\infty)}
=-\left\{\frac{1}{z_\infty} + \frac{1}{z_\infty -1} + \frac{1}{z_\infty -z_s}\right\}
\end{split} 
\end{equation}
Therefore the coefficient of $(z-z_\infty)^{-1}$ in \ref{50} cancels out. Since we still have the freedom in choosing $A_{1}$, $A_{2}$ and $A_{3}$ we make the following choice, 
\begin{equation}
A_{1}^2=-\textrm{C1}~;\qquad A_{2}^2=-\textrm{C2}~;\qquad A_{3}^2=-\textrm{C3}
\end{equation}
Thus the coefficients of $z^{-2}$, $(z-1)^{-2}$ and $(z-z_s)^{-2}$ terms identically vanishes. Hence the final differential equation for the variable $f(z)$ becomes,
\begin{equation}\label {Ang_Eq_06}
\frac{d^2 f}{dz^2}+\left\{\frac{2A_1 +1}{z}+\frac{2A_2 +1}{z-1}+\frac{2A_3+1}{z- z_s}\right\}\frac{df}{dz} 
+\left\{\frac{\cal M}{z}+\frac{\cal N}{z-1}+\frac{\cal P}{z-z_{s}}\right\}f=0
\end{equation}
where, the constants ${\cal M}$, ${\cal N}$ and ${\cal P}$ introduced above has the following expressions,
\begin{align}\label{def_MNP}
\mathcal{M}&\equiv \textrm{C4}-\left(2A_{1}A_{2}+A_{1}+A_{2}\right)-\frac{(2A_{1}A_{3}+A_{1}+A_{3})}{z_{s}}-\frac{1}{z_\infty}+\left\{1-\frac{N}{a}\right\}^{2} \frac{2}{(1-x_{+})(1+x_{-})} 
\nonumber
\\
\mathcal{N}&\equiv \textrm{C5}+\left(2A_{1}A_{2}+A_{1}+A_{2}\right)-\frac{(2A_{2}A_{3}+A_{2}+A_{3})}{z_{s}-1}
-\frac{1}{z_\infty-1}+2\left\{1+\frac{N}{a}\right\}^2 \frac{1}{(x_{+}+1)(x_{-}-1)}
\nonumber
\\
\mathcal{P}&\equiv \textrm{C6}+\frac{\left(2A_{1}A_{3}+A_{1}+A_{3}\right)}{z_{s}} +\frac{(2A_{2}A_{3}+A_{2}+A_{3})}{z_{s}-1}
-\frac{1}{z_\infty-z_{s}} 
-\frac{4(x_{+}-x_{-})}{x_{+}^{2}-1}\left[1+\frac{(1+\frac{N}{a})^2} {x_{-}^2-1}+\frac{2(\frac{N}{a})}{x_{-}+1}\right] 
\end{align}
Expanding each of these terms, after extensive calculations we obtain from \ref{AppB}, the result $\mathcal{M+N+P}=0$. Hence the angular equation can be expressed in the following form,
\begin{equation}\label{52}
\begin{split}
\frac{d^2 f}{dz^2}+\left[\frac{2A_1 +1}{z} +\frac{2A_2 +1}{z-1} + \frac{2A_3+1}{z- z_s}\right]\frac{df}{dz} 
+\left\{\frac{\mathcal{M}z_s - [\mathcal{M}(1+ z_s) + \mathcal{N}z_s +\mathcal{P}]z}{z(z-1)(z-z_s)}\right\}f =0
\end{split}
\end{equation}
In this form, the above differential equation looks very much similar to the Heun differential equation, but there is one crucial difference. In Heun equation, the coefficient of $z$ term associated with the numerator of the quantity multiplying $f(z)$, must be expressible as product of two quantities, such that addition of them can be expressed in terms of the parameters in $(df/dz)$. For this to hold, we must be able to show that $\mathcal{M}(1+ z_s) + \mathcal{N}z_s +\mathcal{P}=\Upsilon \times \epsilon$, such that $\Upsilon+\epsilon+1=2A_{1}+2A_{2}+2A_{3}+3$. To see whether such a decomposition can be performed in the present scenario, we note that $\mathcal{M}(1 +z_s) + \mathcal{N}z_s + \mathcal{P}= \mathcal{M}z_s + \mathcal{N}(z_s-1)$, where we have used the result that $\mathcal{M+N+P}=0$. This combination can be expressed, using \ref{def_MNP} as,
\begin{align}\label{new_n_01}
\mathcal{M}z_s &+ \mathcal{N}(z_s-1)=\left\{\textrm{C4}z_s + \textrm{C5}(z_s-1)\right\} 
-\left\{\frac{z_s}{z_\infty}+\frac{z_s-1}{z_\infty-1}\right\} 
-2\left\{A_1 +A_2 + A_3 + A_1A_2 + A_2A_3 + A_3 A_1\right\}
\nonumber
\\
&+\left[\left(1-\frac{N}{a}\right)^{2} \frac{2z_s}{(1-x_{+})(1+x_{-})} 
+\left(1+\frac{N}{a}\right)^2 \frac{z_s -1}{(x_{+}+1)(x_{-}-1)}\right] 
\end{align}
The above equation can be simplified further, since each of the above terms will cancel among themselves, such that we have, 
\begin{align}
\frac{z_s}{z_\infty}+\frac{z_s-1}{z_\infty-1}&=\frac{2x_-}{(x_+ -x_-)}
\\
\left(1-\frac{N}{a}\right)^{2} \frac{2z_s}{(1-x_{+})(1+x_{-})}&+\left(1+\frac{N}{a}\right)^2 \frac{z_s -1}{(x_{+}+1)(x_{-}-1)}
=\frac{4 (N/a)}{x_+ - x_-}
\end{align}
Using these identities in \ref{new_n_01}, we obtain,
\begin{align}
\mathcal{M}z_s &+ \mathcal{N}(z_s-1)=\textrm{C4}z_s + \textrm{C5}(z_s -1) 
+ \frac{2x_-}{(x_+ -x_-)} + \frac{4 (N/a)}{x_+ - x_-}  
- 2\left\{A_1 +A_2 + A_3 + A_1A_2 + A_2A_3 + A_3 A_1\right\}
\nonumber
\\
&=\textrm{C4} z_s + \textrm{C5} (z_s -1)  
+\frac{2x_-}{(x_+ -x_-)}+\frac{4 (N/a)}{x_+ - x_-} + \left(1 + A_1^2 + A_2^2 + A_3^2\right) 
\nonumber
\\
&-\left[2A_1 +2A_2 + 2A_3 + 2A_1A_2 +2 A_2A_3 +2 A_3 A_1+ 1 + A_1~^2 + A_2~^2 + A_3~^2\right]
\nonumber
\\
&=\textrm{C4} z_s + \textrm{C5} (z_s -1) 
+ \frac{2x_-}{(x_+ -x_-)} +\frac{4 (N/a)}{x_+ - x_-} 
-\left(A_1 + A_2 +A_3 +1\right)^2 
-\left(\textrm{C1 + C2 +C3}\right)+1 
\nonumber
\\
&\equiv A_4^2 - (A_1 + A_2 +A_3 +1 )^2\equiv  -\sigma_+ \sigma_-~,
\end{align}
where we have used the following results, $A_1^2 =-\textrm{C1}$, $A_2^2=-\textrm{C2}$ and $A_3^2=-\textrm{C3}$. We have also defined two new quantities, $A_{4}$ and $\sigma_{\pm}$, such that, $A_4^2\equiv \textrm{C4} z_s + \textrm{C5}(z_s -1) +\{2x_{-}/(x_+ -x_-)\}+\{4(N/a)/(x_+ - x_-)\} -(\textrm{C1 + C2 +C3})+1$ and $\sigma_{\pm} = 1 + A_1 + A_2 +A_3 \pm A_4$. Thus \ref{52} takes the following form,
\begin{equation}\label{final_ang_eq}
\frac{d^2 f}{dz^2}+\left\{\frac{2 A_1 +1}{z} + \frac{2 A_2+1}{z-1} + \frac{2 A_3 +1}{z-z_s}\right\}\frac{df}{dz} 
+\left\{\frac{\sigma_+ \sigma_- z -\left(-\mathcal{M}z_s\right)}{z(z-1)(z-z_s)}\right\} f = 0
\end{equation}
Using the definition of $\sigma_{\pm}$ presented above, it immediately follows that,
\begin{equation}
\sigma_+ + \sigma_- +1 =\left(1 + A_1 + A_2 +A_3 + A_4\right)+\left(1 + A_1 + A_2 +A_3 - A_4\right)+1=(2 A_1 +1)+(2 A_2+1)+ (2 A_3 +1)
\end{equation}
Thus \ref{final_ang_eq} is indeed Heun's equation, since the parameters satisfy the desired relation. Thus for Kerr-de Sitter-NUT spacetime as well the angular equation can be reduced to Heun's equation and hence the angular equation will have regular singular points at $z=0,1,z_{s}$ and at $z=\infty$. Thus one can employ the Frobenius, or, series solution method to determine the solution of the above differential equation. It is well known that the series expansion of the solution around $z=0$ yields a three term recursion relation which can be solved by the method of continued fraction numerically.   

As a final understanding of this result, consider the limit of vanishing NUT charge for which we obtain, $\gamma=0=\eta$. Thus one obtains, $x_{\pm}=(i/\sqrt{\delta})$, which yields the following result, $A_{4}=A_{3}^{*}$, which matches with the results for Kerr-de Sitter spacetime \cite{Suzuki:1998vy}. Hence, in the Kerr-de Sitter-NUT spacetime the decomposition of the angular equation, though highly non-trivial, yields the same differential equation. This may hint towards some underlying symmetry associated with the angular equations in stationary spacetimes, which we hope to address in future.
\subsection{Radial equation in Kerr-de Sitter-NUT spacetime}\label{SecRad}

In this section, we consider the radial equation in the Kerr-de Sitter-NUT spacetime in order to get the quasi-normal modes associated with the scalar perturbation using numerical methods. Formally, the quasi-normal modes are defined as the solution of the perturbation equation subjected to the boundary conditions that at the cosmological horizon $r_{\rm c}$, only outgoing modes are present, whereas at the event horizon $r_{+}$, there are only ingoing modes. An important step in determining the quasi-normal modes using numerical method is to single out the diverging contribution of the radial function $R(r)$, or the behaviour near the regular singular points, so that the radial equation, given by \ref{scalar_radial_01}, can be written in terms of a Frobenius series. For our calculation, we choose the transformation of the radial function as given by \ref{Frobenious_series}, where, $B_{1}(r)$ and $B_{2}(r)$ has identical expressions as in the case of \RN-NUT-\dS spacetime with the following expression for the potential: $V_{r}(r)=(\Sigma+a\chi) \omega -a m$. Substituting the expansion written down in \ref{Frobenious_series}, in \ref{scalar_radial_01}, we obtain a differential equation for the unknown function $y(r)$. Then $y(r)$ can be expanded in a Frobenius series, which can be solved using numerical techniques.

\begin{figure}\label{kerr_numeric}
	\minipage{0.32\textwidth}
	\includegraphics[width=\linewidth]{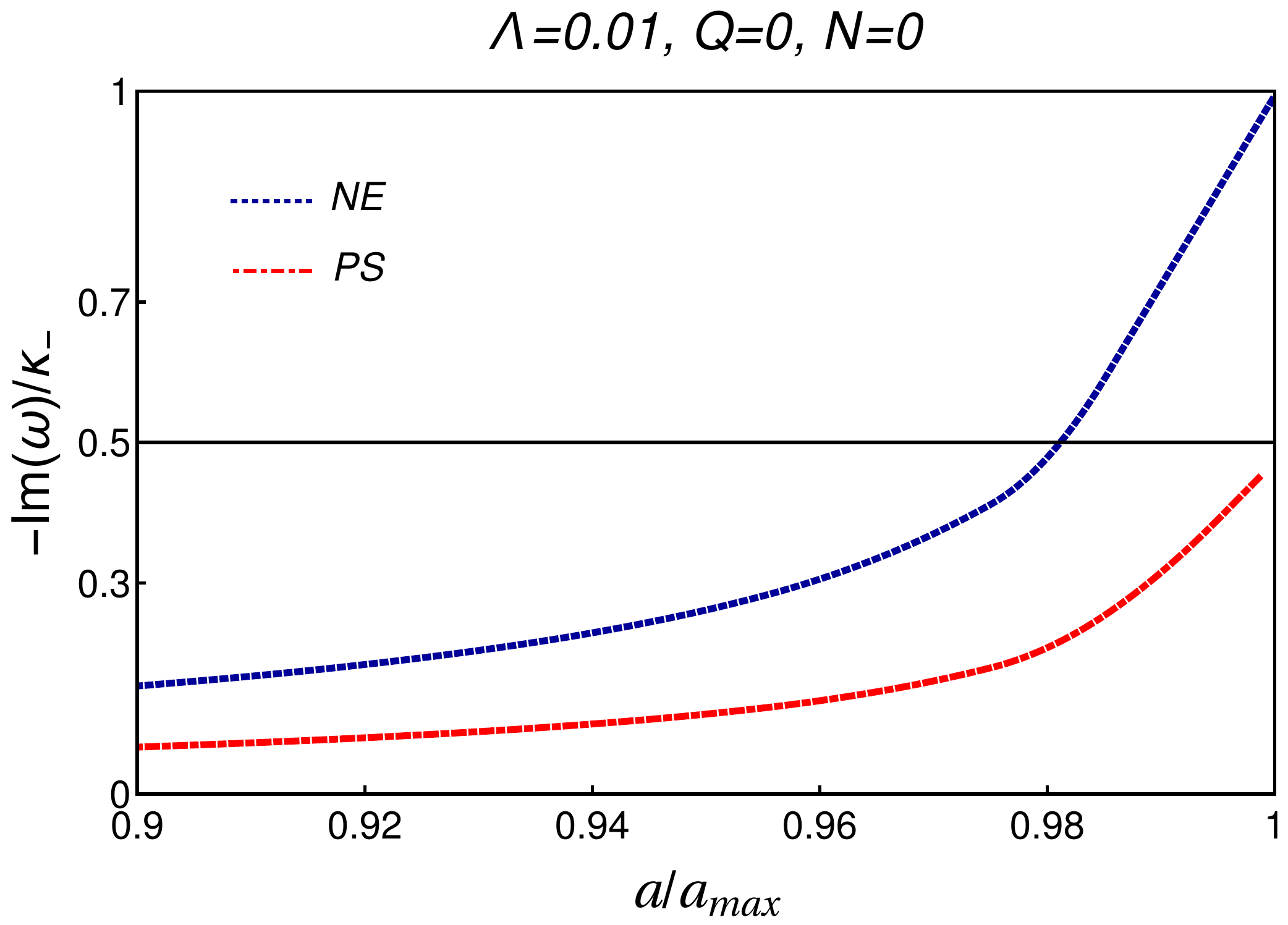}
	\endminipage\hfill
	\minipage{0.32\textwidth}
	\includegraphics[width=\linewidth]{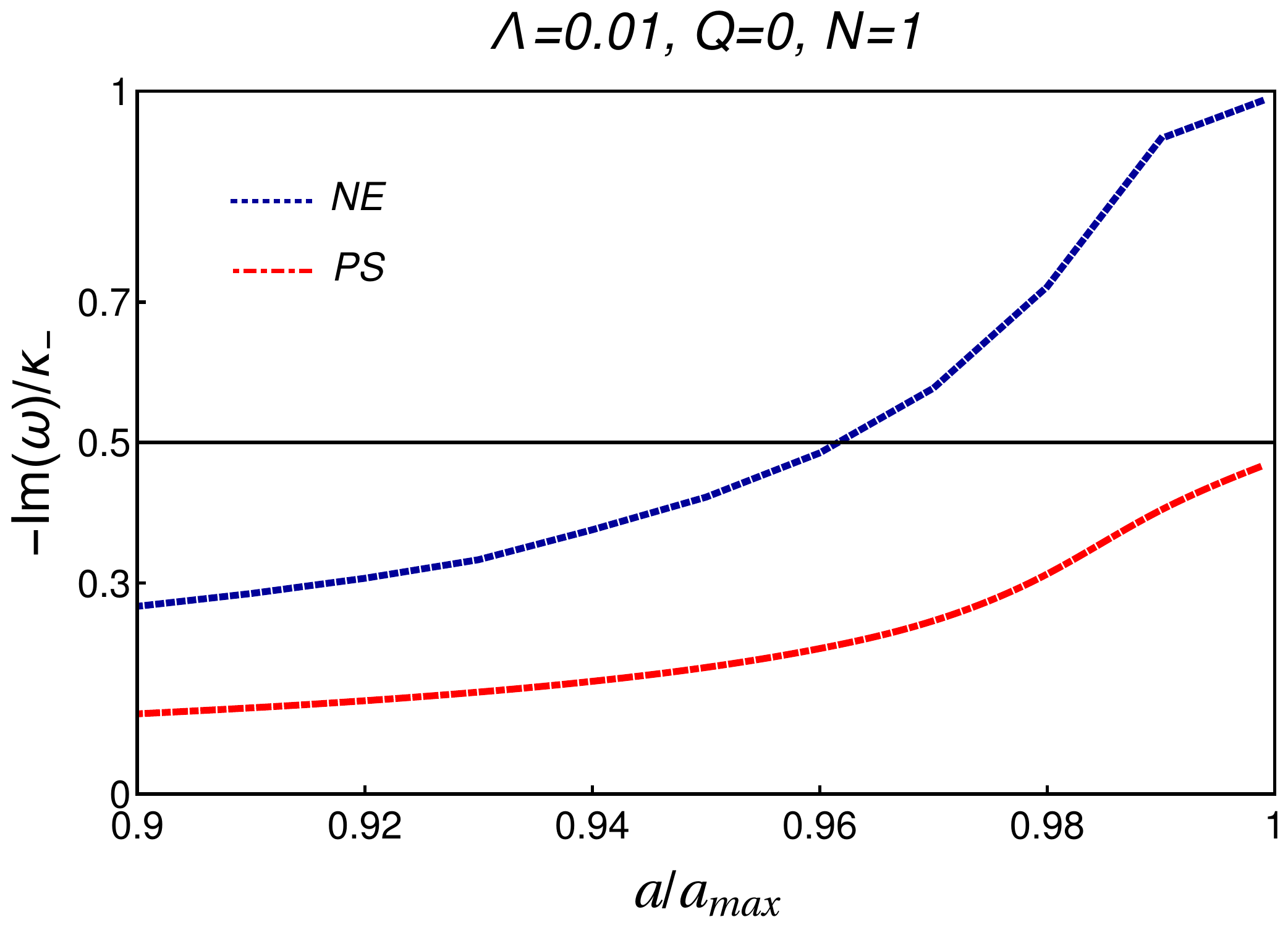}
	\endminipage\hfill
	\minipage{0.32\textwidth}%
	\includegraphics[width=\linewidth]{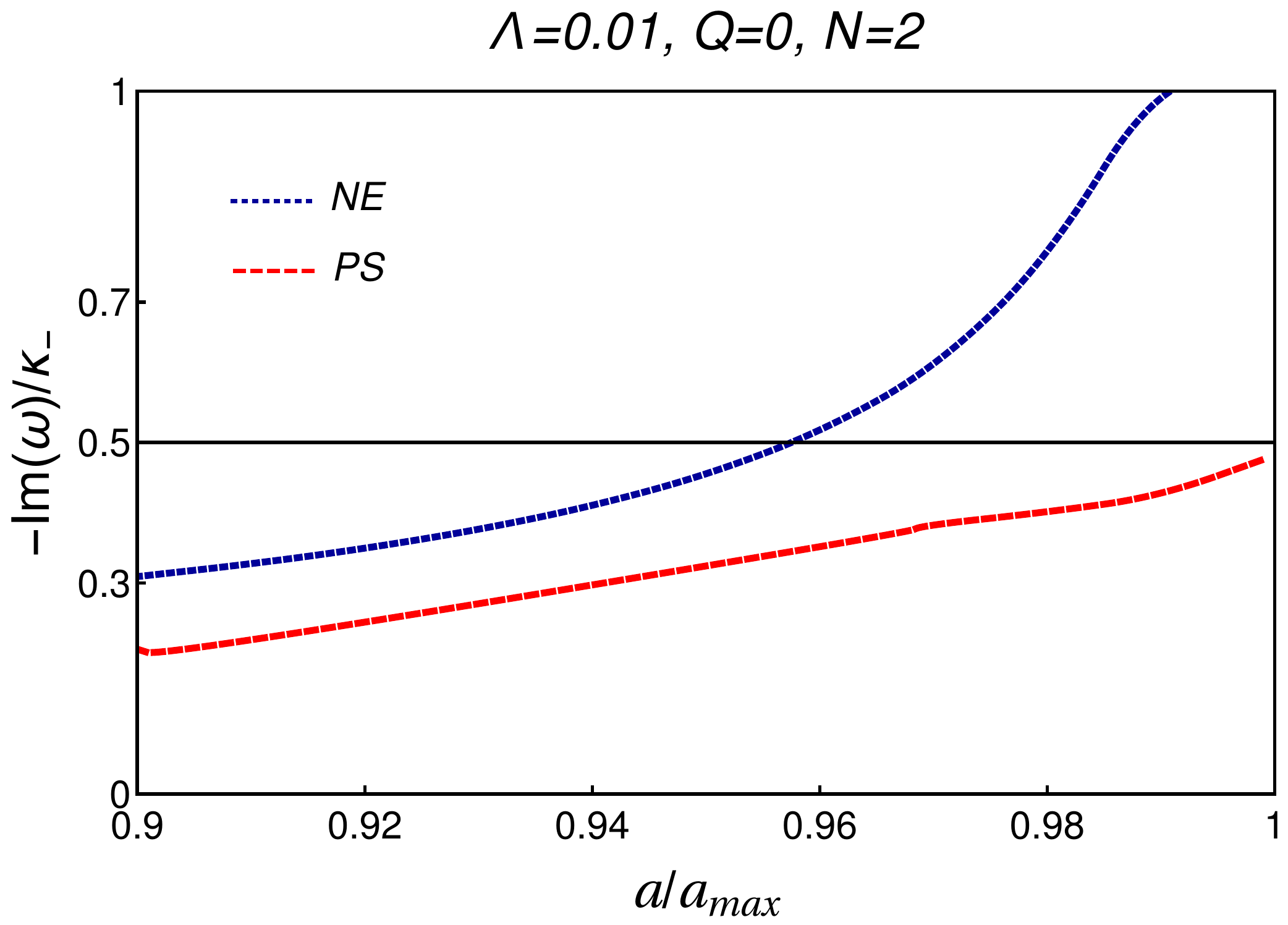}
	\endminipage\hfill
	\minipage{0.32\textwidth}
	\includegraphics[width=\linewidth]{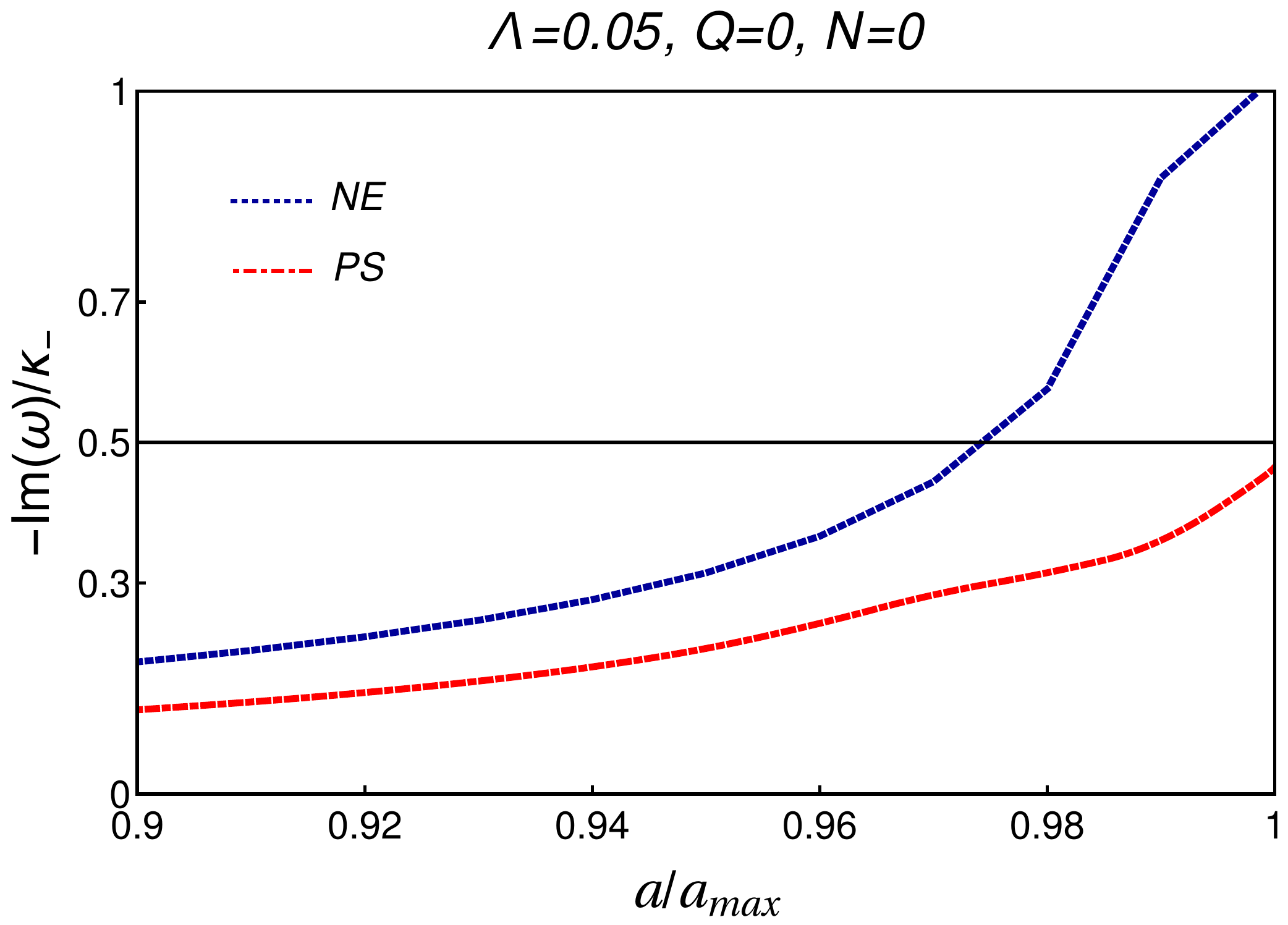}
	\endminipage\hfill
	\minipage{0.32\textwidth}
	\includegraphics[width=\linewidth]{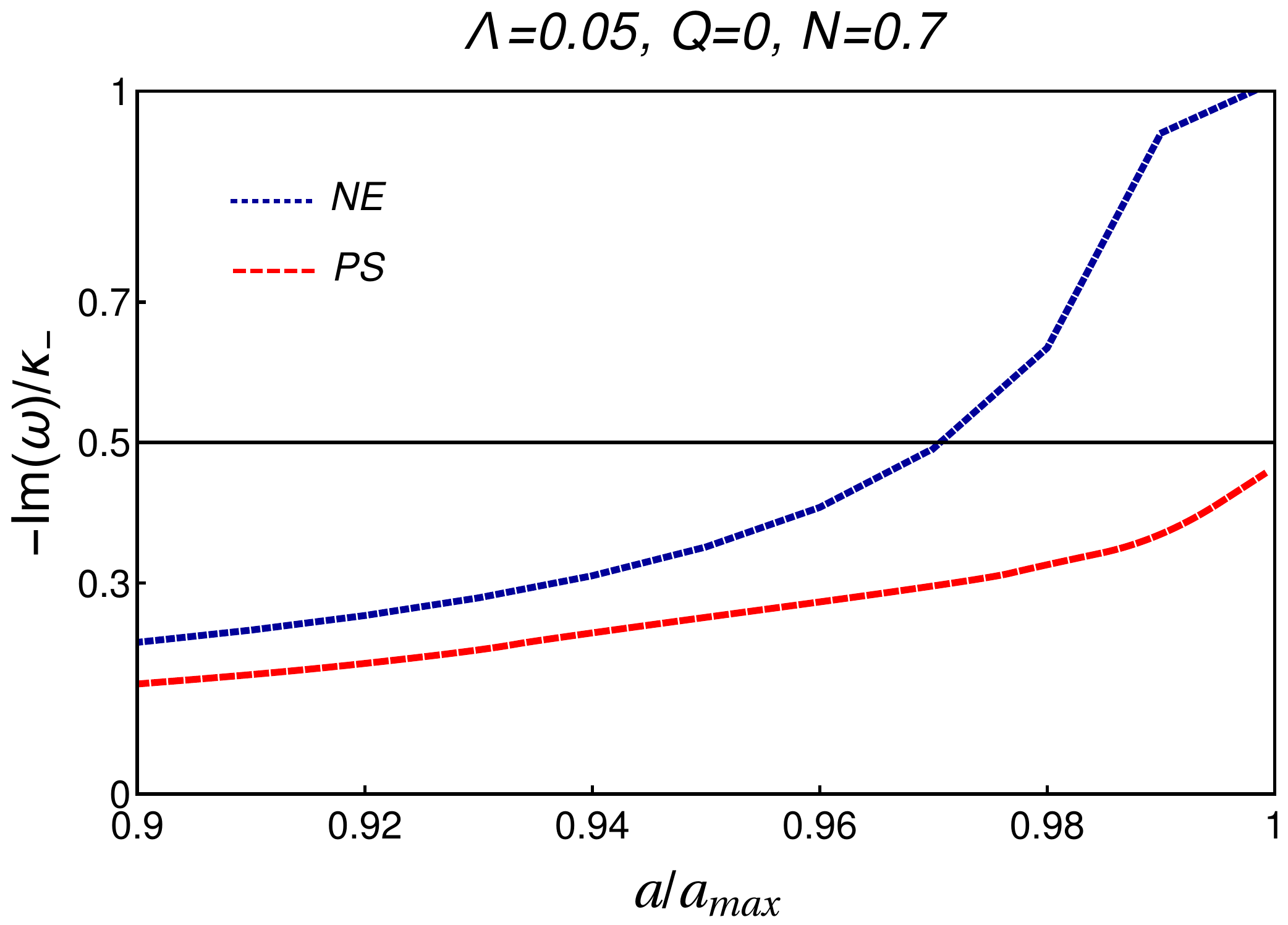}
	\endminipage\hfill
	\minipage{0.32\textwidth}%
	\includegraphics[width=\linewidth]{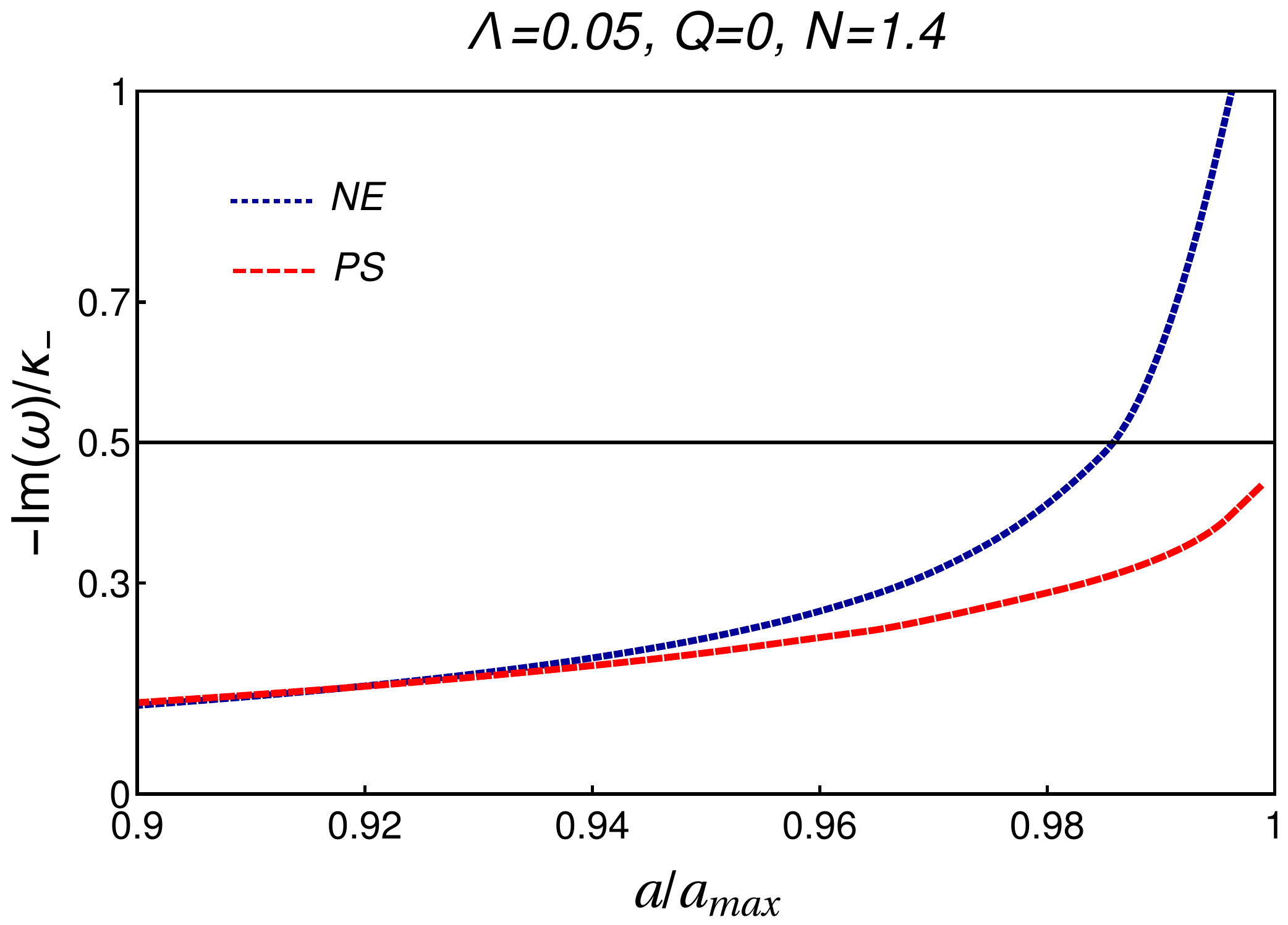}
	\endminipage\hfill
	\minipage{0.32\textwidth}
	\includegraphics[width=\linewidth]{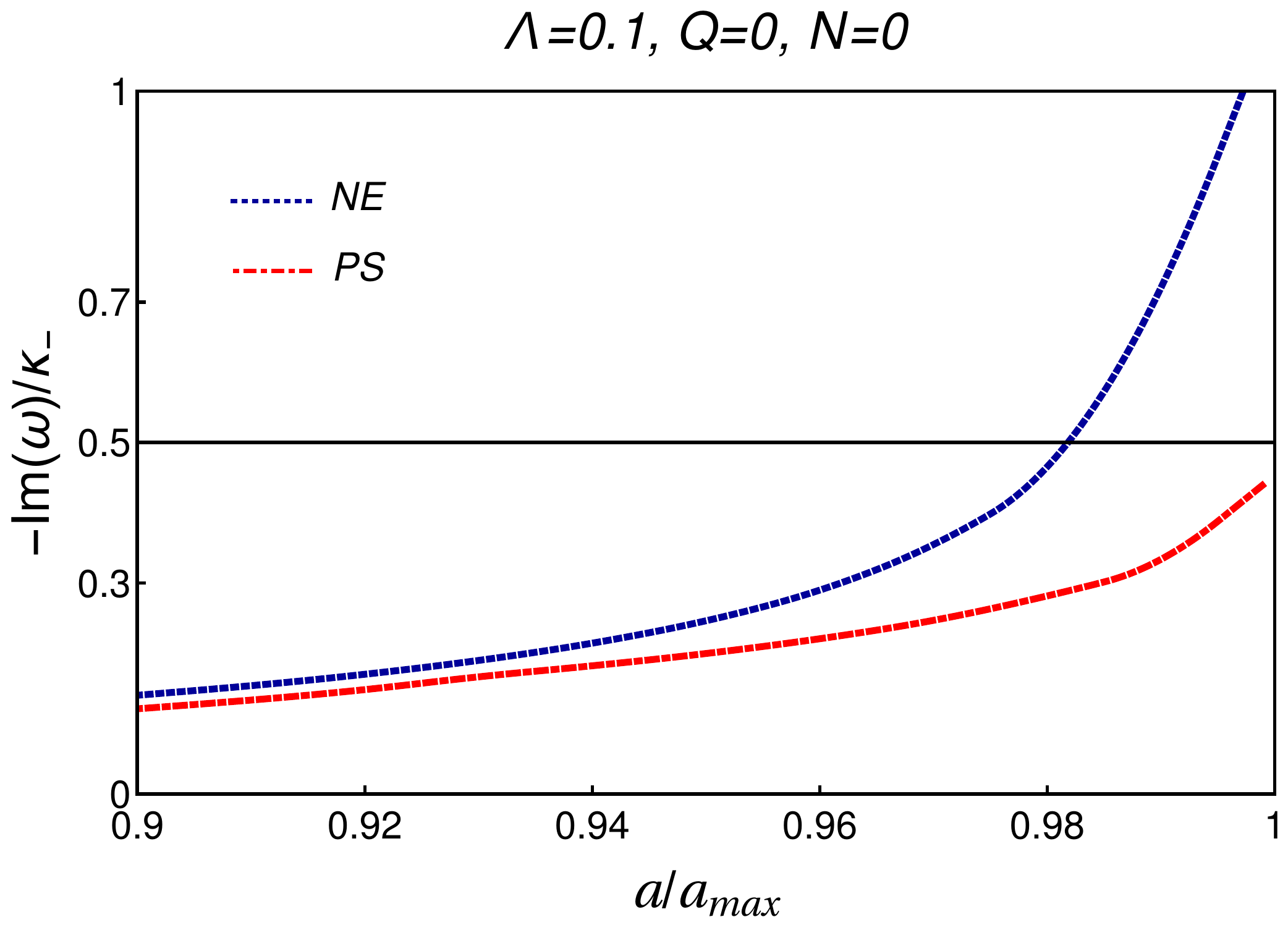}
	\endminipage\hfill
	\minipage{0.32\textwidth}
	\includegraphics[width=\linewidth]{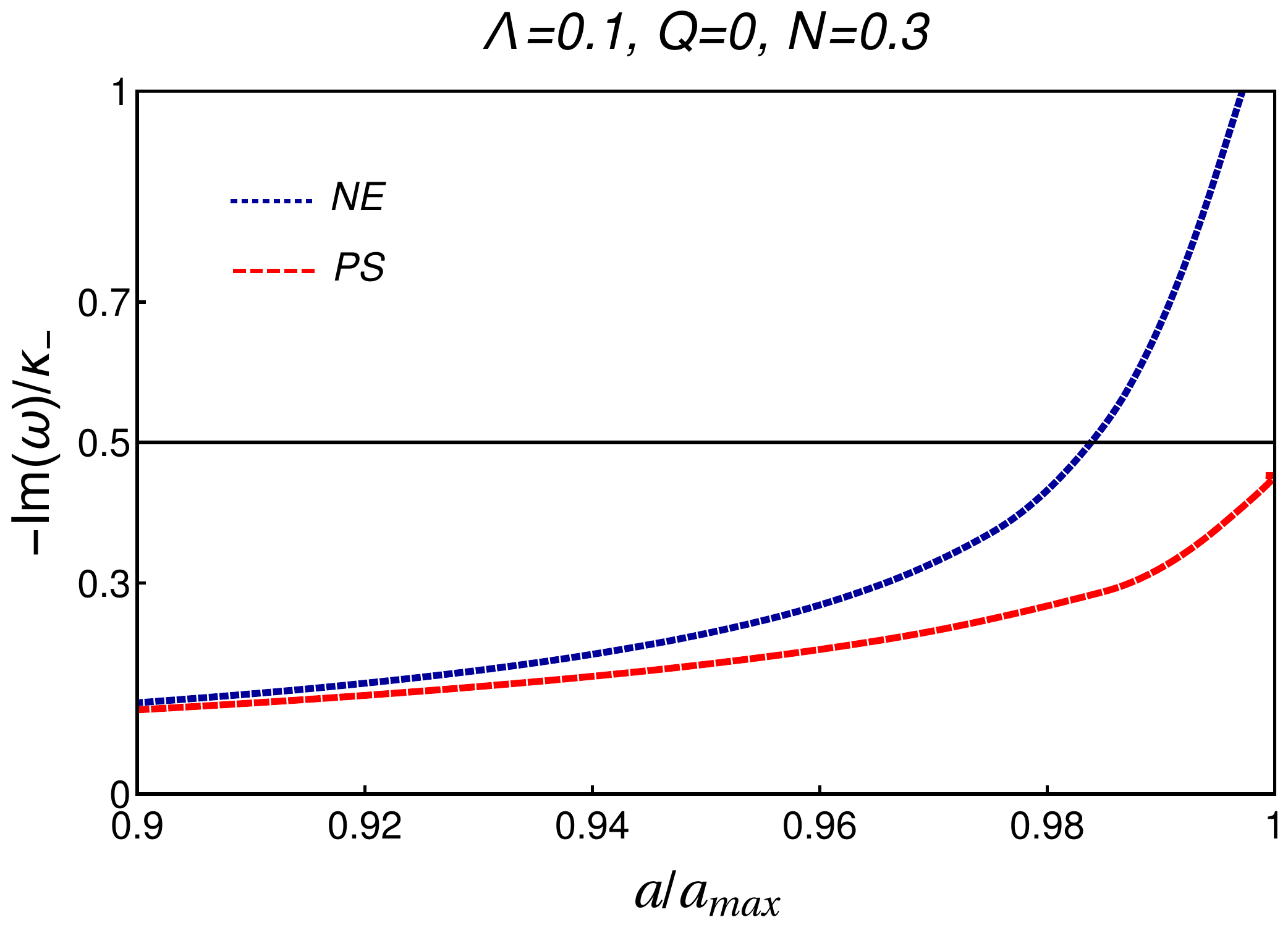}
	\endminipage\hfill
	\minipage{0.32\textwidth}%
	\includegraphics[width=\linewidth]{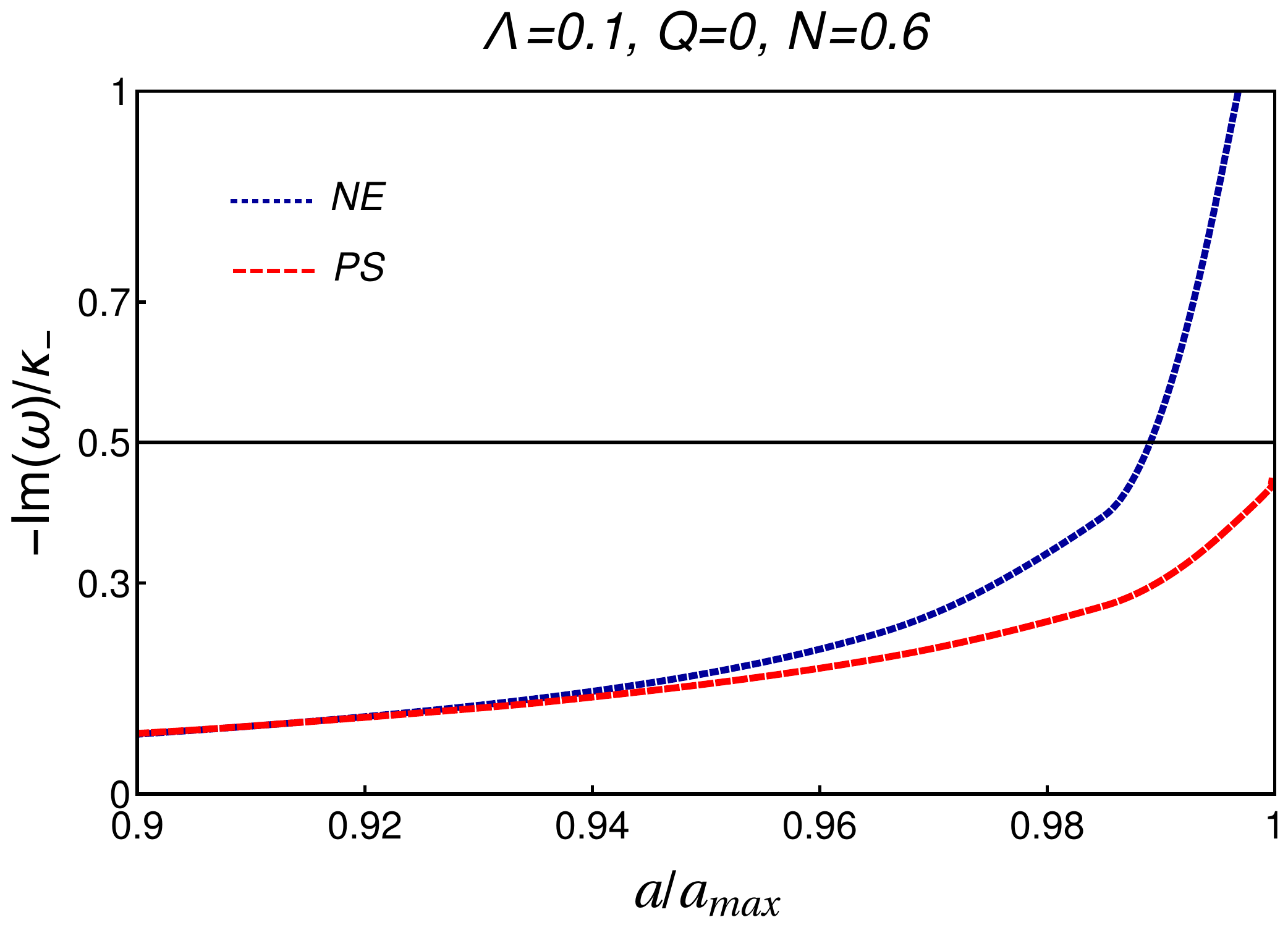}
	\endminipage
	\caption{The plot of $-\{\textrm{Im}(\omega)/\kappa_{-}\}$ as a function of $(a/a_{\rm max})$ for different values of $N$ and $\Lambda$ in a Kerr-NUT-\dS black hole is presented. Here, $a_{\rm max}$ corresponds to the maximum allowed value of the rotation parameter leading to an extremal configuration. The value of $\beta$ is determined by the lowest lying quasi-normal modes, i.e., the mode for which the value of $-\textrm{Im}(\omega)$ is the smallest. In the above plots, the blue lines denote the near extremal modes (corresponds to $\ell=0$) whereas the red lines present the photon sphere modes (corresponds to large $\ell$ value). The plots demonstrate that the photon sphere modes are dominant in all the parameter range. Furthermore, there exist no dominant mode for which the value of $\beta$ becomes greater than $(1/2)$, which means that the Kerr-NUT-\dS spacetime respects the strong cosmic censorship conjecture.}\label{fig_kerr_numeric}
\end{figure}

Note that the solution of the radial equation demands knowledge of the separation constant $k$ appearing in \ref{scalar_radial_01}. Thus we need to solve the angular equation first, given by \ref{final_ang_eq}, and then the radial equation in order to find the quasi-normal modes. In this context the fact that the angular equation can be expressed as Heun equation plays the most important role. Then we can solve the Heun equation, following \cite{Suzuki:1998vy}, leading to a three term recurrence relation. This recurrence relation can be solved using the continued fraction method in order to obtain the separation constant $k$, appearing in the angular equation, as a function of the quasi-normal mode frequency. We have also used Nollart's method to ensure convergence of the angular equation \cite{Nollert:1993zz}. Finally, by substituting the value of the separation constant in the radial equation and using the package \textit{QNMspectral} \cite{Jansen:2017oag}, we obtain the desired quasi-normal modes. The outcome of such a numerical analysis are presented in \ref{fig_kerr_numeric}, where we have depicted two different family of modes namely, the photon sphere modes and the near extremal modes, contributing towards determining the fate of the Cauchy horizon. The former are the eikonal modes which corresponds to larger values of angular momenta (see \ref{kerr_modes} for further discussion) whereas the later corresponds to the $\ell=0$ modes. Note that, the \dS family of modes are absent in the quasi-normal spectrum, due to the conformal coupling of the massless scalar field leading to a \qnm mode spectrum coinciding with the \NE modes. Since the imaginary part of the \NE modes are smaller, they will provide the dominating contribution and hence \dS modes are absent.

This has been explicitly demonstrated in \ref{fig_kerr_numeric}, which shows that the spectrum of the quasi-normal modes is always dominated by the photon sphere modes, depicted by the red curve, while the near extremal modes (blue curves) always decay faster. Interestingly, we find that there exist no dominant modes in the \qnm mode spectrum for which the value of $\beta$ becomes greater than $(1/2)$ in the full parameter space. As a result, the Kerr-NUT-\dS black holes always respect the strong cosmic censorship conjecture. This results are in accord with the previous conclusion that it is impossible to smoothly extend a scalar field across the Cauchy horizon for rotating black holes \cite{PhysRevD.97.104060,Rahman:2018oso}.   
\section{Conclusion}

The existence of Cauchy horizon is one of the most intriguing problem in classical general relativity. Since beyond the Cauchy horizon, the notion of predictability is lost. Whether a voyage to such region is possible, depends highly on the lowest lying quasi-normal modes of the black hole spacetime inheriting the Cauchy horizon. Such a voyage beyond the Cauchy horizon is indeed a violation of the deterministic nature of \gr. To protect the deterministic nature of \gr, it has been conjectured that the spacetime cannot be extended beyond the Cauchy horizon with square integrable connection, known as Christodoulou's version of \scc. In recent years, it has been found that for certain spacetimes, which are asymptotically de Sitter, there is indeed violation of the \scc. Following which there have been several studies to check this issue in different black hole spacetimes and for different fundamental fields. Following Christodoulou's treatment, we have checked whether an extension of this idea is possible for conformally coupled scalar field in the case of \RN-NUT-\dS and Kerr-NUT-\dS black holes. Recent studies have shown that such an extension is allowed in \RN-\dS black holes and its variants \cite{Cardoso:2018nvb, Dias:2018ufh,Dias:2018etb,PhysRevD.99.064014,Ge:2018vjq,Destounis:2018qnb,Rahman:2018oso, Liu:2019lon,Liu:2019rbq}, whereas Kerr-\dS black holes respects the \scc\ \cite{PhysRevD.97.104060, Rahman:2018oso}. Since, these black hole solutions are a subclass of a more general \KN\-NUT-\dS black holes, it is important to check whether extension of spacetime beyond Cauchy horizon is possible in this more general setting. In addition, we discuss the case of a conformally coupled scalar to see its implication on the \scc, in particular, whether such conformal coupling can lead to violation of the $C^{1}$ version of \scc, i.e., whether the scalar field can be extended in a regular manner across the Cauchy horizon, which is much more severe.

Following which we have first considered the effect of a conformally coupled scalar field on a \RN-NUT-\dS black hole spacetime. To assess whether there is any violation of the \scc\ in this spacetime, we have used both analytical and numerical techniques in order to determine the lowest lying quasi-normal modes which is essential to check the validity of the conjecture. In this scenario, two different family of \qnm modes, namely, the near extremal and photon sphere modes determine the fate of extendibility of metric beyond the Cauchy horizon. Interestingly, we found that in the presence of NUT charge and conformal coupling, there is no separate contribution from the \dS mode, in contrast to the case of a minimally coupled scalar field. The reason behind this being the presence of conformal coupling, which modifies the potential of the scalar field. As a result, the \dS modes coincide with the near extremal modes and hence are absent in the spectrum of dominant \qnm modes. This is evident from the numerical estimations for the \qnm modes and are presented in \ref{fig_rn_numric}.

It also follows that the analytical estimation of the quantity $\beta\equiv -\{\textrm{Im}\omega/\kappa_{-}\}$, for photon sphere modes matches quiet well with the numerical estimation as well. Furthermore, as \ref{fig_rn_numric} suggests, for a given $\Lambda$ as the NUT charge increases the dominant mode crosses the $\beta=(1/2)$ line at smaller and smaller values of $(Q/Q_{\rm max})$. Thus the presence of NUT charge leads to a stronger violation of Christodoulou's version of \scc. Surprisingly, for conformal coupling even the $C^{1}$ version of \scc\ is being violated as the dominant \qnm mode crosses the line $\beta=1$ in the near extremal region. Thus even the scalar field can be extended across the Cauchy horizon, leading to a catastrophic failure of determinism in \gr. Note that in this violation the cosmological constant and the NUT charge both plays a crucial role, for a given $\Lambda$ as the NUT charge increases, the dominant \qnm mode crosses $\beta=1$ line for smaller and smaller values of $(Q/Q_{\rm max})$. The same conclusion holds for a fixed $N$ but increasing $\Lambda$ as well, which follows directly from \ref{fig_rn_numric}.  

We have also extended our study by considering the effect of black hole rotation on violation of \scc. The presence of Killing-Yano tensor in \KN\-NUT-\dS spacetime allows us to separate the angular equation for conformally coupled scalar field into radial and angular parts. Intriguingly it turns out that it is possible to express the angular perturbation equation for \KN\-NUT-\dS in a more tracktable form, i.e., Heun equation. The Heun equation can be solved using the Frobenius method, which yields a three term recurrence relation for the angular equation. This can be solved along with the radial equation in order to determine the quasi-normal modes numerically. Following which we have presented the numerical estimation of $\beta$ in \ref{fig_kerr_numeric}.

From the plots, it is clear that the analytical results are compatible with the numerical estimation of $\beta$ for photon sphere modes. Further, from the numerical results presented in \ref{fig_kerr_numeric} it follows that the photon sphere modes are the dominant ones throughout the parameter space, i.e., for all values of $(a/a_{\rm max})$ and the value of $\beta$ never crosses $(1/2)$ for these modes. This suggests that \scc\ is respected in the rotating black hole spacetimes and the photon sphere modes play the central role in restoring the \scc\ in a \KN\-NUT-\dS black hole. Our results are in accord with the previous claim that rotating black holes always respect the conjecture when it get perturbed by scalar fields. Note that this result is independent of the fact that whether the scalar field is conformally coupled or not. The results presented in this work also asks for further investigation along various directions. For example, implications when quantum effects are taken into account needs to be studied, also for rotating spacetime the Dirac field may still pose a serious concern, which is worth studying in the present context. Besides, we have considered the effects of linear perturbation and it is instructive to check the contribution from nonlinear effects as well. These we leave for the future. 
\section*{Acknowledgments}

M.R. thanks INSPIRE-DST, Government of India for a Senior Research Fellowship (Reg.No.DST/INSPIRE\\/03/2015/003030). Research of SC is funded by the INSPIRE Faculty fellowship (Reg. No. DST/INSPIRE\\/04/2018/000893) from Department of Science and Technology, Government of India. SC thanks Albert Einstein Institute, Golm, Germany for warm hospitality where a part of this work was carried out.

\appendix

\labelformat{section}{Appendix #1}
\labelformat{subsection}{Appendix #1}
\labelformat{subsubsection}{Appendix #1}
\section*{Appendices}
\section{Quasi-normal modes with conformal coupling and photon sphere}\label{App_QNM}

Since we will work with conformally coupled scalar field it is not guaranteed that Lyapunov exponent associated with the instability of the photon sphere will have any connection to the \qnm modes. However we will explicitly demonstrate that for both rotating as well as non-rotating case, the Lyapunov exponent indeed captures the \qnm mode frequencies in the eikonal approximation. We will work with the static and spherically symmetric case first and shall then turn over to the case of rotating solutions. 

\subsection{Quasi-normal modes in static and spherically symmetric spacetime}

Let us start with generic static and spherically symmetric spacetime, for which the line element reads,
\begin{equation}
\begin{split}    
ds^2 = - f(r)dt^2 +\frac{dr^{2}}{g(r)}+r^{2} d\Omega^{2}
\end{split}
\end{equation}  
We will assume that the above metric depicts a spacetime with constant scalar curvature. Given the metric, it is straightforward to determine the equation for null geodesics, which reads,
\begin{equation}
\begin{split}
\label{6}
\dot{r}^{2}=\frac{g}{f}\Bigg[E^2 - f(r)\frac{L^2}{r^2}\Bigg]
\end{split}
\end{equation} 
Let us compare this equation for null geodesics with the field equation for conformally coupled scalar field, which takes the following form $\Box{\phi}=(\mathfrak{R}/6)\phi$, where $\mathfrak{R}$ is a constant. The field equation associated with the static and spherically symmetric metric ansatz, presented above, can be written as,
\begin{equation}
\begin{split}
\label{7}
\frac{1}{r^{2}} \frac{d}{dr_\ast}\left[r^{2}\frac{d\Phi_{\ell m}(r)}{dr_\ast}\right]+\left[\omega^{2}-\frac{\ell(\ell+1)}{r^2} f(r)\right]\Phi_{\ell m}(r)=(\mathfrak{R}/6)\Phi_{\ell m}(r)
\end{split}
\end{equation} 
where, $dr_{\ast}=(dr/\sqrt{fg})$ defines the tortoise coordinate and $\ell(\ell+1)$ is the separation constant originating from the angular equation. In order to arrive at the above equation, the original scalar field has been decomposed as, $\phi=\sum_{\ell m}e^{i\omega t}\Phi_{\ell m}(r)Y_{\ell m}(\theta \phi)$. Redefining the variable $\Phi_{\ell m}(r)$ as, $\psi_{\ell m}(r)= r\Phi_{\ell m}(r)$, we can rewrite \ref{7} as,
\begin{equation}
\begin{split}
\frac{d^{2}\psi_{\ell m}}{dr_{\ast}^{2}}+\left[\omega^{2}-\frac{\ell(\ell+1)}{r^2}f(r)\right]\psi_{\ell m}-\frac{1}{r}\frac{d^{2}r}{dr_{\ast}^{2}} \psi_{\ell m}=(\mathfrak{R}/6)\psi_{\ell m}(r)
\end{split}
\end{equation}
In the eikonal approximation, i.e., in the large $\ell$ limit the above differential equation reduces to,
\begin{equation}
\begin{split}
\label{10}
\frac{d^{2}\psi_{\ell m}}{d{r_\ast}^2}+\left[\omega^2 - \frac{\ell^{2}}{r^2} f(r)\right]\psi_{\ell m}=0
\end{split}
\end{equation}
The effective potential appearing in the above equation matches exactly to the potential in the null geodesic equation one given in \ref{6}. Thus even in the presence of conformal coupling, the scalar field experiences the potential of a massless particle. Thus Lyapunov exponent works well in determining the \qnm mode frequencies for eikonal modes. An identical consideration applies to the case of Reissner-Nordstr\"{o}m-NUT-de Sitter spacetime. 

\subsection{Quasi-normal modes in Kerr-NUT-\dS spacetime}

Let us explore the associated correspondence between null geodesic equation and the wave equation for a conformally coupled scalar field in the case of a rotating black hole spacetime. The metric being independent of time $t$ and azimuthal coordinate $\phi$, we have the energy and the angular momentum as the two constants of motion. Thus using the Hamilton-Jacobi equation, it immediately follows that the null geodesics satisfy the following equation, in the Kerr-NUT-dS spacetime,
\begin{align}
-\left[\frac{(\Sigma+ a\chi)^2 \Delta_\theta \sin^2 \theta - \Delta_r \chi^2}{\Sigma \Delta_r \Delta_\theta \sin^2 \theta}\right]E^{2} 
&-2EL \left[\frac{\Delta_r \chi-a (\Sigma+a\chi) \Delta_\theta \sin^2 \theta}{\Sigma \Delta_r \Delta_\theta \sin^2 \theta}\right] 
+L^{2}\left[\frac{\Delta_r -  \Delta_\theta a^2 \sin^2 \theta }{\Sigma \Delta_r \Delta_\theta \sin^2 \theta}\right] 
\nonumber
\\
&+ \frac{\Delta_r}{\Sigma}\left(\frac{dR(r)}{dr}\right)^2 
+\frac{\Delta_\theta}{\Sigma}\left(\frac{d\Theta(\theta)}{d\theta}\right)^2 = 0
\end{align}  
Multiplying throughout by $\Sigma$, the above equation can be further decomposed as,
\begin{equation}
\begin{split}   
\Delta_r\left(\frac{dR}{dr}\right)^{2}&+\Delta_\theta \left(\frac{d\Theta}{d\theta}\right)^2 
-\left\{\frac{(\Sigma + a\chi)^2}{\Delta_r}-\frac{\chi^2}{\Delta_\theta \sin^2 \theta}\right\}E^2 
\nonumber
\\
&+\left\{\frac{a (\Sigma + a\chi)}{\Delta_r}-\frac{\chi}{\Delta_\theta \sin^2 \theta} \right\}2EL  
- \left\{\frac{a^2}{\Delta_r} - \frac{1}{\Delta_\theta \sin^2 \theta}\right\}L^{2}=0
\end{split}
\end{equation}  
Thus the equation for radial null geodesic takes the following form 
\begin{equation}
\begin{split}   
\label {4}
{\Delta_r}^{2}\left(\frac{dR}{dr}\right)^2=-K \Delta_{r}+\left[(\Sigma + a\chi) E-aL\right]^{2}
\end{split}
\end{equation}  
where $K$ is the Carter constant associated with the separability of the radial and angular equations. On the other hand, the field equation for conformally coupled scalar field in the Kerr-NUT-dS background becomes,
\begin{equation}
\begin{split}   
\frac{d}{dr}\left(\Delta_r \frac{dR_{\ell m}}{dr}\right)
+\frac{(\Sigma + a\chi)^2}{\Delta_r}\left[\frac{m^2 a^2}{(\Sigma + a\chi)^2} - \frac{2 m \omega a }{(\Sigma + a\chi)} +\omega^2\right] R_{\ell m}-KR_{\ell m}=0
\end{split}
\end{equation}  
Introducing the tortoise coordinate $r_{\ast}$ through the following differential equation, $dr_{\ast}=dr \{(\Sigma + a\chi)/\Delta_r\}$, we obtain the following equation for the radial part of the scalar field,
\begin{equation}
\begin{split}   
\label{5}
\frac{1}{(\Sigma + a\chi)} \frac{d}{dr_{\ast}}\left[(\Sigma +a\chi) \frac{dR_{\ell m}}{dr_\ast}\right] 
+\left[\frac{m^2 a^2}{(\Sigma + a\chi)^2} - \frac{2 m \omega a }{(\Sigma + a\chi)} +\omega^2 - \frac{K\Delta_r}{(\Sigma + a\chi)^2}\right]R_{\ell m}=0
\end{split}
\end{equation}  
For large separation constant K, which in turn yields the eikonal limit, it immediately follows that the effective potential appearing in \ref{5} matches exactly with the potential the radial null geodesic experiences as given by \ref{4}. Hence the \qnm frequencies for the photon sphere modes are related to the Lyapunov exponent associated with the instability of photon sphere. This is the result we have used in the main text.

\section{Near-Extremal modes for Reissner-Nordstr\"{o}m-de Sitter-NUT spacetime}\label{AppN}

In this section, we will consider the near extremal modes associated with the Reissner-Nordstr\"{o}m-de Sitter-NUT spacetime for smaller values of the cosmological constant $\Lambda$. The near extremal modes can be determined by considering the spacetime structure when the event and the Cauchy horizon coincide. In this situation, the effect of the cosmological constant term can be neglected and hence the coefficient of the $dt^{2}$ term appearing in \ref{2} becomes,
\begin{equation}
\label{17}
\textrm{Coefficient~of~}dt^{2}\textrm{~term}\equiv -f(r)=-\left\{\frac{r^{2}-2Mr+Q^{2}-N^{2}}{r^{2}+N^{2}}\right\}
\end{equation}
The roots of the equation $f(r)=0$ are given by $r=r_{-}$ and $r=r_{+}$, with $r_{+}>r_{-}$, where $r_{-}$ and $r_{+}$ are the Cauchy and the event horizon respectively. In the near extremal limit, these two horizons coincide and hence the near extremal limit is given by, $M \rightarrow \sqrt{Q^2-N^2}$. We now introduce the following coordinate transformation,
\begin{align}
r&=\sqrt{Q^2-N^2}+\epsilon \rho=r_{\rm NE} + \epsilon \rho
\nonumber
\\
M&= \sqrt{Q^2-N^2}+\frac{\epsilon^2 B^2}{2\sqrt{Q^2-N^2}}
\nonumber
\\
t&=\frac{\tau}{\epsilon}~,
\end{align}
where, $\epsilon$ is a small parameter measuring the separation between event and Cauchy horizon. Similarly, $r_{\rm NE}=M=\sqrt{Q^2-N^2}$ is the horizon radius for the extremal black hole and $B$ denotes the deviation of the black hole from extremal configuration. Plugging the above transformation back at the Reissner-Nordstr\"{o}m-de Sitter metric and retaining the lowest order term in $\epsilon$ we get the near extremal line element to read,
\begin{equation}
\label{19}
ds^{2}=f(\rho)d\tau^{2}+\frac{1}{f(\rho)}d\rho^{2}+Q^{2}d\Omega^{2}
\end{equation}
where, $f(\rho)=(\rho^2 - B^2)/Q^2$. Interestingly, the structure of the metric for Reissner-Nordstr\"{o}m-NUT-dS black hole in the near extremal limit turns out to be the same as that of the Reissner-Nordstr\"{o}m-dS metric in the near extremal limit \cite{Rahman:2018oso}. Thus following \cite{Rahman:2018oso}, we briefly sketch the derivation of the \NE \qnm modes here.
   
For this purpose, we consider the perturbation due to a scalar field on the background metric given by \ref{19}, where the scalar perturbation $\Phi$, in the near extremal limit, satisfies a massless Klein-Gordon equation $\square \Phi =0$. Pertaining to the rotational and time translational symmetry of the spacetime, we can rewrite our scalar perturbation as, $\Phi = e^{i \omega \tau} R(\rho) Y_{\ell m}(\theta,\phi)$, where $Y_{lm}(\theta,\phi)$ are the spherical harmonics. When the above ansatz for the scalar perturbation is substituted in the Klein-Gordon equation, one obtains the following differential equation for $R(\rho)$,
\begin{equation}
\begin{split}
\frac{d}{d\rho}\left[\left(\rho^{2}-B^{2}\right) \frac{dR(\rho)}{d\rho}\right] 
+\left[\frac{\omega^2 Q^4}{\rho^2 -B^2} - \ell(\ell+1)\right] R(\rho)=0~.
\end{split}
\end{equation}
The solution of the above differential equation can be immediately obtained in terms of the hypergeometric functions $_{2}F_{1}(a,b,c;z)$, yielding \cite{Rahman:2018oso}
\begin{equation}
\begin{split}
R(x)&=C_{1}\left(x^2-1\right)^{i \mu/2}~_{2}F_{1}\left(1+ i \mu+\sigma, i \mu-\sigma, i \mu+1; \frac{1-x}{2}\right) 
\nonumber
\\
&\hskip 2 cm +C_{2}\left(\frac{x+1}{x-1}\right)^{i\mu/2}~_{2}F_{1}\left(-\sigma, \sigma+1, 1- i\mu; \frac{1-x}{2}\right)~,
\end{split}
\end{equation}
where, $C_1$ and $C_2$ are two arbitrary constants, to be fixed by appropriate boundary conditions and $x\equiv (\rho/B)$ and $\mu=\omega(Q^{2}/B)$, with $\sigma$ satisfying the equation $\sigma(\sigma +1)=\ell(\ell+1)$. In order to determine the associated \qnm modes, we impose the following boundary conditions: a) purely ingoing modes at the event horizon, and b) purely outgoing mode at the cosmological horizon, which gives the following expression for $R(x)$,  
\begin{equation}
\begin{split}
R(x) = {C_{\rm B}}^{\rm in} \left(\frac{x-1}{2}\right)^{\sigma}+{C_{\rm B}}^{\rm out}\left(\frac{x-1}{2}\right)^{-\sigma-1}~.
\end{split}
\end{equation}
Here the two constants ${C_{\rm B}}^{\rm in}$ and ${C_{\rm B}}^{\rm out}$ has the following expressions,
\begin{align}
{C_{\rm B}}^{\rm in} = C_2 \frac{\Gamma(1- i \mu) \Gamma(2\sigma+1)}{\Gamma(1 - i\mu +\sigma) \Gamma(\sigma +1)}~,
\qquad
{C_{\rm B}}^{\rm out} = C_2 \frac{\Gamma(1- i \mu) \Gamma(-2\sigma-1)}{\Gamma(- i\mu -\sigma) \Gamma(\sigma +1)}~.
\end{align}
Demanding that there will be only outgoing modes near the outer boundary, we obtain,
\begin{equation}
\begin{split}
\frac{1}{\Gamma(1 - i\mu +\sigma)}=0~,~\textrm{which implies}~ 1 - i\mu +\sigma =-n~,
\end{split}
\end{equation}
where $n$ is a positive integer. After rearranging and substituting the expression for $\mu$ in terms of the \qnm mode frequency $\omega$, we obtain the quasi-normal frequencies of a near extremal black hole to yield, $\omega_{\rm NE}=-i(n+\sigma+1)\kappa_+$. This is the expression we have used in the main text. As evident form the above discussion, these near extremal modes are purely imaginary.
\section{ADM Decomposition of the Kerr-dS-NUT Spacetime}\label{AppA}

In this appendix we will provide an ADM decomposition for the Kerr-dS-NUT metric, which will be useful for our later purposes while solving for the field equation of a scalar field living in this spacetime. To start with, we would like to express the metric in \ref{29} representing Kerr-dS-NUT spacetime in the ADM form, which reads,
\begin{equation}\label{31}
ds^2 = - \mathbb{N}^2 dt^2 + h_{\mu\nu}\left(dx^{\mu} + \mathbb{N}^{\mu}dt\right)\left(dx^{\nu} + \mathbb{N}^{\nu}dt\right)~.
\end{equation}
We can make a one-to-one map between the above system of coordinates and the one describing the Kerr-dS-NUT spacetime, such that $(t,x^{\mu})\rightarrow (t,r,\theta,\phi)$, i.e., the ADM time coordinate is mapped to the time coordinate in the Kerr-dS-NUT metric. With this identification of the coordinates one can compare \ref{29} and \ref{31}, which yields the following choices for various combinations of $\mathbb{N}$, $\mathbb{N}^{\mu}$ and $h_{\mu \nu}$ as,
\begin{align}\label{32}
-g_{tt}&=\mathbb{N}^{2}-h_{\phi\phi}(\mathbb{N}^{\phi})^{2}=\frac{\Delta_{r}-a^{2}\Delta_{\theta}\sin^2\theta}{\Sigma}~,
\nonumber
\\
g_{rr}&=h_{rr}=\frac{\Sigma}{\Delta_r};\qquad 
g_{\theta\theta}=h_{\theta\theta}=\frac{\Sigma}{\Delta_{\theta}};\qquad 
g_{\phi\phi}=h_{\phi\phi}=\frac{(\Sigma+ a\chi)^2 \Delta_{\theta} \sin^2\theta - \Delta_r\chi^{2}}{\Sigma}
\nonumber
\\
g_{t\phi}&=h_{\phi\phi}N^{\phi} = \frac{\Delta_r \chi - a(\Sigma + a\chi)\Delta_{\theta} \sin^2\theta}{\Sigma}
\end{align}
Using the expression for $h_{\phi \phi}$ presented above one obtains $\mathbb{N}^{\phi}$ in a straightforward manner. Finally, using both $h_{\phi \phi}$ and $\mathbb{N}^{\phi}$ in the above equations, one can get the following expression for $\mathbb{N}^{2}$,
\begin{align}\label{36}
\mathbb{N}^{2}&=\frac{\Delta_r - a^2 \Delta_{\theta} \sin^2\theta}{\Sigma} +\frac{(\Sigma+ a\chi)^2 \Delta_{\theta} \sin^2\theta - \Delta_r\chi^2}{\Sigma} \left[\frac{\Delta_r \chi - a(\Sigma + a\chi)\Delta_{\theta}\sin^2\theta}{(\Sigma+ a\chi)^2 \Delta_{\theta} \sin^2\theta - \Delta_r\chi^2}\right]^2
\nonumber
\\
&=\frac{(\Sigma+ a\chi)^{2}\Delta_r\Delta_{\theta} \sin^2\theta+a^{2}\chi^2 \Delta_r \Delta_{\theta}\sin^2\theta - 2a\chi(\Sigma + a\chi)\Delta_r\Delta_{\theta} \sin^2\theta}{\Sigma \left[(\Sigma + a\chi)^2 \Delta_{\theta} \sin^2\theta - \Delta_r \chi^{2}\right]} 
\nonumber
\\
&=\frac{\Sigma^2 \Delta_r\Delta_{\theta} \sin^2\theta}{\Sigma[(\Sigma + a\chi)^2 \Delta_{\theta} \sin^2\theta - \Delta_r \chi^2]} = \frac{\Sigma \Delta_r\Delta_{\theta} \sin^2\theta}{\left[(\Sigma + a\chi)^2 \Delta_{\theta} \sin^2\theta - \Delta_r \chi^2\right]}
\end{align}
Thus collecting together all the ADM parameters for the Kerr-dS-NUT metric we have the following expressions,
\begin{align}
\mathbb{N}^{2}&=\frac{\Sigma \Delta_r\Delta_{\theta} \sin^2\theta}{\left[(\Sigma + a\chi)^2 \Delta_{\theta} \sin^2\theta - \Delta_r \chi^2\right]}~,
\nonumber
\\
\mathbb{N}^{\phi}&=\frac{\Delta_r \chi - a(\Sigma + a\chi)\Delta_{\theta} \sin^2\theta}{[(\Sigma + a\chi)^2 \Delta_{\theta} \sin^2\theta - \Delta_r \chi^2]}~,
\nonumber
\\
h_{rr}&=\frac{\Sigma}{\Delta_r};\qquad
h_{\theta\theta}=\frac{\Sigma}{\Delta_{\theta}};\qquad
h_{\phi\phi}=\frac{\left[(\Sigma + a\chi)^2 \Delta_{\theta} \sin^2\theta - \Delta_r \chi^2\right]}{\Sigma}
\end{align}
The inverse metric elements can be expressed in terms of the ADM variables as follows,
\begin{align}
g^{tt}&=-\frac{1}{\mathbb{N}^2}~;\qquad
g^{rr}=\frac{1}{h_{rr}}~;\qquad 
g^{\theta\theta}=\frac{1}{h_{\theta\theta}}~;
\nonumber
\\
g^{t\phi}&=\frac{\mathbb{N}^{\phi}}{\mathbb{N}^{2}}~;\qquad 
g^{\phi\phi}=h^{\phi\phi}-\frac{\left(\mathbb{N}^{\phi}\right)^{2}}{\mathbb{N}^{2}}~.
\end{align}
Since all the other inverse metric elements, except for $g^{\phi\phi}$, involves simple multiplicative operation with inversion, they can be easily determined. The only nontrivial component is that of $g^{\phi \phi}$, which gives,
\begin{align}\label{38}
g^{\phi\phi}&=\frac{\Sigma}{(\Sigma + a\chi)^2 \Delta_{\theta} \sin^2\theta - \Delta_r \chi^2}
-\frac{\left\{\Delta_r \chi-a(\Sigma + a\chi) \Delta_{\theta}\sin^2\theta\right\}^2}{\left\{(\Sigma + a\chi)^2 \Delta_{\theta}\sin^2\theta - \Delta_r \chi^2\right\}^2} \frac{(\Sigma + a\chi)^2 \Delta_{\theta}\sin^2\theta - \Delta_r \chi^2}{\Sigma \Delta_r\Delta_{\theta}\sin^2\theta} 
\nonumber
\\
&=\frac{\Sigma^2\Delta_r\Delta_{\theta}\sin^2\theta-\left\{\Delta_r \chi -a(\Sigma + a\chi)\Delta_{\theta}\sin^2\theta\right\}^2}
{\left\{(\Sigma + a\chi)^2 \Delta_{\theta}\sin^2\theta - \Delta_r \chi^2\right\}\Sigma \Delta_r\Delta_{\theta}\sin^2\theta} 
\end{align}
The numerator of the above equation can be further expanded, yielding,
\begin{align}
\Sigma^2\Delta_r\Delta_{\theta} \sin^2\theta&-\left\{\Delta_r \chi -a(\Sigma + a\chi) \Delta_{\theta}\sin^2\theta\right\}^{2}
\nonumber
\\
&=\Sigma^2\Delta_r\Delta_{\theta} \sin^2\theta - \Delta_r^2 \chi^2 -a^2(\Sigma + a\chi)^2 \Delta_{\theta}^2\sin^4\theta
+2a\Delta_r \chi (\Sigma + a\chi) \Delta_{\theta}\sin^2\theta
\nonumber
\\
&=\left\{(\Sigma + a\chi)^{2}\Delta_{\theta}\sin^2\theta-\Delta_r\chi^2\right\}\left\{-a^2 \Delta_{\theta}\sin^2\theta + \Delta_r\right\}
\nonumber
\\
&-\Delta_r a^2 \chi^2 \Delta_{\theta}\sin^2\theta - (\Sigma + a\chi)^2 \Delta_r\Delta_{\theta} \sin^2\theta  + \Sigma^2 \Delta_r \Delta_{\theta} \sin^2\theta
+2a \Delta_r \chi (\Sigma + a\chi) \Delta_{\theta} \sin^2{\theta} 
\nonumber
\\
&=\left\{(\Sigma + a\chi)^2 \Delta_{\theta}\sin^2\theta - \Delta_r\chi^{2}\right\}\left\{-a^2 \Delta_{\theta}\sin^2\theta + \Delta_{r}\right\}
\end{align}
Hence the inverse metric components take the following form,
\begin{align}
g^{tt}&=-\frac{(\Sigma + a\chi)^2 \Delta_{\theta}\sin^2\theta - \Delta_r \chi^2}{\Sigma \Delta_r\Delta_{\theta} \sin^2\theta}~; \qquad
g^{t\phi}=\frac{\Delta_r \chi - a(\Sigma + a\chi)\Delta_{\theta} \sin^2\theta}{\Sigma \Delta_r\Delta_{\theta} \sin^2\theta}
\nonumber
\\
g^{rr}&=\frac{\Delta_r}{\Sigma}~;\qquad 
g^{\theta\theta}=\frac{\Delta_{\theta}}{\Sigma}~;\qquad
g^{\phi\phi}=\frac{\Delta_r - a^2\Delta_{\theta}\sin^2\theta}{\Sigma \Delta_r\Delta_{\theta} \sin^2\theta}
\end{align}
Using these ADM variables we can also write down the following expression for $\sqrt{-g}$, which reads,
\begin{align}\label{39}
\sqrt{-g}&=\mathbb{N} \sqrt{h}
= \frac{\sqrt{\Sigma \Delta_r\Delta_{\theta} \sin^2\theta}}{\sqrt{(\Sigma + a\chi)^2 \Delta_{\theta} \sin^2\theta - \Delta_r \chi^2}}.\frac{\Sigma\sqrt{(\Sigma + a\chi)^2\Delta_{\theta} \sin^2\theta-\Delta_r \chi^2}}{\sqrt{\Sigma \Delta_r\Delta_{\theta}}} 
\nonumber
\\
&= \Sigma \sin\theta
\end{align}
These are the expressions we have used in the main text while computing the field equation for the scalar field living on the Kerr-dS-NUT spacetime. 

\section{Simplifications of the angular equation of a scalar field in Kerr-dS-NUT spacetime}\label{AppB}

In this appendix, we will discuss various simplifications done to the scalar perturbation equation as well as to the angular equation in the Kerr-dS-NUT spacetime. First of all, the scalar perturbation equation for a conformally coupled scalar field takes the following form in generic spacetimes,
\begin{equation}
\left(\Box -\frac{\mathfrak{R}}{6}\right)\Phi= \frac{1}{\sqrt{-g}} \partial_i \left(\sqrt{-g} g^{ij} \partial_j \Phi\right)-\frac{2\Lambda}{3}\Phi=0~,
\end{equation}
where we have used the result that for Kerr-dS-NUT spacetime, $\mathfrak{R}=4\Lambda$. Explicitly writing the inverse metric components derived in \ref{AppA} for Kerr-dS-NUT spacetime in the above scalar perturbation equation, we obtain,
\begin{align}
0&=\frac{1}{\Sigma \sin\theta} \partial_t (\Sigma \sin\theta g^{tt}\partial_t \Phi)
+\frac{1}{\Sigma \sin\theta} \partial_r (\Sigma \sin\theta g^{rr}\partial_r \Phi) 
+\frac{1}{\Sigma \sin\theta} \partial_\theta (\Sigma \sin\theta g^{\theta\theta}\partial_\theta \Phi) 
\nonumber
\\
&+\frac{1}{\Sigma \sin\theta} \partial_\phi (\Sigma \sin\theta g^{\phi\phi}\partial_\phi \Phi) 
+\frac{1}{\Sigma \sin\theta} \partial_t (\Sigma \sin\theta g^{t\phi}\partial_\phi \Phi) 
+\frac{1}{\Sigma \sin\theta} \partial_\phi (\Sigma \sin\theta g^{\phi t}\partial_t \Phi)-\frac{2\Lambda}{3}\Phi
\end{align}
Multiplying both sides by $\Sigma \sin \theta$ we obtain,
\begin{align}\label{scalar_eq_a1}
0&=\partial_t\left[\Sigma \sin\theta \left\{\frac{\Delta_r \chi^2 - (\Sigma + a\chi)^2 \Delta_\theta \sin^2 \theta}{\Sigma \Delta_r \Delta_{\theta} \sin^2\theta}\right\} \partial_ t \Phi \right] 
+\partial_r\left[\Sigma \sin\theta \left\{\frac{\Delta_r}{\Sigma}\right\}\partial_r \Phi\right] 
+\partial_{\theta} \left[\Sigma \sin\theta \left\{\frac{\Delta_\theta}{\Sigma}\right\}\partial_\theta \Phi\right]
\nonumber
\\
&+ \partial_{\phi}\left[\Sigma \sin\theta \left\{\frac{\Delta_{r}-a^2\Delta_\theta \sin^2\theta}{\Sigma \Delta_r \Delta_{\theta} \sin^2\theta}\right\}\partial_\phi \Phi\right] 
+2\partial_t\left[\Sigma \sin\theta \left\{\frac{\Delta_r \chi - a(\Sigma + a\chi)\Delta_{\theta} \sin^2\theta}{\Sigma \Delta_r\Delta_{\theta} \sin^2\theta} \right\} \partial_\phi \Phi\right]-\frac{2\Lambda}{3}\Sigma \sin \theta \Phi
\end{align}
This is the expression we have used to arrive at \ref{scalar_eq_01}. 

Introducing, $\Phi=e^{i\omega t}e^{-im\phi}R(r)\Theta(\theta)$, the differential equation for the scalar field presented as in \ref{scalar_eq_a1} can be expressed as,
\begin{align}
0&=\left(-\omega^2\right)R \Theta~\left\{\frac{\Delta_r\chi^2 -(\Sigma + a\chi)^2 \Delta_\theta \sin^2\theta}{\Delta_r \Delta_\theta \sin\theta}\right\}
+\left(-m^{2}\right)R \Theta~\left\{\frac{\Delta_r- a^2 \Delta_\theta \sin^2\theta}{\Delta_r \Delta_\theta \sin\theta}\right\}
-\frac{2\Lambda}{3}\Sigma \sin \theta R\Theta
\nonumber
\\
&+2\left(\omega m\right) R \Theta~\left\{\frac{\Delta_r\chi - a(\Sigma + a\chi) \Delta_\theta \sin^2\theta}{\Delta_r \Delta_\theta \sin\theta}\right\}
+\partial_r \left\{\Delta_r \sin\theta \left(\partial_r R\right)\right\} \Theta 
+\partial_\theta \left\{\Delta_\theta \sin\theta \left(\partial_\theta \Theta\right)\right\}R
\nonumber
\\
&=\left[\frac{\chi^2}{\Delta_\theta \sin\theta} - \frac{(\Sigma + a\chi)^2 \sin\theta}{\Delta_r}\right]\left(-\omega^2\right) R \Theta 
+\left[\frac{1}{\Delta_\theta \sin\theta} - \frac{a^2 \sin\theta}{\Delta_r}\right]\left(-m^2\right)R\Theta 
-\frac{2\Lambda}{3}\Sigma \sin \theta R\Theta
\nonumber
\\
&+ 2\left[\frac{\chi}{\Delta_\theta \sin\theta} - \frac{a(\Sigma + a\chi) \sin \theta}{\Delta_r}\right]\left(\omega m\right) R\Theta 
+\partial_r \left\{\Delta_r \sin\theta \left(\partial_r R\right)\right\}\Theta 
+\partial_\theta \left\{\Delta_\theta \sin\theta \left(\partial_\theta \Theta\right)R\right\}
\end{align}
Dividing the above equation throughout by $\sin\theta$ and $R\Theta$, we obtain,
\begin{align}
\Bigg[\frac{\chi^2}{\Delta_\theta \sin^2\theta}&-\frac{(\Sigma + a\chi)^2 }{\Delta_r}\Bigg]\left(-\omega^2\right)
+\left[\frac{1}{\Delta_\theta \sin^2\theta} - \frac{a^2}{\Delta_r}\right]\left(-m^2\right)
+2\left[\frac{\chi}{\Delta_\theta \sin^2\theta} - \frac{a(\Sigma + a\chi)}{\Delta_r}\right]\left(\omega m\right)
\nonumber
\\
&+\frac{d}{dr}\left(\Delta_r  \frac{dR}{dr}\right) \dfrac{1}{R} 
+\dfrac{1}{\sin\theta}\frac{d}{d\theta}\left(\Delta_\theta \sin\theta \frac{d\Theta}{d\theta}\right)\frac{1}{\Theta}
-\frac{2\Lambda}{3}\left\{r^{2}+\left(N+a\cos\theta\right)^{2}\right\}= 0
\end{align}
This is the expression we have used in order to arrive at the separated radial and angular equations in the main text, i.e., \ref{scalar_radial_01} and \ref{scalar_angular_01}. 

In the simplification of the angular equation, the first term one encounters is the potential in \ref{angular_03}, which can be expanded to the following form, 
\begin{equation}
\begin{split}
&\frac{\left[\xi (1 - x^2) - (m + 2NC \omega) - 2N\omega x\right]^2}{(1 - x^2) (1+ \gamma x + \delta x^2)} 
\nonumber
\\
&= \frac{\xi^2(1-x^2)^2+4N^2\omega^2 x^2+(m+2NC\omega)^2+4\omega Nx(m+2NC\omega)-4\omega N\xi x(1-x^2) 
-2\xi (1-x^2)(m + 2NC \omega)}{(1 - x^2) (1+ \gamma x + \delta x^2)}
\nonumber
\\
&=\frac{\xi^2 (1-x^2)}{(1+ \gamma x + \delta x^2)} - \frac{2\xi (m + 2NC \omega)}{(1+ \gamma x + \delta x^2)} - \frac{4\omega N\xi x}{(1+ \gamma x + \delta x^2)} + \frac{(m+2NC\omega)^2}{(1 - x^2)(1+ \gamma x + \delta x^2)}+\frac{4\omega Nx(m+2NC\omega )+4N^2\omega^2 x^2}{(1-x^2)(1+ \gamma x + \delta x^2)}
\nonumber
\\
&=\frac{\xi^2}{\delta}\left\{\frac{\delta+1+\gamma x-(1+ \gamma x + \delta x^2)}{1+ \gamma x + \delta x^2}\right\}-\frac{2\xi (m+2NC\omega)}{1+\gamma x+\delta x^2} 
- \frac{4 \omega N\xi x}{1+ \gamma x + \delta x^2}+ \frac{(m+2NC\omega)^2}{(1-x^2) (1+ \gamma x + \delta x^2) } 
\nonumber
\\
&+ \frac{4\omega Nx(m+2NC\omega) + 4N^2\omega^2 x^2}{(1 - x^2) (1+ \gamma x + \delta x^2)}
\nonumber
\\
&=-\frac{\xi^2}{\delta}+\frac{1}{(1+ \gamma x + \delta x^2)}\left\{\frac{(\delta + 1) \xi^2}{\delta} - 2\xi(m + 2C\eta)\right\}
+\frac{x}{(1+ \gamma x + \delta x^2)}\left\{\frac{\gamma \xi^2}{\delta} - 4\xi \eta\right\} 
+\frac{(m + 2C\eta)^2}{(1-x^2) (1+ \gamma x + \delta x^2)} 
\nonumber
\\
&+ \frac{- 4 \eta^2 (1-x^2) + 4 \eta^2 + 4\eta x (m + 2C\eta)}{(1 - x^2) (1+ \gamma x + \delta x^2)} 
\nonumber
\\
&= - \frac{\xi^2}{\delta}+\frac{1}{(1+ \gamma x + \delta x^2)}\left\{\frac{(\delta + 1) \xi^2}{\delta} - 2\xi(m + 2C\eta)-4 \eta^2\right\}
+\frac{x}{(1+ \gamma x + \delta x^2)}\left\{\frac{\gamma \xi^2}{\delta} - 4\xi \eta\right\} 
+\frac{(m + 2C\eta)^2 + 4 \eta^2}{(1-x^2) (1+ \gamma x + \delta x^2)} 
\nonumber
\\
&+ \frac{ 4\eta x (m + 2C\eta)}{(1-x^2) (1+ \gamma x + \delta x^2)} 
\end{split}
\end{equation}
This is the expression we have used in order to obtain \ref{angular_04}. 

Following the definition of the variable $z$ in \ref{ang_coord}, let us try to express the term involving derivative in the angular equation. For this purpose, we can use the following relation, 
\begin{equation}\label{46}
\frac{dz}{dx} =\frac{1- x_+}{2(x - x_+)}-\frac{( x+1)(1 - x_+)}{2(x-x_+)^{2}}
=\left(\frac{1- x_+}{2}\right)\frac{x-x_+ - (x+1)}{(x - x_+)^2}= -\frac{\{1 - (x_+)^2\}}{2(x - x_+)^2}
\end{equation}
Using the above equation the derivative part of the angular equation can be simplified further. For this purpose, we use the result $(1+\gamma x+\delta x^2)=\delta(x - x_{+})(x - x _{-})$, along with \ref{46}, such that the derivative part of the angular equation reads,
\begin{align}\label{App_ang_01}
\frac{d}{dx}&\left[(1 - x^2) \delta (x - x_+)(x - x _-)\frac{dz}{dx}\frac{d\Theta}{dz}\right]
=\frac{dz}{dx}\frac{d}{dz}\left[(1 - x^2) \delta (x - x_+)(x - x _-) \frac{(x_{+}^{2} -1)}{2(x - x_+)^2} \frac{d\Theta}{dz}\right]
\\
&=\delta(1 - x^2)(x - x_+)(x - x _-) \frac{(x_{+}^{2}-1)}{2(x - x_{+})^2}\times\frac{(x_{+}^2 - 1)}{2(x - x_+)^2} \frac{d^2 \Theta}{dz^2} 
\nonumber
\\
&+\frac{\delta(x_{+}^2-1)}{2(x - x_+)^2} \frac{d\Theta}{dz} 
\nonumber
\\
&\times \Big[-(1+x)(x-x_-)(x-x_+)+(1-x)(x-x_-)(x-x_+)+(1-x^2)(x-x_-)+(1-x^2)(x-x_+)\Big] 
\nonumber
\\
&+\left[\frac{\delta(1-x^2)(x-x_-)(x-x_+)}{2}\times(x_{+}^2 -1) \times \left\{\frac{-2 }{(x - x_+)^3}\right\}\right] \frac{d\Theta}{dz}
\nonumber
\\
&=\frac{ \delta(1 - x^2)(x - x_+)(x - x _-)(x_{+}^2 - 1)^2}{4 (x - x_+)^4}\Bigg[\frac{d^2\Theta}{dz^2} 
+\frac{2(x -x_+)^2}{(x_{+}^2-1)}\left\{-\frac{1}{1-x} + \frac{1}{1+x} + \frac{1}{x -x_-} +\frac{1}{x- x_+}\right\}\frac{d\Theta}{dz} 
\nonumber
\\
&\qquad \qquad \qquad + \frac{2 (x -x_+)^4}{x_{+}^2 -1}\left\{\frac{-2 }{(x - x_+)^3}\right\}\frac{d\Theta}{dz}\Bigg] 
\nonumber
\\
&=\frac{\delta(1 - x^2) (x - x_+)(x - x _-)({x_+}^2 - 1)^2}{4 (x - x_+)^4}
\Bigg[\frac{d^2\Theta}{dz^2} 
\nonumber
\\
&\qquad \qquad \qquad+\frac{2}{x_{+}^2 -1}\left\{- \frac{(x- x_+)^2}{1-x} + \frac{(x- x_+)^2}{1 + x} +\frac{x- x_+)^2}{x - x_-} - (x -x_+)\right\}\frac{d\Theta}{dz}\Bigg]
\end{align}
Using the definition of $z$ presented in \ref{ang_coord}, we have the following result, $2z(x - x_+)=(x +1)(1 - x_+)$. This can be inverted yielding $x$ in terms $z$ through the following relation,
\begin{equation}
x=\frac{(1- x_+) + 2z x_+}{2z - (1 -x_+)}
\end{equation}
Given the above relation where $x$ is a function of $z$, one can immediately obtain the following identities involving $x$ as,
\begin{align}\label{Imp_Ident}
x -1&=\frac{2(x_+ -1)(z-1)}{2z + (x_+ -1)}~;\qquad x + 1 = \frac{2 z (x_+ +1)}{2z + (x_+ -1)} 
\nonumber
\\
x -x_{+}&=-\frac{x_{+}^2 -1}{2z + (x_+ -1)}~;\qquad x - x_{-}=\frac{2z (x_+ -x_-) + (1 - x_+)(1 + x_-)}{2z + (x_+ -1)}
\end{align}
These expressions can be rewritten in terms of the variables defined in \ref{definition_new}, such that, 
\begin{align}
x- 1&=\frac{(x_+ - 1)(z-1)}{z - z_\infty}~;\qquad x+ 1=\frac{z(1 + x_+)}{z - z_\infty}~, 
\nonumber
\\
x -x_{+}&=-\frac{(x_{+}^2 - 1)}{2(z - z_\infty)}~;\qquad  x-x_{-}=(x_+ - x_-)\left\{\frac{z-z_s}{z - z_\infty}\right\}~.
\end{align}
Using these results, the coefficient of $(d\Theta/dz)$ term in \ref{App_ang_01} can be expressed as,
\begin{align}\label{App_ang_02}
\textrm{coefficient~of~}\left(\frac{d\Theta}{dz}\right)~\textrm{term}&=\frac{(x -x_+)^{2}}{x - 1}+\frac{(x -x_+)^2}{x + 1}+\frac{(x -x_+)^2}{x - x_-}-(x - x_+)
\nonumber
\\
&=\frac{(x_{+}^2 -1)^2}{4(z- z_\infty)} \times \frac{1}{(x_+ -1) (z -1)}
+\frac{(x_{+}^2 -1)^2}{4(z- z_\infty)} \times \frac{1}{z(1 + x_+)} 
\nonumber
\\
&+\frac{(x_{+}^2 -1)^2}{4(z- z_\infty)} \times \frac{1}{(x_+ - x_-) (z-z_s)} 
+\frac{(x_{+}^2 -1)}{2(z - z_\infty)}
\nonumber
\\
&=\frac{(x_{+}^2 -1)^2}{4(x_+ -1)}\times \frac{1}{(z - 1)(z -z_\infty)}
+\frac{(x_{+}^2 -1)^2}{4(1 + x_+)}\times \frac{1}{z(z -z_\infty)}
\nonumber
\\
&+ \frac{(x_{+}^2 -1)^2}{4(x_+ -x_-)}\frac{1}{(z-z_\infty)(z-z_s)}
+\frac{(x_{+}^2 -1)}{2(z -z_\infty)}
\end{align}
To simplify the above term further, let us work out the above expression term by term, which yields,
\begin{align}
\textrm{Term~1}&=\frac{(x_{+}^2 -1)^2}{4(x_+ -1)} \times \frac{1}{(z - 1)(z -z_\infty)} 
\nonumber
\\
&=\frac{(x_{+}^2 -1)^2}{4(x_+ -1)} \times \frac{(z - z_\infty)-(z -1)}{(z - 1)(z -z_\infty)(1 - z_\infty)} 
\nonumber
\\
&=\frac{(x_{+}^2 -1)^2}{4(x_+ -1)} \times \frac{2}{x_{+}+1} \times \left(\frac{1}{z-1} - \frac{1}{z-z_\infty}\right) 
\nonumber
\\
&=\frac{(x_{+}^2 -1)}{2} \times \left(\frac{1}{z-1} - \frac{1}{z-z_\infty}\right)
\end{align}
where we have used the result, $1-z_{\infty}=(1+x_{+})/2$. Further, the second term in \ref{App_ang_02} takes the following form,
\begin{align}
\textrm{Term~2}&=\frac{(x_{+}^2 -1)^2}{4(1 + x_+)} \times \frac{1}{z(z -z_\infty)}
\nonumber
\\
&=\frac{(x_{+}^2 -1)^2}{4(1+x_{+})} \times \frac{z - (z - z_\infty)}{z_\infty z(z -z_\infty)}
\nonumber
\\
&=\frac{(x_{+}^2 -1)}{2} \frac{(z -z_\infty) - z}{z(z -z_\infty)}
\nonumber
\\
&=\frac{(x_{+}^2 -1)}{2}\left(\frac{1}{z} - \frac{1}{z -z_\infty}\right)
\end{align}
where expression for $z_{\infty}$ has been used. At this stage note the following identity, 
\begin{align}
z_s - z_\infty&=\frac{x_+ -1 }{2} - \frac{(1 - x_+)(1 + x_-)}{2 (x_+ - x_-)} 
\nonumber
\\
&=\frac{(x_+ -1)(x_+ - x_-) - 1 +x_+ - x_- + x_+ x_-}{2 (x_+ - x_-)}
\nonumber
\\
&=\frac{x_{+}^2 -1}{2 (x_+ - x_-)}
\end{align}
Proceeding further, the third term in \ref{App_ang_02} takes the following form,
\begin{align}
\textrm{Term~3}&=\frac{(x_{+}^2 -1)^2}{4(x_+ -x_-)}\frac{1}{(z-z_\infty)(z-z_s)}
\nonumber
\\
& = \frac{(x_{+}^2 -1)^2}{4(x_+ -x_-)} \frac{(z -z _\infty) - (z -z_s)}{(z -z _\infty) (z -z_s)} \frac{1}{z_s - z_\infty} 
\nonumber
\\
&=\frac{(x_{+}^2 -1)}{2} \left(\frac{1}{z-z_s} - \frac{1}{z-z_\infty}\right)
\end{align}
where the above identity for $(z_{s}-z_{\infty})$ has been used. Finally, the last term in \ref{App_ang_02} can be expressed as,
\begin{equation}
\textrm{Term~4}=\frac{x_{+}^2 -1}{2} \frac{1}{z - z_\infty}
\end{equation}
Thus we obtain the coefficient of $(d\Theta/dz)$ term to yield,
\begin{align}\label{App_ang_04}
\textrm{coefficient~of~}\left(\frac{d\Theta}{dz}\right)~\textrm{term}&=\textrm{Term~1}+\textrm{Term~2}+\textrm{Term~3}+\textrm{Term~4}
\nonumber
\\
&=\frac{(x_{+}^2 -1)}{2} \times \left(\frac{1}{z-1} - \frac{1}{z-z_\infty}\right)+\frac{(x_{+}^2 -1)}{2}\left(\frac{1}{z} - \frac{1}{z -z_\infty}\right)
\nonumber
\\
&+\frac{(x_{+}^2 -1)}{2} \left(\frac{1}{z-z_s} - \frac{1}{z-z_\infty}\right)+\frac{(x_{+}^2 -1)}{2} \frac{1}{z - z_\infty}
\nonumber
\\
&=\frac{(x_{+}^2 -1)}{2}\left\{-\frac{2}{z - z_\infty}+\frac{1}{z-1}+\frac{1}{z}+\frac{1}{z-z_s}\right\}
\end{align}
Thus the angular equation becomes,
\begin{align}
\label{Ang_Eq_01}
0&=\frac{\delta (1 - x^2) (x - x_+)(x - x _-)(x_{+}^2 - 1)^2}{4 (x - x_+)^4}
\Bigg[\frac{d^2 \Theta}{dz^2}+\left(\frac{1}{z}+\frac{1}{z - 1}+\frac{1}{z - z_s}-\frac{2}{z - z_\infty}\right)\frac{d\Theta}{dz}\Bigg]
\nonumber
\\
&+\Bigg[k+ \frac{\xi^2}{\delta}+\frac{1}{(1+ \gamma x + \delta x^2)}\left\{\frac{-(\delta + 1) \xi^2}{\delta} + 2\xi(m + 2C\eta) + 4 \eta^2\right\}
\nonumber
\\
&-\frac{x}{(1+ \gamma x + \delta x^2)}\left\{\frac{\gamma \xi^2}{\delta}-4\xi \eta\right\}
-\frac{(m + 2C\eta)^2 + 4 \eta^2}{(1-x^2) (1+ \gamma x + \delta x^2)} - \frac{ 4\eta x (m + 2C\eta)}{(1-x^2) (1+ \gamma x + \delta x^2)}
-2\delta\left(x+\frac{N}{a}\right)^{2}\Bigg] \Theta
\end{align}
Multiplying this result on both sides by $(4/\delta)[(x - x_+)^4/\{(1 - x^2)(x - x_+)(x - x _-)(x_{+}^2 - 1)^2\}]$ we obtain \ref{Ang_Eq_02}.

Let us start by discussing the term in the potential appearing due to conformal coupling of the scalar field, which is the last term in \ref{Ang_Eq_03}, which can be decomposed as,
\begin{equation}
\left(x +\frac{N}{a}\right)^{2} = x^2+2\frac{N}{a}x+\left(\frac{N}{a}\right)^2 
=1-\left(1 -x^2\right)+2\frac{N}{a}\left\{1- (1- x)\right\}+\left(\frac{N}{a}\right)^2
\end{equation}
Thus multiplying the above term with the overall scaling factor appearing in \ref{Ang_Eq_02} we obtain,
\begin{align}
V_{\rm conformal}^{\rm modified}&=\frac{-2\delta \left[1-(1-x^2)+ 2\frac{N}{a}\left\{1-(1-x)\right\}+\left(\frac{N}{a}\right)^2\right]}
{\delta(1 - x^2)(x - x _-)(x_{+}^2 - 1)^2} \times 4(x - x_+)^3 
\nonumber
\\
&=-\frac{8(x - x_+)^3 }{(1 - x^2)(x - x _-)(x_{+}^2 - 1)^2} 
+\frac{8(x - x_+)^3 }{(x - x _-)(x_{+}^2 - 1)^2}
-\frac{16 (N/a)(x - x_+)^3}{(1 - x^2)(x - x _-)(x_{+}^2 - 1)^2}
\nonumber
\\
&+\frac{16 (N/a)(x - x_+)^3}{(1+x)(x - x _-)(x_{+}^2 - 1)^2}
-\frac{8 (N/a)^2 (x - x_+)^3}{(1 - x^2)(x - x _-)(x_{+}^2 - 1)^2}
\nonumber
\\
&=-\frac{8\left(1+\frac{N}{a}\right)^2(x - x_+)^3}{(1 - x^2)(x - x _-)(x_{+}^2 - 1)^2} 
+\frac{8(x - x_+)^3 }{(x - x _-)(x_{+}^2 - 1)^2} 
+\frac{16 (N/a) (x - x_+)^3}{(1+x)(x - x _-)(x_{+}^2 - 1)^2}
\end{align}
Expressing in terms of z coordinate, we obtain,
\begin{equation}
\begin{split}
V_{\rm conformal}^{\rm modified}(z)&=-\frac{\left(1+\frac{N}{a}\right)^2 (x_{+}^2-1)^3}
{(z -z_\infty)^3 (x_{+}^2-1)^2} \times \frac{z-z_\infty}{(x_+ -x_-) (z-z_s)} \times \frac{z-z_\infty}{z(1+x_+)}\times \frac{z-z_\infty}{(z-1)(x_+ -1)}  
\nonumber
\\
&-\frac{8(x_{+}^2 - 1)^3}{8(z-z_\infty)^3(x_{+}^2 - 1)^2} \times \frac{z -z_\infty}{(x_+ -x_-) (z-z_s)} 
-\frac{16 (N/a) (x_{+}^2-1)^3}{8(z -z_\infty)^3(x_{+}^2 - 1)^2} \times \frac{(z -z_\infty)}{(x_+ - x_-)(z-z_s)} \times \frac{(z -z_\infty)}{z(1 +x_+)} 
\nonumber
\\
&= -\frac{8\left(1+\frac{N}{a}\right)^2(x_{+}^2 - 1)}{8(x_+ - x_-)(x_{+}^2 - 1)z (z-1)(z-z_s)} 
-\frac{(x_{+}^2 - 1)}{(z-z_\infty)^2 (z-z_s)} \frac{1}{(x_+ -x_-)}
\nonumber
\\
&-\frac{2 (N/a) (x_{+}^2 - 1)}{z (z -z_\infty) (z-z_s)} \frac{1}{(x_+ -x_-) (1 + x_+)}
\nonumber
\\
&=-\frac{\left(1+\frac{N}{a}\right)^2}{(x_+ -x_-)} \frac{1}{z (z -1) (z-z_s)}
-\frac{(x_{+}^2 - 1)}{x_+ -x_-} \frac{1}{(z-z_\infty)^2 (z-z_s)}
-\frac{2(N/a) (x_{+}^2 - 1)}{(x_+ -x_-) (1 +x_+)}\frac{1}{z (z -z_\infty) (z-z_s)} 
\end{split}
\end{equation}
Decomposing the above expression, so that each of the individual terms have singularity at one and only one value of $z$, we obtain,
\begin{equation}
\begin{split}
V_{\rm conformal}^{\rm modified}(z)&=- \frac{\left(1+\frac{N}{a}\right)^2}{(x_+ -x_-)} 
\left[\frac{1}{1-z_s} \frac {1}{z-1} + \frac{1}{z_s}\frac{1}{z} + \frac{1}{z_s(z_s -1)} \frac{1}{z-z_s}\right]
\nonumber
\\
&-\frac{(x_{+}^2 - 1)}{x_+ -x_-}\left[\left(\frac{1}{z_\infty -z _s}\right)\frac{1}{(z-z_\infty)^2}
-\left(\frac{1}{z_\infty -z _s}\right)^2 \frac{1}{(z-z_\infty)}
+\left(\frac{1}{z_\infty -z _s}\right)^2 \frac{1}{(z-z_s)}\right] 
\nonumber
\\
&-\frac{2(N/a)(x_{+}^2 - 1)}{(x_+ -x_-)(1 +x_+)} 
\left[\left(\frac{1}{z_s z_\infty}\right)\frac{1}{z}+\left(\frac{1}{z_s(z_s-z_\infty)}\right)\frac {1}{z-z_s}
+\left(\frac{1}{z_\infty(z_\infty-z_s)}\right)\frac {1}{z-z_\infty}\right]
\end{split} 
\end{equation}
Thus we have the following expression for coefficients of various powers of $z$ dependent terms,
\begin{align}
\textrm{Coefficient~of}~(z-z_\infty)^{-2}&=-\frac{(x_{+}^2-1)}{x_+ -x_-}\frac{1}{z_\infty -z _s} 
=\frac{(x_{+}^2-1)}{x_+ -x_-}\frac{2(x_+ -x_-)}{(x_{+}^2 - 1)}=2
\nonumber
\\
\textrm{Coefficient~of}~(z-z_\infty)^{-1}&=\frac{(x_{+}^2 - 1)}{x_+ -x_-}\frac{1}{(z_\infty -z _s)^2}
-\frac{2(N/a) (x_{+}^2 - 1)}{(x_+ -x_-) (1 +x_+)}\frac{1}{z_\infty(z_\infty-z_s)} 
\nonumber
\\
&=\frac{(x_{+}^2 - 1)}{x_+ -x_-} \frac{1}{z_\infty-z_s}\left[\frac{1}{z_\infty-z_s} - \frac{2(N/a)}{x_+ +1} \frac {1}{z_\infty}\right] 
\nonumber
\\
&=-2\left[\frac{2(x_+ -x_-)}{1 - x_{+}^2} + \frac{2(N/a)}{x_+ +1} \frac{2}{x_+-1} \right]  
\nonumber
\\
& = 4\left[\frac{x_+ -x_-}{x_{+}^2 - 1} - \frac{2(N/a)}{x_{+}^2 - 1}\right]
\nonumber
\\
&=4\left[\frac{x_+ -x_-}{x_{+}^2 - 1}+\frac{(1/2)(x_+ + x_-)}{x_{+}^2 - 1}\right] 
=\frac{2(3 x_+ - x_-)}{x_{+}^2 - 1} 
\end{align}
where we have used the result, $N/a=(\gamma/4\delta)=-(1/4)(x_+ -x_-)$. Proceeding further we obtain, 
\begin{equation}
\begin{split}
\textrm{coefficient~of}~z^{-1}&=-\frac{\left(1+\frac{N}{a}\right)^2}{(x_+ -x_-)}\frac{1}{z_s}
-\frac{2(N/a)(x_{+}^2 - 1)}{(x_+ -x_-) (1 +x_+)}\left(\frac{1}{z_\infty z_s}\right) 
\nonumber
\\
& = \frac{\left(1+\frac{N}{a}\right)^2}{(x_+ -x_-)}\frac{2 (x_+ -x_-)}{(1+x_-)(1-x_+)}
-\frac{2(N/a)(x_{+}^2 - 1)}{(x_+ -x_-) (1 +x_+)}\times\frac{2(x_+ -x_-)}{(1+x_-)(1-x_+)}\times\frac{2}{x_+ -1} 
\nonumber
\\
&=\frac{2\left(1+\frac{N}{a}\right)^2}{(1+x_-)(1-x_+)}+\frac{8(N/a) (x_{+}^2 - 1)}{(x_{+}^2 - 1)(1 +x_-)(x_+ - 1)} \\
& = \frac{2  \left(1-\frac{N}{a}\right)^2} {(1+x_-)(1-x_+)}
\end{split} 
\end{equation}
Along identical lines we obtain,
\begin{equation}
\begin{split}
\textrm{Coefficient~of}~(z-1)^{-1}&=-\frac{\left(1+\frac{N}{a}\right)^2}{x_+ -x_-}\frac{1}{1 -z_s} 
=\frac{\left(1+\frac{N}{a}\right)^2}{x_+ -x_-} \frac{2 (x_+ -x_-)}{(x_+ +1)(x_- -1)}
=\frac{2 \left(1+\frac{N}{a}\right)^2}{(x_+ +1)(x_{-}-1)}
\end{split} 
\end{equation}
Finally, the coefficient of the term involving $(z-z_s)^{-1}$ takes the following form,
\begin{equation}
\begin{split}
\textrm{Coefficient~of}~(z-z_{s})^{-1}&=-\frac{\left(1+\frac{N}{a}\right)^2}{x_+ -x_-}\frac{1}{z_s(z_s -1)}
-\frac{(x_{+}^2 - 1)}{x_+ -x_-}\frac{1}{(z_\infty -z_s)^2} 
-\frac{2(N/a)(x_{+}^2 - 1)}{(x_+ -x_-)(1+ x_+)}\frac{1}{z_s(z_s -z_\infty)} 
\nonumber
\\
&= -\frac{\left(1+\frac{N}{a}\right)^2}{x_+ -x_-}\frac{2(x_+ -x_-)}{(1 + x_+)(x_- -1)} \frac{2(x_+ -x_-)}{(x_+ -1)(x_- +1)} 
-\frac{(x_{+}^2 - 1)}{x_+ -x_-}\times \frac{4(x_+ -x_-)^2}{(x_{+}^2 - 1)^2} 
\nonumber
\\
&-\frac{2(N/a)(x_{+}^2 - 1)}{(x_+ - x_-)(1+x_+)}\times \frac{2(x_+ -x_-)}{(x_{+}^2 - 1)}\times \frac{2(x_+ -x_-)}{(x_+ -1)(x_- +1)} \\
& \\
&= -4\frac{x_+ -x_-}{(x_{+}^2 - 1)}\left[1+\frac{\left(1+\frac{N}{a}\right)^2}{(x_{-}^2-1)}+2\frac{(N/a)}{x_- +1}\right]
\end{split} 
\end{equation}
Thus the contribution to the potential appearing in the angular equation due to the conformal coupling as well as the pre-factor appearing in \ref{Ang_Eq_03} takes the form,
\begin{equation}
\begin{split}
V_{\rm conformal}^{\rm modified}(z)&=\frac{2}{(z-z_\infty)^{2}}
+\left\{\frac{2(3 x_+ - x_-)}{x_{+}^2 - 1}\right\}\frac{1}{(z-z_\infty)}
+\left\{\frac{2  \left(1-\frac{N}{a}\right)^2} {(1+x_-)(1-x_+)}\right\}\frac{1}{z}
\nonumber
\\
&+\left\{\frac{2 \left(1+\frac{N}{a}\right)^2}{(x_+ +1)(x_{-}-1)}\right\}\frac{1}{z-1}
-4\frac{x_+ -x_-}{(x_{+}^2 - 1)}\left[1+\frac{\left(1+\frac{N}{a}\right)^2}{(x_{-}^2-1)}+2\frac{(N/a)}{x_- +1}\right]\frac{1}{z-z_{s}}
\end{split} 
\end{equation}
For vanishing NUT charge, the above conformal factor would become,
\begin{equation}
\begin{split}
V_{\rm conformal}^{\rm modified(KdS)}(z)&=\frac{2}{(z-z_\infty)^{2}}
-\left\{\frac{8i\sqrt{\delta}}{(1 + \delta)}\right\}\frac{1}{(z-z_\infty)}
-\left\{\frac{2\delta}{(1+i\sqrt{\delta})^2}\right\}\frac{1}{z}
\nonumber
\\
&+\left\{\frac{2 \delta}{(1-i\sqrt{\delta})^2}\right\}\frac{1}{z-1}
+\left\{\frac{8 i \sqrt{\delta}}{(1+\delta)^2}\right\}\frac{1}{z-z_{s}}
\end{split} 
\end{equation}
One can explicitly check that the form of the potential due to the conformal factor matches with the result of  \cite{Suzuki:1998vy}.

Let us now simplify the other terms in the potential by adopting the same procedure as followed above. We will start by expressing the overall factor in the front of the potential in \ref{Ang_Eq_03} in terms of the $z$ coordinate as follows,
\begin{equation}
\begin{split}
\label{1}
\frac{\delta}{4}(1 -x^2)&(x-x_-)(x-x_+)^{-3}({x_+}^2 -1)^2
\nonumber
\\
&=\frac{\delta}{4} \frac{(x_+ -1)(z-1)}{(z-z_\infty)}\times (-1)\times\frac{z(1+x_+)}{z-z_\infty} \times {(x_+ -x_-)} \frac{z-z_s}{z-z_\infty} 
\times (-1) \times \frac{8(z-z_\infty)^3}{(x_{+}^2 -1)^3} (x_{+}^2 -1)^2 
\nonumber
\\
&=\frac{\delta}{4} \frac{(x_{+}^2 -1) z (z-1) (z-z_s) (x_+ -x_-)}{(z-z_\infty)^3} \frac{8(z-z_\infty)^3}{(x_{+}^2 -1)}  
\nonumber
\\
&=2 \delta (x_+ -x_-) z(z-1) (z-z_s)
\end{split}
\end{equation}
Thus the potential appearing in the angular equation without the conformal part reads,
\begin{align}
V_{\rm non-conformal}&=K + \dfrac{\xi^2}{\delta} + \frac{\mathcal{A}}{\delta(x-x_+)(x-x_-)} + \frac{x \mathcal{B}}{\delta(x-x_+)(x-x_-)} 
\nonumber
\\
&+\frac{\mathcal{C}}{\delta (1 -x^2)(x-x_+)(x-x_-)}+\frac{\mathcal{D} x}{\delta (1 -x^2)(x-x_+)(x-x_-)}
\end{align}
where, the constants $\mathcal{A}$, $\mathcal{B}$, $\mathcal{C}$ and $\mathcal{D}$ defined above has the following expressions,
\begin{align}
\mathcal{A}&\equiv -(\delta+1) \frac{\xi^2}{\delta}+2 \xi (m +2C\eta) + 4 \eta^2~;\qquad \mathcal{B}\equiv -\frac{\gamma\xi^2}{\delta}+4 \xi \eta 
\nonumber
\\
\mathcal{C}&\equiv-(m + 2 C \eta)^{2}-4 \eta ^2~;\qquad \mathcal{D}\equiv- 4 \eta (m+ 2C\eta)
\end{align}
The above potential can also be broken into individual terms in the following fashion,
\begin{equation}
\begin{split}
V(x)_{\rm non-conformal}&=K+\frac{\xi^2}{\delta}+\frac{(\mathcal{A}/\delta)}{(x-x_+)(x-x_-)}+\frac{(\mathcal{B}/\delta)}{x_+ -x_-}
\left(\frac{x_+}{x-x_+} - \frac{x_-}{ x-x_-}\right) 
+ \frac{(\mathcal{C}/\delta)}{(1 -x^2)(x-x_+)(x-x_-)} 
\nonumber
\\
&+\frac{\mathcal{D}}{\delta}\Bigg[-\frac{x_+}{(x_+ -1)(x_+ +1)(x_+ -x_-)(x-x_+)}-\frac{x_-}{(x_- -1)(x_- +1)(x_- -x_+)(x-x_-)} 
\nonumber
\\
&-\frac{1}{2 (x_+ -1)(x_- -1)(x-1)}-\frac{1}{2 (x_+ +1)(x_- +1)(x+1)}\Bigg]
\end{split}
\end{equation}
Each of these terms can be expressed in term of the new variable $z$ following \ref{Imp_Ident} and hence the potential takes the following form,
\begin{equation}
\begin{split}
V_{\rm non-conformal}(z)&=K+\frac{\xi^2}{\delta}-\frac{\mathcal{A}}{\delta}\frac{2 (z-z_\infty)^2}{(x_{+}^2 -1)(x_+ - x_-)}\frac{1}{z-z_s}
+ \frac{(\mathcal{B}/\delta)}{x_+ -x_-}\left[-\frac{2(z-z_\infty)x_+}{(x_{+}^2 -1)} - \frac{x_-(z-z_\infty)}{(x_+ -x_-)(z-z_s)}\right] 
\nonumber
\\
&+\frac{\mathcal{C}}{\delta}\left[\frac{z-z_\infty}{(x_+ -1) (z-1)} \times \frac{z-z_\infty}{z(x_+ +1)} \times \frac{2(z-z_\infty)}{(x_{+}^2 -1)}\times \frac{z-z_\infty}{(x_+ - x_-)(z-z_s)}\right]
\nonumber
\\
&-\frac{(\mathcal{D}/\delta)}{x_+ -x_-} [-\frac{2 x_+(z-z_\infty)}{(x_{+}^2 -1)}-\frac{x_- (z-z_\infty)}{(x_+ -x_-)(z-z_s)}] \times \frac{z-z_\infty}{(x_+ -1)(z-1)} \times \frac{z-z_\infty}{z(x_+ +1)} 
\nonumber
\\
&=K+\frac{\xi^2}{\delta} - \frac{(2\mathcal{A}/\delta)}{(x_{+}^2 -1)} \frac{1}{x_+ - x_-} \frac{(z-z_\infty)^2}{z-z_s} 
-\frac{(\mathcal{B}/\delta)}{x_+ - x_-} \left[-\frac{2 x_+(z-z_\infty)}{(x_{+}^2 -1)}-\frac{x_-}{x_+ -x_-} \frac{z-z_\infty}{z-z_s}\right]
\nonumber
\\
&+\frac{(2\mathcal{C}/\delta)}{(x_{+}^2 -1)^2} \frac{1}{x_+ -x_-} \frac{(z-z_\infty)^4}{z(z-1)(z -z_s)}
\nonumber
\\
&+\frac{(\mathcal{D}/\delta)}{(x_+ -x_-)}\frac{1}{(x_{+}^2 -1)} \frac{(z-z_\infty)^2}{z(z-1)}\left[\frac{2 x_+(z-z_\infty)}{(x_{+}^2 -1)} + \frac{x_-}{x_+ -x_-} \frac{z-z_\infty}{z-z_s}\right]
\end{split}
\end{equation}
Thus after the division of the above potential by the overall factor, which we expressed in terms of the $z$ variable in \ref{1}, the contribution to the angular equation can be divided into five parts, such that,
\begin{equation}
\begin{split}
V_{\rm non-conformal}^{\rm modified}(z)&= \frac{V_{\rm non-conformal}(z)}{2 \delta(x_+ -x_-)(z-1) z (z-z_s)} 
\nonumber
\\
&=\textrm{Term~1}+\textrm{Term~2}+\textrm{Term~3}+\textrm{Term~4}+\textrm{Term~5}
\end{split}
\end{equation}
Let us now investigate the potential term by term, which we will decompose into individual terms depending on a single singular value of $z$, to start with the first term yields,
\begin{equation}
\begin{split}
\textrm{Term~1}&=\frac{K+(\xi^2/\delta)}{2 \delta (x_+ -x_- )}
\left[\frac{1}{z_s}\frac{1}{z} + \frac{1}{1-z_s}\frac{1}{z-1} - \frac{1}{z_s (1-z_s)}\frac{1}{z-z_s}\right]
\end{split}
\end{equation}
Similarly, the second term gives the following expression,
\begin{equation}
\begin{split}
\textrm{Term~2}&=-\frac{(\mathcal{A}/\delta^2)}{(x_{+}^2-1)} \frac{1}{(x_+ - x_-)^2} 
\Bigg[-\frac{{z_\infty}^2}{z {z_s}^2} - \frac{-1 + 2 z_\infty - {z_\infty}^2}{(z_s -1)^2} \frac{1}{z-1} + \frac{-z_{s}^2 +2 z_{s}^2 {z_\infty} + z_{\infty}^2 - 2 z_{s} z_{\infty}^2}{z_{s}^2 (z_s -1)^2(z-z_s)}
\nonumber
\\
&+\frac{z_{s}^2 - 2 z_s z_\infty + z_{\infty}^2}{z_s (z_s-1)} \frac{1}{(z-z_s)^2}\Bigg]
\end{split}
\end{equation}
Exploring the third term appearing in the potential we obtain,
\begin{equation}
\begin{split}
\textrm{Term~3}&=-\frac{(\mathcal{B}/2 \delta^2)}{(x_+ - x_-)^2}\left[- \frac{2 x_+(z-z_\infty)}{(x_{+}^2 -1)z(z-1)(z-z_s)}-\frac{x_-}{x_+ -x_-} \frac{z-z_\infty}{z(z-z_s)^2(z-1)}\right] 
\nonumber
\\
&=-\frac{(\mathcal{B}/2 \delta^2)}{(x_+ - x_-)^2}\Bigg[-\frac{2 x_+}{(x_{+}^2 -1)}\left\{\frac{z_\infty-1}{(z_s -1)(z-1)} - \frac{1}{z}\frac{z_\infty}{z_s} + \frac{z_s -z_\infty}{z_s (z_s -1) (z-z_s)}\right\}
\nonumber
\\
&- \frac{x_-}{x_+ -x_-}\left\{\frac{z_\infty}{z z_{s}^2} + \frac{1- z_\infty}{(z-1) (z_s-1)^2} + \frac{z_s -z_\infty}{z_s (z_s -1)( z-z_s)^2} + \frac{-z_{s}^2 -z_\infty + 2z_s z_\infty }{z_{s}^2 (z_s -1)^2 (z-z_s)}\right\}\Bigg]
\end{split}
\end{equation}
The fourth term appearing in the potential of the angular equation, has the following decomposition,
\begin{equation}
\begin{split}
\textrm{Term~4}&=\frac{(\mathcal{C}/\delta^2)}{(x_{+}^2 -1)^2} \frac{1}{(x_+ -x_-)^2} \frac{(z-z_\infty)^4}{z^2 (z-1)^2 (z -z_s)^2}
\nonumber
\\
&=\frac{(\mathcal{C}/\delta^2)}{(x_{+}^2 -1)^2}\frac{1}{(x_+ -x_-)^2}\left[\frac{a}{(z-1)^2} + \frac{b}{z-1} + \frac{c}{z^2} + \frac{d}{z} + \frac{e}{(z-z_s)^2} + \frac{f}{z-z_s}\right]
\end{split}
\end{equation}
where the constants appearing above has the following expressions,
\begin{equation}
\begin{split}
a&\equiv\frac{1 -4 z_\infty + 6 z_{\infty}^2 - 4 z_{\infty}^3 + z_{\infty}^4}{(z_s -1)^2}~,
\nonumber
\\
b&\equiv \frac{-2 (-z_s + 2 z_\infty + 2 z_s z_\infty - 6 z_{\infty}^2 + 6 z_{\infty}^3 -2 z_s z_{\infty}^3 -2 z_{\infty}^4 + z_s z_{\infty}^4)}{(z_s-1)^3}~,
\nonumber
\\
c&\equiv\frac{z_{\infty}^4}{z_{s}^2}~;\qquad 
d\equiv \frac{2(-2 z_s z_{\infty}^3 + z_{\infty}^4 + z_s z_{\infty}^4)}{z_{s}^3}~,
\nonumber 
\\
e&\equiv\frac{z_{s}^4 - 4 z_{s}^3 {z_\infty} +6 z_{s}^2 z_{\infty}^2 -4z_{s} z_{\infty}^3 + z_{\infty}^4}{z_s^2 (z_s -1) ^2}~,
\nonumber
\\
f&\equiv\frac{2\left(-z_s^4 + 2 z_{s}^3 {z_\infty} + 2 z_{s}^4 {z_\infty} - 6 z_{s}^3 z_{\infty}^2 -2 {z_s} z_{\infty}^3 + 6 z_{s}^2 z_{\infty}^3 + z_{\infty}^4 -2 {z_s} z_{\infty}^4\right)}{z_s^3 (z_s -1) ^3}
\end{split}
\end{equation}
and finally the fifth term can be decomposed as,
\begin{equation}
\begin{split}
\textrm{Term~5}&= \frac{(\mathcal{D}/2 \delta^2)}{(x_+ - x_-)^2} \frac{1}{x_{+}^2 -1 }  
\left[\frac{2 x_+}{x_+^2 -1} \frac{(z-z_\infty^3)}{z^2 (z-1) ^2 (z-z_s)} + \frac{x_-}{(x_+ -x_-)} \frac{(z-z_\infty^3)}{z^2 (z-1) ^2 (z-z_s)^2}\right] 
\nonumber
\\
&=\frac{(\mathcal{D}/2 \delta^2)}{ (x_+ - x_-)^2} \frac{1}{x_+^2 -1 } 
\Bigg[\frac{2 x_+}{x_+^2 -1}\Bigg\{ \frac{-1 + 3 z_\infty -3 z_\infty^2 + z_\infty^3}{(z_s-1)(z-1)^2} 
+\frac{-z_s + 3 z_\infty -6 z_\infty^2 + 3 z_s z_\infty + 3 z_\infty^3 -2 z_s z_\infty^3}{(z_s-1)^2(z-1)}   
\nonumber
\\
&+ \frac{z_\infty^3}{z_s z^2}
+\frac{-3 {z_s} z_{\infty}^2 + 2 {z_s} z_{\infty}^3 + z_\infty^3}{z_s^2 z} 
+\frac{z_s^3 -3 z_{s}^2 {z_\infty} +3 {z_s} z_{\infty}^2 - z_\infty^3}{z_s^2 (z_s-1)^2 (z-z_s)}\Bigg\}
\nonumber
\\
&+\frac{x_-}{x_+ -x_-}\Bigg\{\frac{(1 -z_\infty)^3}{(z_s -1 )^2(z-1)^2} 
+\frac{ 1 + z_s -6 z_\infty + 9z_\infty^2 -3 {z_s} z_{\infty}^2 +2 {z_s} z_{\infty}^3 -4 z_\infty^3}{(z-1) (z_s-1 )^3}  
-\frac{z_\infty^3}{z_s^2 z^2} 
\nonumber
\\
&+\frac{3 z_s z_\infty^2 -2 z_\infty^3 -2 z_s z_\infty^3}{z_s^3 z} 
+\frac{z_s^3 -3 z_{s}^2 {z_\infty} +3 {z_s} z_{\infty}^2 - z_\infty^3}{z_s^2 (z_s-1)^2 (z-z_s)^2}
\nonumber
\\
&+\frac{-z_s^3 -z_s^4 + 6 z_{s}^3 {z_\infty} + 3 {z_s} z_{\infty}^2 - 9 z_{s}^2 z_{\infty}^2 -2 z_\infty^3 + 4 {z_s} z_{\infty}^3  }{(z_s-1)^3 z_s^3 (z-z_s)}\Bigg\}\Bigg] 
\end{split}
\end{equation}
Thus the non-conformal part of the potential appearing in the angular equation can be written as,
\begin{equation}
\begin{split}
V_{\rm non-conformal}^{\rm modified}(z)=\frac{\textrm{C1}}{z^2}+\frac{\textrm{C2}}{(z-1)^2} 
+\frac{\textrm{C3}}{(z-z_s)^2}+\frac{\textrm{C4}}{z}+\frac{\textrm{C5}}{z-1}+\frac{\textrm{C6}}{z-z_s}
\end{split}
\end{equation}
where, the constant coefficients introduced above has the following expressions,
\begin{align}\label{Ang_Coeff_Def}
\textrm{C1}&\equiv \frac{(\mathcal{C}/ \delta^2)}{(x_+^2 -1)^2}\frac{1}{(x_+ -x_-)^2}\frac{z_\infty^4}{z_s^2} 
+ \frac{(\mathcal{D}/2\beta^2)}{ (x_+ - x_-)^2} \frac{1}{x_+^2 -1 } \times \left[\frac{2x_+}{x_+^2 -1} \frac{z_\infty^3}{z_s} -\frac{x_-}{x_+ -x_-} \frac{z_\infty^3}{z_s^2}\right]
\nonumber
\\
\textrm{C2}&\equiv \frac{(\mathcal{C}/\delta^2)}{(x_{+}^2 -1)^2}\frac{1}{(x_+ -x_-)^2} \frac{(1-z_\infty)^4}{(1-z_s)^2}  
+\frac{(\mathcal{D}/2\beta^2)}{(x_+ - x_-)^2} \frac{1}{x_+^2 -1 } 
\left[ \frac{2x_+}{x_+^2 -1}\frac{(z_\infty-1)^3}{z_s-1} -\frac{x_-}{x_+ -x_-} \frac{(1-z_\infty)^3}{(z_s-1)^2}\right]
\nonumber
\\
\textrm{C3}&\equiv - \frac{(\mathcal{A}/\delta^2)}{(x_{+}^2 -1)} \frac{1}{(x_+ - x_-)^2}\frac{(z_\infty-z_s)^2}{z_s(z_s-1)} 
+\frac{(\mathcal{B}/2\delta^2)}{(x_+ - x_-)^2} \frac{x_-}{x_+ -x_-}\frac{(z_s -z_\infty)}{z_s(z_s-1)} 
\nonumber
\\
&+ \frac{(\mathcal{C}/\delta^2)}{(x_{+}^2 -1)} \frac{1}{(x_+ - x_-)^2} \frac{(z_s -z_\infty)^4}{z_s^2(z_s-1)^2} 
+ \frac{(\mathcal{D}/2\delta^2)}{(x_+ - x_-)^2} \frac{1}{x_+^2 -1 } \frac{ x_-}{x_+ -x_-}  \frac{(z_s -z_\infty)^3}{z_s^2(z_s-1)^2}
\nonumber
\\
\textrm{C4}&\equiv\frac{K+(\xi^2/\delta)}{2\delta (x_+ -x_- )} \frac{1}{z_s}
-\frac{(\mathcal{A}/\delta^2)}{(x_{+}^2 -1)}\frac{1}{(x_+ - x_-)^2} \left(\frac{-{z_\infty}^2}{ {z_s}^2}\right) 
- \frac{(\mathcal{B}/2\delta^2)}{(x_+ - x_-)^2}\left[\frac{- x_-}{x_+ -x_-} \frac{z_\infty}{z_s^2} + \frac{2 x_+}{({x_+}^2 -1)} \frac{z_\infty}{z_s } \right] 
\nonumber
\\
&+ \frac{(\mathcal{C}/\delta^2)}{(x_{+}^2 -1)^2} \frac{1}{(x_+ -x_-)^2} \left\{\frac{2(-2 z_s z_\infty^3 + z_\infty^4 + z_s z_\infty^4)}{z_s^3}\right\} 
\nonumber
 \\
&+\frac{(\mathcal{D}/2\delta^2)}{(x_+ - x_-)^2} \frac{1}{x_+^2 -1 }\Big[\frac{2 x_+}{({x_+}^2 -1)} \frac{-3 z_s z_\infty^2 + z_\infty^3 + 2 z_s z_\infty^3}{z_s^2}
+ \frac{x_-}{x_+ -x_-} \frac{3 z_s z_\infty^2 -2 z_\infty^3 - 2 z_s z_\infty^3 }{z_s^3} \Big]
\nonumber
\\
\textrm{C5}&\equiv  \frac{K+(\xi^2/\delta)}{2 \delta (x_+ -x_- )}\frac{1}{1-z_s} 
-\frac{(\mathcal{A}/\delta^2)}{(x_{+}^2 -1)}\frac{1}{(x_+ - x_-)^2} \frac{(z_\infty-1)^2}{(z_s-1)^2}
\nonumber
\\
&-\frac{(\mathcal{B}/2\delta^2)}{(x_+ - x_-)^2} \left[- \frac{2 x_+}{(x_{+}^2 -1)}\frac{z_\infty-1}{(z_s -1)} 
+\frac{x_-}{x_+ -x_-} \frac{ z_\infty -1}{ (z_s-1)^2} \right]
\nonumber
\\
&+\frac{(\mathcal{C}/\delta^2)}{(x_{+}^2 -1)^2}\frac{1}{(x_+ -x_-)^2}\frac{-2 (-z_s + 2 z_\infty + 2 z_s z_\infty - 6 z_\infty^2 + 6 z_\infty^3 -2 z_s z_\infty^3 -2 z_\infty^4 + z_s z_\infty^4)}{(z_s-1)^3}
\nonumber
\\
&+\frac{(\mathcal{D}/2\delta^2)}{(x_+ - x_-)^2} \frac{1}{x_+^2 -1 }\Bigg[\frac{2 x_+}{x_+^2 -1}\times
\frac{-z_s + 3 z_\infty -6 z_\infty^2 + 3 z_s z_\infty + 3 z_\infty^3 -2 z_s z_\infty^3}{(z_s-1)^2} 
\nonumber
\\
&+\frac{x_-}{x_+ -x_-}\frac{ 1 + z_s -6 z_\infty + 9z_\infty^2 -3 {z_s} z_{\infty}^2 +2 {z_s} z_{\infty}^3 -4 z_\infty^3}{(z_s-1 )^3}\Bigg]
\nonumber
\\
\textrm{C6}&\equiv\frac{K+(\xi^2/\delta)}{2 \delta (x_+ -x_- )}\frac{1}{z_s(z_s-1)} 
-\frac{(\mathcal{A}/\delta^2)}{(x_{+}^2 -1)} \frac{1}{(x_+ - x_-)^2} \frac{-z_s^2 + 2 z_s^2 z_\infty + z_\infty^2 -2 z_s z_\infty^2}{z_s^2 (z_s-1)^2} 
\nonumber
\\
&-\frac{(\mathcal{B}/2\delta^2)}{(x_+ - x_-)^2}\left[- \frac{2 x_+}{(x_{+}^2 -1)}\frac{z_s -z_\infty}{z_s (z_s-1)} +  \frac{x_-}{x_+ -x_-} \frac{(z_s -z_\infty)^2}{z_s^2 (z_s-1)^2}\right]
\nonumber
\\
&+\frac{(\mathcal{C}/\delta^2)}{(x_{+}^2 -1)^2}\frac{1}{(x_+ -x_-)^2} \frac{2 (- z_s^4 + 2 z_{s}^3 {z_\infty} + 2 z_{s}^4 {z_\infty} - 6 z_{s}^3 z_{\infty}^2 -2 {z_s} z_{\infty}^3 + 6 z_{s}^2 z_{\infty}^3 - z_\infty^4 -2 {z_s} z_{\infty}^4 )}{z_s^3 (z_s -1)^3} 
\nonumber
\\
&+ \frac{(\mathcal{D}/2\delta^2)}{ (x_+ - x_-)^2} \frac{1}{x_+^2 -1 }\Bigg[\frac{2 x_+}{(x_{+}^2 -1)} \frac{(z_s -z_\infty)^3}{z_s^2 (z_s-1)^2}
\nonumber
\\
&+\frac{x_-}{x_+ -x_-}\frac{-z_s^3 -z_s^4 + 6 z_{s}^3 {z_\infty} + 3 {z_s} z_{\infty}^2 - 9 z_{s}^2 z_{\infty}^2 -2 z_\infty^3 + 4 {z_s} z_{\infty}^3  }{(z_s-1)^3 z_s^3 }\Bigg]
\end{align}
This is the result we have used in \ref{Ang_Eq_05}, in the main text. To check the validity of this expression, one can again consider the Kerr-dS limit, i.e., with vanishing NUT charge. Then we have, $\mathcal{D}=0=\gamma$ in the above expressions. Thus the coefficient of $(1/z^{2})$ term becomes,
\begin{equation}
\begin{split}
\textrm{C1}_{\rm KdS}&=\frac{(\mathcal{C}/\delta^{2})}{(x_+^2 -1)^2}\frac{1}{(x_+ -x_-)^2}\frac{z_\infty^4}{z_s^2} 
\nonumber
\\
&=-\frac{m^2}{\delta^2} \frac{1}{ (x_+^2 -1)^2 (x_+ -x_-)^2} \times\frac{(x_+ -1)^4}{ 16} \times \frac{4(x_+ -x_-)^2}{ (1- x_+)^2 (1 + x_-)^2} 
\nonumber
\\
&=- \frac{m^2}{4 \delta^2} \frac{1}{(x_+ +1)^2 (x_- +1)^2}= \frac{-m^2}{4} \times\frac{1}{(1+\delta)^2} 
\end{split}
\end{equation}
Similarly, the coefficient of $(z-1)^{-2}$ term becomes,
\begin{equation}
\begin{split}
\textrm{C2}_{\rm KdS}&=\frac{(\mathcal{C}/\delta^2)}{(x_{+}^2 -1)^2}
\frac{1}{(x_+ -x_-)^2} \frac{(1-z_\infty)^4}{(1-z_s)^2} 
\nonumber
\\
&=\frac{-m^2}{\delta^2}\frac{1}{(x_+ +1)^2 (x_+ -1)^2 (x_+ -x_-)^2} \times \frac{(x_+ +1)^4}{16} \times \frac{4 (x_+ -x_-)^2}{(1 +x_+)^2 (1 -x_-)^2}
\nonumber
\\
&=-\frac{m^2}{4 \delta^2} \frac{1}{(x_+ -1)^2 (1-x_-)^2} 
=-\frac{m^2}{4} \frac{1}{(1+\delta)^2} 
\end{split}
\end{equation}
As one can explicitly verify these two coefficients matches with the result for Kerr-dS spacetime presented in \cite{Suzuki:1998vy}.

Following the computations presented above, we have been able to rewrite the angular equation in a form similar to the Heun's equation. However, there is one last hurdle in this derivation, i.e., we must be able to demonstrate that sum of the three coefficients $\mathcal{M}$, $\mathcal{N}$ and $\mathcal{P}$ appearing in \ref{Ang_Eq_06} vanishes. For this purpose, let us start with the sum $\mathcal{M}+\mathcal{N}+\mathcal{P}$, which yields,
\begin{equation}
\begin{split}
{\mathcal{M}+\mathcal{N}+\mathcal{P}}&=\textrm{C4+C5+C6}-\frac{1}{z_\infty}-\frac{1}{z_\infty-1}-\frac{1}{z_\infty-z_{s}} 
+\left(1-\frac{N}{a}\right)^{2}\frac{2}{(1-x_{+})(1+x_{-})} +\frac{2\left(1+\frac{N}{a}\right)^{2}}{(x_{+}+1)(x_{-}-1)}
\nonumber
\\
&-\frac{4(x_+ -x_-)}{x_{+}^2 -1}\left[1+\frac{\{1+(N/a)\}^2}{x_{-}^2 -1}+\frac{2(N/a)}{x_- +1}\right]
\end{split}
\end{equation}
The last and last but one term appearing in the first line of the above expression can be simplified, yielding,
\begin{align}
&\frac{2(1+\frac{N}{a})^2}{(x_{+}+1)(x_{-}-1)}-\frac{2 [\{1+(N/a)\}^2-4(N/a)]}{(x_- +1)(x_+ -1)}
\nonumber
\\
&=\frac{2(1+\frac{N}{a})^2}{(x_{+}+1)(x_{-}-1)} -\frac{2(1+\frac{N}{a})^2}{(x_{+}-1)(x_{-} + 1)} 
+\frac{8(N/a)}{(x_{-}+1)(x_{+}-1)}
\nonumber
\\
&=\frac{2(1+\frac{N}{a})^2}{(x_{+}^{2}-1)(x_{-}^{2}-1)} \times \left[x_{+}x_{-}-1-x_{-}+x_{+}-x_{+}x_{-}-x_{-}+x_{+}+1\right] 
+\frac{8(N/a)}{(x_{-}+1)(x_{+}-1)}
\nonumber
\\
&=\frac{2\{1+(N/a)\}^2 \times 2(x_+ -x_-)} {(x_{+}^2-1)(x_{-}^2-1)} + \frac{8(N/a)}{(x_- +1)(x_+ -1)}
\end{align}
Hence, the summation of the coefficients appearing in \ref{Ang_Eq_06}, i.e., the quantity $(\mathcal{M+N+P})$ can be expressed as,
\begin{equation}
\begin{split}
\mathcal{M+N+P}&=\textrm{C4+C5+C6} 
-\left(\frac{1}{z_\infty} + \frac{1}{z_\infty -1} + \frac{1}{z_\infty -z_s}\right)  
+\frac{4(1+N/a)^2 (x_+ -x_-)}{(x_{+}^2 -1)(x_{-}^2 -1)} 
+\frac{8(N/a)}{(x_- +1)(x_+ -1)} -\frac{4 (x_+ -x_-)}{x_{+}^2 -1}
\nonumber
\\
&-\frac{4(x_+ -x_-)\{1+(N/a)\}^2}{(x_{+}^2 -1)(x_{-}^2 -1)} 
-\frac{8(N/a)(x_+ -x_-)}{(x_{+}^2 -1)(x_{-} +1)} 
\nonumber
\\
&=\textrm{C4+C5+C6}-\left(\frac{1}{z_\infty} + \frac{1}{z_\infty -1} + \frac{1}{z_\infty -z_s}\right)
-\frac{4 (x_+ -x_-)}{x_{+}^2 -1}+\frac{8(N/a)}{(x_{+}^2 -1)(x_{-} +1)}\left[x_+ +1 -x_{+}+x_-\right]
\nonumber
\\
&=\textrm{C4+C5+C6}-\left(\frac{1}{z_\infty} + \frac{1}{z_\infty -1} + \frac{1}{z_\infty -z_s}\right)
-\frac{4(x_+ -x_-)}{x_{+}^2 -1}+\frac{8(N/a)(x_- + 1)}{(x_{+}^2 -1)({x_-} +1)}
\end{split}
\end{equation}
At this stage we may recall the following identity involving $x_{\pm}$ as well as $z_{s}$ and $z_{\infty}$, which reads,
\begin{equation}
\begin{split}
-\frac{4(x_+ -x_-)}{x_{+}^2 -1}+\frac{8(N/a)}{(x_{+}^2 -1)}
=\frac{2}{x_{+}^2 -1}\left[-2(x_+ -x_-)+(-x_+ - x_-)\right] 
=\frac{1}{z_\infty} + \frac{1}{z_\infty -1} + \frac{1}{z_\infty -z_s}
\end{split}
\end{equation}
Thus the sum $(\mathcal{M+N+P})$ becomes dependent only on the coefficients $\textrm{C4}$, $\textrm{C5}$ and $\textrm{C6}$ such that 
\begin{align}
\mathcal{M+N+P}&=\textrm{C4+C5+C6} 
\nonumber
\\
& = \frac{K + \xi^2/ \delta}{2 \delta(x_+ -x_-)}\left[\frac{1}{z_s} + \frac{1}{1 - z_s } + \frac{1}{z_s(z_s-1)}\right]
-\frac{(\mathcal{A}/\delta)^2}{x_{+}^2 -1}\times \frac{1}{(x_+ -x_-)^2}\Bigg[-\left(\frac{z_\infty}{z_s}\right)^2 
+\left(\frac{z_\infty -1}{z_s-1}\right)^2
\nonumber
\\
&+\frac{-z_{s}^2 + 2 z_{s}^2 z_\infty +z_{\infty}^2 - 2z_s z_{\infty}^2}{z_{s}^2(z_s-1)^2}\Bigg]
-\frac{(\mathcal{B}/2 \delta^2)}{(x_+ -x_-)^2}
\Bigg[-\frac{x_-}{x_+ -x_-}\frac{z_\infty}{z_{s}^2} 
+ \frac{2 x_+}{x_{+}^2 -1}\frac{z_\infty}{z_s} 
-\frac{2 x_+}{x_{+}^2 -1}\frac{z_\infty - 1}{z_s-1} 
\nonumber
\\
&+\frac{x_-}{x_+ -x_-} \frac{z_\infty - 1}{(z_s -1)^2} 
-\frac{2 x_+}{x_{+}^2 -1}\frac{z_s - z_\infty}{z_s (z_s-1)} 
+\frac{x_-}{x_+ -x_-}.\frac{z_{s}^2 + z_\infty - 2 z_s z_\infty}{z_{s}^2(z_s -1)^2}\Bigg]
\nonumber
\\
&+ \frac{(\mathcal{C}/\delta^2)}{(x_{+}^2 -1)^2} \frac{1}{(x_+ -x_-)^2} \Bigg[\frac{2(z_{\infty}^4+ z_sz_{\infty}^4 - 2z_s z_{\infty}^3)}{z_{s}^3}
\nonumber
\\
&-2 \frac{(-z_s +2z_\infty + 2z_s z_\infty-6 z_{\infty}^2 + 6z_{\infty}^3 - 2z_s z_{\infty}^3 -2 z_{\infty}^4 
+z_s z_{\infty}^4)}{(z_s-1)^3}
\nonumber
\\
&+ \frac{2(- {z_s}^4 + 2z_{s}^3 z_\infty + 2z_{s}^4 z_\infty -6 z_{s}^3z_{\infty}^2 -2 z_{\infty}^3 z_s + 6 z_{s}^2z_{\infty}^3 
+z_{\infty}^4  - 2 z_s z_{\infty}^4 )}{z_{s}^3 (z_s -1)^3}\Bigg]
\nonumber
\\
&+\frac{(\mathcal{D}/ 2\beta^2)}{(x_{+}^2 -1)(x_+ -x_-)^2} \Bigg[\frac{2 x_+}{x_{+}^2 -1}
\left\{\frac{-3 z_s z_{\infty}^2 + z_{\infty}^3 + 2 z_s z_{\infty}^3}{z_{s}^2}\right\} 
+\frac{x_-}{(x_+ -x_-)} \frac{3 z_s z_{\infty}^2 -2 z_{\infty}^3 -2z_s z_{\infty}^3}{z_{s}^3}
\nonumber
\\
&+\frac{2x_+}{x_{+}^2 -1}\left\{\frac{-z_s + 3z_\infty -6z_{\infty}^2  + 3 z_s z_{\infty}^2 +3 z_{\infty}^3 -2 z_s z_{\infty}^3}{(z_s -1)^2}\right\} 
\nonumber
\\
&+\frac{x_-}{x_+ -x_-}(\frac{1 +z_s -6 z_\infty +9 z_{\infty}^2 -3 z_s z_{\infty}^2 -4 z_{\infty}^3 +2 z_s z_{\infty}^3}{(z_s-1)^3})
\nonumber
\\
&+\frac{2 x_+}{x_{+}^2 -1}\frac{(z_s -z_\infty)^3}{z_{s}^2 (z_s -1)^2} 
+\frac{x_-}{x_+ - x_-}\left\{\frac{-z_{s}^3 -z_{s}^4 +6z_{s}^3 z_\infty + 3z_s z_{\infty}^2  
-9 z_{s}^2 z_{\infty}^2 -2 z_{\infty}^3 +4 z_s z_{\infty}^3}{z_{s}^3 (z_s-1)^3}\right\}\Bigg]
\end{align}
Now let us take a look at the coefficients of various terms appearing in the above expression. Let us start with the coefficient of $\{K +(\xi^2/ \delta)\}$, which can be simplified as,
\begin{align}
\textrm{Coefficient~of}~\left\{\frac{K + \xi^2/ \delta}{2 \delta(x_+ -x_-)}\right\}= \frac{1}{z_s} - \frac{1}{1 - z_s } + \frac{1}{z_s(z_s-1)} = \frac{z_s -1-z_s +1}{z_s(z_s-1)} =0  \\
\end{align}
Proceeding further, the coefficient of $(\mathcal{A}/\delta^{2})$ term reads,
\begin{align}
\textrm{Coefficient~of}~ (A/\delta^2)&=-\left(\frac{z_\infty}{z_s}\right)^2 
+\left(\frac{z_\infty -1}{z_s-1}\right)^2 + \frac{-z_{s}^2 + 2 z_{s}^2 z_\infty +z_{\infty}^2 - 2z_s z_{\infty}^2}{z_{s}^2(z_s-1)^2}
\nonumber
\\
&= \frac{-z_{\infty}^2 (z_s-1)^2 + z_{s}^2(z_\infty-1)^2 - z_{s}^2 + 2 z_{s}^2 z_\infty +z_{\infty}^2 - 2z_s z_{\infty}^2}{z_{s}^2(z_s-1)^2}=0
\end{align}
Subsequently, the coefficient of $(\mathcal{B}/2 \delta^2)$ term yields,
\begin{equation}
\begin{split}
\textrm{Coefficient~of}~ (\mathcal{B}/2\beta^2)&=-\frac{x_-}{x_+ -x_-}\frac{z_\infty}{z_{s}^2}
+\frac{2 x_+}{x_{+}^2 -1}\frac{z_\infty}{z_s} 
-\frac{2 x_+}{x_{+}^2 -1}\frac{z_\infty - 1}{z_s-1} 
+ \frac{x_-}{x_+ -x_-}\frac{z_\infty - 1}{(z_s -1)^2}
\nonumber
\\
&-\frac{2 x_+}{x_{+}^2 -1}\frac{z_s - z_\infty}{z_s (z_s-1)} 
+\frac{x_-}{x_+ -x_-}\frac{z_{s}^2 + z_\infty - 2 z_s z_\infty}{z_{s}^2(z_s -1)^2 }
\nonumber
\\
&=-\frac{x_-}{x_+ -x_-}\Bigg[\frac{z_\infty(z_s -1)^2 - (z_\infty -1)z_{s}^2 +(2 z_s z_\infty -z_{s}^2 - z_\infty)}{z_{s}^2 (z_s -1)^2}\Bigg] 
\nonumber
\\
&+\frac{2 x_+}{x_{+}^2 -1}\Bigg[\frac{z_\infty(z_s-1) -z_s(z_\infty -1) -z_s +z_\infty}{z_s (z_s-1)}\Bigg]
\nonumber
\\
&=-\frac{x_-}{x_+ -x_-}\Bigg[\frac{z_\infty z_{s}^2 - 2z_s z_\infty + z_\infty -z_{s}^2 z_\infty + z_{s}^2 +2z_s z_\infty - z_{s}^2 -z_\infty}{z_{s}^2 (z_s -1)^2}\Bigg] 
\nonumber
\\
&+\frac{2 x_+}{{x_+}^2 -1}\Bigg[\frac{z_\infty z_s -z_\infty -z_s z_\infty +z_s-z_s + z_\infty}{z_s (z_s-1)}\Bigg]=0
\end{split}
\end{equation}
The coefficient of $(\mathcal{C}/ \delta^2)$ term has the following expression,
\begin{align}
\textrm{Coefficient~of}~ (\mathcal{C}/\delta^2)&=\frac{2(z_{\infty}^4 + z_sz_{\infty}^4 - 2z_s z_{\infty}^3)}{z_{s}^3} 
- 2\frac{(-z_s +2z_\infty + 2z_s z_\infty-6 z_{\infty}^2 + 6z_{\infty}^3 - 2z_s z_{\infty}^3 -2 z_{\infty}^4 +z_s z_{\infty}^4)}{(z_s-1)^3} 
\nonumber
\\
&+\frac{2 (- z_{s}^4 + 2z_{s}^3 z_\infty + 2z_{s}^4 z_\infty -6 z_{s}^3z_{\infty}^2 -2 z_{\infty}^3 z_s + 6 z_{s}^2z_{\infty}^3 +z_{\infty}^4  - 2 z_sz_{\infty}^4 )}{z_{s}^3 (z_s -1)^3}
\nonumber
\\
&=2\Bigg[-z_{s}^4 + 2 z_{s}^3 z_\infty + 2 z_{s}^4 z_\infty -6 z_{s}^3z_{\infty}^2 +6z_{s}^2z_{\infty}^3 
\nonumber
\\
&- 2 z_s z_{\infty}^3 + z_{\infty}^4 -2z_s z_{\infty}^4 + z_{s}^4 -2 z_\infty z_{s}^3  - 2 z_{s}^4 z_\infty + 6 z_{s}^3z_{\infty}^2
-6 {z_s}^3{z_\infty}^3 + 2 {z_s}^4{z_\infty}^3 
\nonumber
\\
&+2 z_{s}^3z_{\infty}^4 -z_{s}^4 z_{\infty}^4 +(z_{s}^3 - 3 z_{s}^2  + 3z_s -1)(z_{\infty}^4 + z_sz_{\infty}^4 -2 z_s z_{\infty}^3)\Bigg] 
\nonumber
\\
&=2\Bigg[6 z_{s}^2 z_{\infty}^3  -2 z_s z_{\infty}^3 + z_{\infty}^4 - 2 z_sz_{\infty}^4 - 6 z_{s}^3 z_{\infty}^3 
+ 2 z_{s}^4 z_{\infty}^3 +2 z_{s}^3 z_{\infty}^4 - z_{s}^4 z_{\infty}^4 
\nonumber
\\
&+ z_{s}^3 z_{\infty}^4 -3 z_{s}^2 z_{\infty}^4 + 3 z_s z_{\infty}^4 - z_{\infty}^4 + z_{s}^4 z_{\infty}^4 
-3 z_{s}^3 z_{\infty}^4 + 3z_{s}^2 z_{\infty}^4 - z_{s}z_{\infty}^4 
\nonumber
\\
&-2 z_{s}^4 z_{\infty}^3 
+ 6 z_{s}^3 z_{\infty}^3 - 6 z_{s}^2 z_{\infty}^3 + 2 z_{s}^2 z_{\infty}^3\Bigg]=0
\end{align}
Finally the coefficient of $(\mathcal{D}/ 2 \delta^2)$ term involves two contributions, these yield,
\begin{align}
\textrm{coefficient~of}&~\{2x_+/(x_{+}^2 -1)\}=\frac{-3z_sz_{\infty}^2 + z_{\infty}^3 + 2 z_s z_{\infty}^3}{z_{s}^2} 
+\frac{(z_s - z_\infty)^3}{z_{s}^2 (z_s -1)^2}
\nonumber
\\
&+\frac{-z_s +3 z_\infty -6 z_{\infty}^2+ 3 z_s z_{\infty}^2 + 3z_{\infty}^3 - 2z_s z_{\infty}^3}{(z_s -1)^2} 
\nonumber
\\
&=\frac{(z_s - z_\infty)^3 + z_{s}^2(-z_s +3 z_\infty-6z_{\infty}^2 + 3 z_s z_\infty + 3z_{\infty}^3 -2 z_s z_{\infty}^3 )}{z_{s}^2 (z_s -1)^2} 
+ (z_s -1)^2 (- 3 z_s z_{\infty}^2 + z_{\infty}^3 + 2z_s z_{\infty}^3)
\nonumber
\\
&=\frac{3 z_sz_{\infty}^2 - z_{\infty}^3 -  6 z_{s}^2z_{\infty}^2  + 3 z_{s}^3z_{\infty}^2 +3 z_{s}^2 z_{\infty}^3 -2z_{s}^3z_{\infty}^3 - 3 z_{s}^3z_{\infty}^2 + 6 z_{s}^2 z_{\infty}^2 -3 {z_s}z_{\infty}^2}{{z_s}^2 (z_s -1)^2} 
\nonumber
\\
&+\frac{ z_{s}^2 z_{\infty}^3 -2 {z_s}z_{\infty}^3 + z_{\infty}^3 + 2 z_{s}^3z_{\infty}^3 -4 z_{s}^2z_{\infty}^3 + 2 {z_s}z_{\infty}^3}{z_{s}^2 (z_s -1)^2} =0
\\
\textrm{coefficient~of}&~ \{x_-/( x_+ -x_-)\}=\frac{3 z_s z_{\infty}^2 - 2 z_{\infty}^3 -2 z_s z_{\infty}^3}{z_{s}^3}  
+\frac{1+ z_s -6 z_\infty + 9 z_{\infty}^2 - 3 z_s z_{\infty}^2 -4 z_{\infty}^3 +2z_s z_{\infty}^3}{(z_s -1)^3}
\nonumber
\\
& + \frac{- z_{s}^3- z_{s}^4 + 6 z_{s}^3{z_\infty} + 3 {z_s}z_{\infty}^2 -9 z_{s}^2 z_{\infty}^2 
-2z_{\infty}^3 + 4 z_s z_{\infty}^3}{z_{s}^3(z_s -1)^3} 
\\
&=(z_s)^{-3}(z_s -1)^{-3} \Bigg[- z_{s}^3 - z_{s}^4 +6 z_{s}^3{z_\infty}+ 3 {z_s}z_{\infty}^2 - 9z_{s}^2z_{\infty}^2  -2 z_{\infty}^3 + 4 {z_s}z_{\infty}^3
\nonumber
\\
&+z_{s}^3 + z_{s}^4 -6 z_{s}^3 z_{\infty} +9z_{s}^3z_{\infty}^2 - 3z_{s}^4 z_{\infty}^2 - 4 z_{s}^3 z_{\infty}^3  + 2 z_{s}^4 z_{\infty}^3
\nonumber
\\
&+\left(z_{s}^3 - z_{s}^2 +3 z_s -1\right)\left(3z_s z_{\infty}^2 -2 z_{\infty}^3 - 2z_s z_{\infty}^3\right)\Bigg] = 0~.
\end{align}
This immediately suggests $\mathcal{M+N+P}=0$ and is the result we have used in the main text.

\bibliography{Reference_1,Reference_2}

\bibliographystyle{utphys1}
\end{document}